%% file: Main.tex
\documentclass[longauth]{aa}

\newcommand{\MG}[1]{\textcolor{black}{#1}}
\newcommand{\brown}[1]{\textcolor{black}{#1}}

\usepackage{euclid}
\usepackage{graphicx}
\usepackage{natbib}
\usepackage{scalerel}
\usepackage{float}
\usepackage{booktabs}
\usepackage{multirow}
\usepackage{adjustbox}
\usepackage{mathtools}

\usepackage[table]{xcolor}

\usepackage{txfonts}
\usepackage{glossaries}
\usepackage[pdfencoding=auto,psdextra]{hyperref}
\hypersetup{
    colorlinks=true,
    linkcolor=blue,
    filecolor=magenta,      
    urlcolor=blue,
    citecolor=blue
}
\makeatletter
\renewcommand*\aa@pageof{, page \thepage{} of \pageref*{LastPage}}
\makeatother

\usepackage[utf8]{inputenc}
\usepackage{lastpage}
\usepackage[switch, modulo]{lineno}

\newacronym{hod}{HOD}{halo occupation distribution}
\newacronym{fob}{FoB}{figure of bias}
\newacronym{dic}{DIC}{Deviance Information Criterion}
\newacronym{spt}{SPT}{Standard Perturbation Theory}
\newacronym{eft}{EFTofLSS}{Effective Field Theory of Large Scale Structure}
\newacronym{bao}{BAO}{baryon-acoustic-oscillation}

\renewcommand{\vec}{\mathbf}

\newcommand{\bdm}{\begin{displaymath}}
\newcommand{\edm}{\end{displaymath}}

\newcommand{\xv}{\vec{x}}
\newcommand{\xr}{\vec{r}}

\newcommand{\kv}{\vec{k}}

\newcommand{\rv}{\vec{r}}
\newcommand{\qv}{\vec{q}}

\newcommand{\Le}{{\mathcal L}}

\newcommand{\rmin}{r_{\rm min}}
\newcommand{\rminthree}{r_{\rm min}^{\rm \,3PCF}}
\newcommand{\rmintwo}{r_{\rm min}^{\rm \,2PCF}}
\newcommand{\rmax}{r_{\rm max}}
\newcommand{\etamin}{\eta_{\rm min}}

\newcommand{\Ha}{H$\alpha$\ }

\newcommand{\Pgg}{P_{\rm g}}
\newcommand{\Plin}{P_{\rm L}}
\newcommand{\Pl}{P_{\rm L}}

\newcommand{\bdtwod}{b_{\nabla^{\,2}\delta}}

\newcommand{\dd}{{\mathrm d}}
\newcommand{\mycomment}[1]{}

\newcommand{\Mpc}{\, h^{-1} \, {\rm Mpc}}

\newcommand{\kMpc}{\, h \, {\rm Mpc}^{-1}}

\begin{document}
%
%
   \title{\Euclid preparation. Full-shape modelling of 2-point and 3-point correlation functions in real space}

\newcommand{\orcid}[1]{} 
\author{Euclid Collaboration: M.~Guidi\orcid{0000-0001-9408-1101}\thanks{\email{massimo.guidi6@unibo.it}}\inst{\ref{aff1},\ref{aff2}}
\and A.~Veropalumbo\orcid{0000-0003-2387-1194}\inst{\ref{aff3},\ref{aff4},\ref{aff5}}
\and A.~Pugno\inst{\ref{aff6}}
\and M.~Moresco\orcid{0000-0002-7616-7136}\inst{\ref{aff7},\ref{aff2}}
\and E.~Sefusatti\orcid{0000-0003-0473-1567}\inst{\ref{aff8},\ref{aff9},\ref{aff10}}
\and C.~Porciani\orcid{0000-0002-7797-2508}\inst{\ref{aff6}}
\and E.~Branchini\orcid{0000-0002-0808-6908}\inst{\ref{aff5},\ref{aff4},\ref{aff3}}
\and M.-A.~Breton\inst{\ref{aff11},\ref{aff12},\ref{aff13}}
\and B.~Camacho~Quevedo\orcid{0000-0002-8789-4232}\inst{\ref{aff9},\ref{aff14},\ref{aff8},\ref{aff15},\ref{aff11}}
\and M.~Crocce\orcid{0000-0002-9745-6228}\inst{\ref{aff11},\ref{aff15}}
\and S.~de~la~Torre\inst{\ref{aff16}}
\and V.~Desjacques\orcid{0000-0003-2062-8172}\inst{\ref{aff17}}
\and A.~Eggemeier\orcid{0000-0002-1841-8910}\inst{\ref{aff6}}
\and A.~Farina\orcid{0009-0000-3420-929X}\inst{\ref{aff5},\ref{aff3},\ref{aff4}}
\and M.~K{\"a}rcher\orcid{0000-0001-5868-647X}\inst{\ref{aff16},\ref{aff18},\ref{aff19}}
\and D.~Linde\orcid{0000-0001-7192-1067}\inst{\ref{aff20}}
\and M.~Marinucci\orcid{0000-0003-1159-3756}\inst{\ref{aff21},\ref{aff22}}
\and A.~Moradinezhad~Dizgah\orcid{0000-0001-8841-9989}\inst{\ref{aff23}}
\and C.~Moretti\orcid{0000-0003-3314-8936}\inst{\ref{aff14},\ref{aff24},\ref{aff8},\ref{aff9},\ref{aff10}}
\and K.~Pardede\orcid{0000-0002-7728-8220}\inst{\ref{aff20}}
\and A.~Pezzotta\orcid{0000-0003-0726-2268}\inst{\ref{aff25},\ref{aff26}}
\and E.~Sarpa\orcid{0000-0002-1256-655X}\inst{\ref{aff14},\ref{aff24},\ref{aff10}}
\and A.~Amara\inst{\ref{aff27}}
\and S.~Andreon\orcid{0000-0002-2041-8784}\inst{\ref{aff3}}
\and N.~Auricchio\orcid{0000-0003-4444-8651}\inst{\ref{aff2}}
\and C.~Baccigalupi\orcid{0000-0002-8211-1630}\inst{\ref{aff9},\ref{aff8},\ref{aff10},\ref{aff14}}
\and D.~Bagot\inst{\ref{aff28}}
\and M.~Baldi\orcid{0000-0003-4145-1943}\inst{\ref{aff1},\ref{aff2},\ref{aff29}}
\and S.~Bardelli\orcid{0000-0002-8900-0298}\inst{\ref{aff2}}
\and P.~Battaglia\orcid{0000-0002-7337-5909}\inst{\ref{aff2}}
\and A.~Biviano\orcid{0000-0002-0857-0732}\inst{\ref{aff8},\ref{aff9}}
\and M.~Brescia\orcid{0000-0001-9506-5680}\inst{\ref{aff30},\ref{aff31}}
\and S.~Camera\orcid{0000-0003-3399-3574}\inst{\ref{aff32},\ref{aff33},\ref{aff34}}
\and G.~Ca\~nas-Herrera\orcid{0000-0003-2796-2149}\inst{\ref{aff35},\ref{aff36},\ref{aff37}}
\and V.~Capobianco\orcid{0000-0002-3309-7692}\inst{\ref{aff34}}
\and C.~Carbone\orcid{0000-0003-0125-3563}\inst{\ref{aff38}}
\and V.~F.~Cardone\inst{\ref{aff39},\ref{aff40}}
\and J.~Carretero\orcid{0000-0002-3130-0204}\inst{\ref{aff41},\ref{aff42}}
\and M.~Castellano\orcid{0000-0001-9875-8263}\inst{\ref{aff39}}
\and G.~Castignani\orcid{0000-0001-6831-0687}\inst{\ref{aff2}}
\and S.~Cavuoti\orcid{0000-0002-3787-4196}\inst{\ref{aff31},\ref{aff43}}
\and K.~C.~Chambers\orcid{0000-0001-6965-7789}\inst{\ref{aff44}}
\and A.~Cimatti\inst{\ref{aff45}}
\and C.~Colodro-Conde\inst{\ref{aff46}}
\and G.~Congedo\orcid{0000-0003-2508-0046}\inst{\ref{aff47}}
\and L.~Conversi\orcid{0000-0002-6710-8476}\inst{\ref{aff48},\ref{aff49}}
\and Y.~Copin\orcid{0000-0002-5317-7518}\inst{\ref{aff50}}
\and F.~Courbin\orcid{0000-0003-0758-6510}\inst{\ref{aff51},\ref{aff52}}
\and H.~M.~Courtois\orcid{0000-0003-0509-1776}\inst{\ref{aff53}}
\and A.~Da~Silva\orcid{0000-0002-6385-1609}\inst{\ref{aff54},\ref{aff55}}
\and H.~Degaudenzi\orcid{0000-0002-5887-6799}\inst{\ref{aff56}}
\and G.~De~Lucia\orcid{0000-0002-6220-9104}\inst{\ref{aff8}}
\and H.~Dole\orcid{0000-0002-9767-3839}\inst{\ref{aff57}}
\and M.~Douspis\orcid{0000-0003-4203-3954}\inst{\ref{aff57}}
\and F.~Dubath\orcid{0000-0002-6533-2810}\inst{\ref{aff56}}
\and X.~Dupac\inst{\ref{aff49}}
\and S.~Dusini\orcid{0000-0002-1128-0664}\inst{\ref{aff22}}
\and S.~Escoffier\orcid{0000-0002-2847-7498}\inst{\ref{aff58}}
\and M.~Farina\orcid{0000-0002-3089-7846}\inst{\ref{aff59}}
\and R.~Farinelli\inst{\ref{aff2}}
\and F.~Faustini\orcid{0000-0001-6274-5145}\inst{\ref{aff39},\ref{aff60}}
\and S.~Ferriol\inst{\ref{aff50}}
\and F.~Finelli\orcid{0000-0002-6694-3269}\inst{\ref{aff2},\ref{aff61}}
\and P.~Fosalba\orcid{0000-0002-1510-5214}\inst{\ref{aff15},\ref{aff11}}
\and S.~Fotopoulou\orcid{0000-0002-9686-254X}\inst{\ref{aff62}}
\and M.~Frailis\orcid{0000-0002-7400-2135}\inst{\ref{aff8}}
\and E.~Franceschi\orcid{0000-0002-0585-6591}\inst{\ref{aff2}}
\and M.~Fumana\orcid{0000-0001-6787-5950}\inst{\ref{aff38}}
\and S.~Galeotta\orcid{0000-0002-3748-5115}\inst{\ref{aff8}}
\and B.~Gillis\orcid{0000-0002-4478-1270}\inst{\ref{aff47}}
\and C.~Giocoli\orcid{0000-0002-9590-7961}\inst{\ref{aff2},\ref{aff29}}
\and J.~Gracia-Carpio\inst{\ref{aff26}}
\and A.~Grazian\orcid{0000-0002-5688-0663}\inst{\ref{aff63}}
\and F.~Grupp\inst{\ref{aff26},\ref{aff64}}
\and L.~Guzzo\orcid{0000-0001-8264-5192}\inst{\ref{aff19},\ref{aff3},\ref{aff65}}
\and S.~V.~H.~Haugan\orcid{0000-0001-9648-7260}\inst{\ref{aff66}}
\and W.~Holmes\inst{\ref{aff67}}
\and F.~Hormuth\inst{\ref{aff68}}
\and A.~Hornstrup\orcid{0000-0002-3363-0936}\inst{\ref{aff69},\ref{aff70}}
\and K.~Jahnke\orcid{0000-0003-3804-2137}\inst{\ref{aff71}}
\and M.~Jhabvala\inst{\ref{aff72}}
\and B.~Joachimi\orcid{0000-0001-7494-1303}\inst{\ref{aff73}}
\and E.~Keih\"anen\orcid{0000-0003-1804-7715}\inst{\ref{aff74}}
\and S.~Kermiche\orcid{0000-0002-0302-5735}\inst{\ref{aff58}}
\and A.~Kiessling\orcid{0000-0002-2590-1273}\inst{\ref{aff67}}
\and B.~Kubik\orcid{0009-0006-5823-4880}\inst{\ref{aff50}}
\and M.~K\"ummel\orcid{0000-0003-2791-2117}\inst{\ref{aff64}}
\and M.~Kunz\orcid{0000-0002-3052-7394}\inst{\ref{aff75}}
\and H.~Kurki-Suonio\orcid{0000-0002-4618-3063}\inst{\ref{aff76},\ref{aff77}}
\and A.~M.~C.~Le~Brun\orcid{0000-0002-0936-4594}\inst{\ref{aff78}}
\and S.~Ligori\orcid{0000-0003-4172-4606}\inst{\ref{aff34}}
\and P.~B.~Lilje\orcid{0000-0003-4324-7794}\inst{\ref{aff66}}
\and V.~Lindholm\orcid{0000-0003-2317-5471}\inst{\ref{aff76},\ref{aff77}}
\and I.~Lloro\orcid{0000-0001-5966-1434}\inst{\ref{aff79}}
\and G.~Mainetti\orcid{0000-0003-2384-2377}\inst{\ref{aff80}}
\and D.~Maino\inst{\ref{aff19},\ref{aff38},\ref{aff65}}
\and E.~Maiorano\orcid{0000-0003-2593-4355}\inst{\ref{aff2}}
\and O.~Mansutti\orcid{0000-0001-5758-4658}\inst{\ref{aff8}}
\and S.~Marcin\inst{\ref{aff81}}
\and O.~Marggraf\orcid{0000-0001-7242-3852}\inst{\ref{aff6}}
\and K.~Markovic\orcid{0000-0001-6764-073X}\inst{\ref{aff67}}
\and M.~Martinelli\orcid{0000-0002-6943-7732}\inst{\ref{aff39},\ref{aff40}}
\and N.~Martinet\orcid{0000-0003-2786-7790}\inst{\ref{aff16}}
\and F.~Marulli\orcid{0000-0002-8850-0303}\inst{\ref{aff7},\ref{aff2},\ref{aff29}}
\and R.~Massey\orcid{0000-0002-6085-3780}\inst{\ref{aff82}}
\and E.~Medinaceli\orcid{0000-0002-4040-7783}\inst{\ref{aff2}}
\and S.~Mei\orcid{0000-0002-2849-559X}\inst{\ref{aff83},\ref{aff84}}
\and M.~Melchior\inst{\ref{aff85}}
\and Y.~Mellier\inst{\ref{aff86},\ref{aff87}}
\and M.~Meneghetti\orcid{0000-0003-1225-7084}\inst{\ref{aff2},\ref{aff29}}
\and E.~Merlin\orcid{0000-0001-6870-8900}\inst{\ref{aff39}}
\and G.~Meylan\inst{\ref{aff88}}
\and A.~Mora\orcid{0000-0002-1922-8529}\inst{\ref{aff89}}
\and B.~Morin\inst{\ref{aff13}}
\and L.~Moscardini\orcid{0000-0002-3473-6716}\inst{\ref{aff7},\ref{aff2},\ref{aff29}}
\and E.~Munari\orcid{0000-0002-1751-5946}\inst{\ref{aff8},\ref{aff9}}
\and C.~Neissner\orcid{0000-0001-8524-4968}\inst{\ref{aff90},\ref{aff42}}
\and S.-M.~Niemi\orcid{0009-0005-0247-0086}\inst{\ref{aff35}}
\and C.~Padilla\orcid{0000-0001-7951-0166}\inst{\ref{aff90}}
\and S.~Paltani\orcid{0000-0002-8108-9179}\inst{\ref{aff56}}
\and F.~Pasian\orcid{0000-0002-4869-3227}\inst{\ref{aff8}}
\and K.~Pedersen\inst{\ref{aff91}}
\and W.~J.~Percival\orcid{0000-0002-0644-5727}\inst{\ref{aff92},\ref{aff93},\ref{aff94}}
\and V.~Pettorino\inst{\ref{aff35}}
\and S.~Pires\orcid{0000-0002-0249-2104}\inst{\ref{aff13}}
\and G.~Polenta\orcid{0000-0003-4067-9196}\inst{\ref{aff60}}
\and M.~Poncet\inst{\ref{aff28}}
\and L.~A.~Popa\inst{\ref{aff95}}
\and F.~Raison\orcid{0000-0002-7819-6918}\inst{\ref{aff26}}
\and R.~Rebolo\orcid{0000-0003-3767-7085}\inst{\ref{aff46},\ref{aff96},\ref{aff97}}
\and A.~Renzi\orcid{0000-0001-9856-1970}\inst{\ref{aff21},\ref{aff22}}
\and J.~Rhodes\orcid{0000-0002-4485-8549}\inst{\ref{aff67}}
\and G.~Riccio\inst{\ref{aff31}}
\and E.~Romelli\orcid{0000-0003-3069-9222}\inst{\ref{aff8}}
\and M.~Roncarelli\orcid{0000-0001-9587-7822}\inst{\ref{aff2}}
\and R.~Saglia\orcid{0000-0003-0378-7032}\inst{\ref{aff64},\ref{aff26}}
\and Z.~Sakr\orcid{0000-0002-4823-3757}\inst{\ref{aff98},\ref{aff99},\ref{aff100}}
\and A.~G.~S\'anchez\orcid{0000-0003-1198-831X}\inst{\ref{aff26}}
\and D.~Sapone\orcid{0000-0001-7089-4503}\inst{\ref{aff101}}
\and B.~Sartoris\orcid{0000-0003-1337-5269}\inst{\ref{aff64},\ref{aff8}}
\and J.~A.~Schewtschenko\orcid{0000-0002-4913-6393}\inst{\ref{aff47}}
\and P.~Schneider\orcid{0000-0001-8561-2679}\inst{\ref{aff6}}
\and T.~Schrabback\orcid{0000-0002-6987-7834}\inst{\ref{aff102}}
\and M.~Scodeggio\inst{\ref{aff38}}
\and A.~Secroun\orcid{0000-0003-0505-3710}\inst{\ref{aff58}}
\and G.~Seidel\orcid{0000-0003-2907-353X}\inst{\ref{aff71}}
\and M.~Seiffert\orcid{0000-0002-7536-9393}\inst{\ref{aff67}}
\and S.~Serrano\orcid{0000-0002-0211-2861}\inst{\ref{aff15},\ref{aff103},\ref{aff11}}
\and P.~Simon\inst{\ref{aff6}}
\and C.~Sirignano\orcid{0000-0002-0995-7146}\inst{\ref{aff21},\ref{aff22}}
\and G.~Sirri\orcid{0000-0003-2626-2853}\inst{\ref{aff29}}
\and A.~Spurio~Mancini\orcid{0000-0001-5698-0990}\inst{\ref{aff104}}
\and L.~Stanco\orcid{0000-0002-9706-5104}\inst{\ref{aff22}}
\and J.~Steinwagner\orcid{0000-0001-7443-1047}\inst{\ref{aff26}}
\and P.~Tallada-Cresp\'{i}\orcid{0000-0002-1336-8328}\inst{\ref{aff41},\ref{aff42}}
\and D.~Tavagnacco\orcid{0000-0001-7475-9894}\inst{\ref{aff8}}
\and A.~N.~Taylor\inst{\ref{aff47}}
\and I.~Tereno\orcid{0000-0002-4537-6218}\inst{\ref{aff54},\ref{aff105}}
\and N.~Tessore\orcid{0000-0002-9696-7931}\inst{\ref{aff73}}
\and S.~Toft\orcid{0000-0003-3631-7176}\inst{\ref{aff106},\ref{aff107}}
\and R.~Toledo-Moreo\orcid{0000-0002-2997-4859}\inst{\ref{aff108}}
\and F.~Torradeflot\orcid{0000-0003-1160-1517}\inst{\ref{aff42},\ref{aff41}}
\and A.~Tsyganov\inst{\ref{aff109}}
\and I.~Tutusaus\orcid{0000-0002-3199-0399}\inst{\ref{aff99}}
\and L.~Valenziano\orcid{0000-0002-1170-0104}\inst{\ref{aff2},\ref{aff61}}
\and J.~Valiviita\orcid{0000-0001-6225-3693}\inst{\ref{aff76},\ref{aff77}}
\and T.~Vassallo\orcid{0000-0001-6512-6358}\inst{\ref{aff64},\ref{aff8}}
\and G.~Verdoes~Kleijn\orcid{0000-0001-5803-2580}\inst{\ref{aff110}}
\and Y.~Wang\orcid{0000-0002-4749-2984}\inst{\ref{aff111}}
\and J.~Weller\orcid{0000-0002-8282-2010}\inst{\ref{aff64},\ref{aff26}}
\and G.~Zamorani\orcid{0000-0002-2318-301X}\inst{\ref{aff2}}
\and F.~M.~Zerbi\inst{\ref{aff3}}
\and E.~Zucca\orcid{0000-0002-5845-8132}\inst{\ref{aff2}}
\and V.~Allevato\orcid{0000-0001-7232-5152}\inst{\ref{aff31}}
\and M.~Ballardini\orcid{0000-0003-4481-3559}\inst{\ref{aff112},\ref{aff113},\ref{aff2}}
\and M.~Bolzonella\orcid{0000-0003-3278-4607}\inst{\ref{aff2}}
\and E.~Bozzo\orcid{0000-0002-8201-1525}\inst{\ref{aff56}}
\and C.~Burigana\orcid{0000-0002-3005-5796}\inst{\ref{aff114},\ref{aff61}}
\and R.~Cabanac\orcid{0000-0001-6679-2600}\inst{\ref{aff99}}
\and M.~Calabrese\orcid{0000-0002-2637-2422}\inst{\ref{aff115},\ref{aff38}}
\and A.~Cappi\inst{\ref{aff2},\ref{aff116}}
\and D.~Di~Ferdinando\inst{\ref{aff29}}
\and J.~A.~Escartin~Vigo\inst{\ref{aff26}}
\and L.~Gabarra\orcid{0000-0002-8486-8856}\inst{\ref{aff117}}
\and J.~Mart\'{i}n-Fleitas\orcid{0000-0002-8594-569X}\inst{\ref{aff118}}
\and S.~Matthew\orcid{0000-0001-8448-1697}\inst{\ref{aff47}}
\and M.~Maturi\orcid{0000-0002-3517-2422}\inst{\ref{aff98},\ref{aff119}}
\and N.~Mauri\orcid{0000-0001-8196-1548}\inst{\ref{aff45},\ref{aff29}}
\and R.~B.~Metcalf\orcid{0000-0003-3167-2574}\inst{\ref{aff7},\ref{aff2}}
\and A.~A.~Nucita\inst{\ref{aff120},\ref{aff121},\ref{aff122}}
\and M.~P\"ontinen\orcid{0000-0001-5442-2530}\inst{\ref{aff76}}
\and I.~Risso\orcid{0000-0003-2525-7761}\inst{\ref{aff123}}
\and V.~Scottez\orcid{0009-0008-3864-940X}\inst{\ref{aff86},\ref{aff124}}
\and M.~Sereno\orcid{0000-0003-0302-0325}\inst{\ref{aff2},\ref{aff29}}
\and M.~Tenti\orcid{0000-0002-4254-5901}\inst{\ref{aff29}}
\and M.~Viel\orcid{0000-0002-2642-5707}\inst{\ref{aff9},\ref{aff8},\ref{aff14},\ref{aff10},\ref{aff24}}
\and M.~Wiesmann\orcid{0009-0000-8199-5860}\inst{\ref{aff66}}
\and Y.~Akrami\orcid{0000-0002-2407-7956}\inst{\ref{aff125},\ref{aff126}}
\and I.~T.~Andika\orcid{0000-0001-6102-9526}\inst{\ref{aff127},\ref{aff128}}
\and S.~Anselmi\orcid{0000-0002-3579-9583}\inst{\ref{aff22},\ref{aff21},\ref{aff12}}
\and M.~Archidiacono\orcid{0000-0003-4952-9012}\inst{\ref{aff19},\ref{aff65}}
\and F.~Atrio-Barandela\orcid{0000-0002-2130-2513}\inst{\ref{aff129}}
\and A.~Balaguera-Antolinez\orcid{0000-0001-5028-3035}\inst{\ref{aff46},\ref{aff130}}
\and D.~Bertacca\orcid{0000-0002-2490-7139}\inst{\ref{aff21},\ref{aff63},\ref{aff22}}
\and M.~Bethermin\orcid{0000-0002-3915-2015}\inst{\ref{aff131}}
\and L.~Blot\orcid{0000-0002-9622-7167}\inst{\ref{aff132},\ref{aff78}}
\and H.~B\"ohringer\orcid{0000-0001-8241-4204}\inst{\ref{aff26},\ref{aff133},\ref{aff134}}
\and S.~Borgani\orcid{0000-0001-6151-6439}\inst{\ref{aff135},\ref{aff9},\ref{aff8},\ref{aff10},\ref{aff24}}
\and M.~L.~Brown\orcid{0000-0002-0370-8077}\inst{\ref{aff136}}
\and S.~Bruton\orcid{0000-0002-6503-5218}\inst{\ref{aff137}}
\and A.~Calabro\orcid{0000-0003-2536-1614}\inst{\ref{aff39}}
\and F.~Caro\inst{\ref{aff39}}
\and C.~S.~Carvalho\inst{\ref{aff105}}
\and T.~Castro\orcid{0000-0002-6292-3228}\inst{\ref{aff8},\ref{aff10},\ref{aff9},\ref{aff24}}
\and F.~Cogato\orcid{0000-0003-4632-6113}\inst{\ref{aff7},\ref{aff2}}
\and S.~Conseil\orcid{0000-0002-3657-4191}\inst{\ref{aff50}}
\and S.~Contarini\orcid{0000-0002-9843-723X}\inst{\ref{aff26}}
\and A.~R.~Cooray\orcid{0000-0002-3892-0190}\inst{\ref{aff138}}
\and O.~Cucciati\orcid{0000-0002-9336-7551}\inst{\ref{aff2}}
\and S.~Davini\orcid{0000-0003-3269-1718}\inst{\ref{aff4}}
\and F.~De~Paolis\orcid{0000-0001-6460-7563}\inst{\ref{aff120},\ref{aff121},\ref{aff122}}
\and G.~Desprez\orcid{0000-0001-8325-1742}\inst{\ref{aff110}}
\and A.~D\'iaz-S\'anchez\orcid{0000-0003-0748-4768}\inst{\ref{aff139}}
\and J.~J.~Diaz\orcid{0000-0003-2101-1078}\inst{\ref{aff46}}
\and S.~Di~Domizio\orcid{0000-0003-2863-5895}\inst{\ref{aff5},\ref{aff4}}
\and J.~M.~Diego\orcid{0000-0001-9065-3926}\inst{\ref{aff140}}
\and P.~Dimauro\orcid{0000-0001-7399-2854}\inst{\ref{aff39},\ref{aff141}}
\and A.~Enia\orcid{0000-0002-0200-2857}\inst{\ref{aff1},\ref{aff2}}
\and Y.~Fang\inst{\ref{aff64}}
\and A.~G.~Ferrari\orcid{0009-0005-5266-4110}\inst{\ref{aff29}}
\and P.~G.~Ferreira\orcid{0000-0002-3021-2851}\inst{\ref{aff117}}
\and A.~Finoguenov\orcid{0000-0002-4606-5403}\inst{\ref{aff76}}
\and A.~Franco\orcid{0000-0002-4761-366X}\inst{\ref{aff121},\ref{aff120},\ref{aff122}}
\and K.~Ganga\orcid{0000-0001-8159-8208}\inst{\ref{aff83}}
\and J.~Garc\'ia-Bellido\orcid{0000-0002-9370-8360}\inst{\ref{aff125}}
\and T.~Gasparetto\orcid{0000-0002-7913-4866}\inst{\ref{aff8}}
\and V.~Gautard\inst{\ref{aff142}}
\and E.~Gaztanaga\orcid{0000-0001-9632-0815}\inst{\ref{aff11},\ref{aff15},\ref{aff143}}
\and F.~Giacomini\orcid{0000-0002-3129-2814}\inst{\ref{aff29}}
\and F.~Gianotti\orcid{0000-0003-4666-119X}\inst{\ref{aff2}}
\and G.~Gozaliasl\orcid{0000-0002-0236-919X}\inst{\ref{aff144},\ref{aff76}}
\and C.~M.~Gutierrez\orcid{0000-0001-7854-783X}\inst{\ref{aff145}}
\and C.~Hern\'andez-Monteagudo\orcid{0000-0001-5471-9166}\inst{\ref{aff97},\ref{aff46}}
\and H.~Hildebrandt\orcid{0000-0002-9814-3338}\inst{\ref{aff146}}
\and J.~Hjorth\orcid{0000-0002-4571-2306}\inst{\ref{aff91}}
\and S.~Joudaki\orcid{0000-0001-8820-673X}\inst{\ref{aff41}}
\and J.~J.~E.~Kajava\orcid{0000-0002-3010-8333}\inst{\ref{aff147},\ref{aff148}}
\and Y.~Kang\orcid{0009-0000-8588-7250}\inst{\ref{aff56}}
\and V.~Kansal\orcid{0000-0002-4008-6078}\inst{\ref{aff149},\ref{aff150}}
\and D.~Karagiannis\orcid{0000-0002-4927-0816}\inst{\ref{aff112},\ref{aff151}}
\and K.~Kiiveri\inst{\ref{aff74}}
\and C.~C.~Kirkpatrick\inst{\ref{aff74}}
\and S.~Kruk\orcid{0000-0001-8010-8879}\inst{\ref{aff49}}
\and M.~Lattanzi\orcid{0000-0003-1059-2532}\inst{\ref{aff113}}
\and L.~Legrand\orcid{0000-0003-0610-5252}\inst{\ref{aff152},\ref{aff153}}
\and M.~Lembo\orcid{0000-0002-5271-5070}\inst{\ref{aff87},\ref{aff113}}
\and F.~Lepori\orcid{0009-0000-5061-7138}\inst{\ref{aff154}}
\and G.~Leroy\orcid{0009-0004-2523-4425}\inst{\ref{aff155},\ref{aff82}}
\and G.~F.~Lesci\orcid{0000-0002-4607-2830}\inst{\ref{aff7},\ref{aff2}}
\and J.~Lesgourgues\orcid{0000-0001-7627-353X}\inst{\ref{aff156}}
\and L.~Leuzzi\orcid{0009-0006-4479-7017}\inst{\ref{aff2}}
\and T.~I.~Liaudat\orcid{0000-0002-9104-314X}\inst{\ref{aff157}}
\and A.~Loureiro\orcid{0000-0002-4371-0876}\inst{\ref{aff158},\ref{aff159}}
\and J.~Macias-Perez\orcid{0000-0002-5385-2763}\inst{\ref{aff160}}
\and G.~Maggio\orcid{0000-0003-4020-4836}\inst{\ref{aff8}}
\and M.~Magliocchetti\orcid{0000-0001-9158-4838}\inst{\ref{aff59}}
\and F.~Mannucci\orcid{0000-0002-4803-2381}\inst{\ref{aff161}}
\and R.~Maoli\orcid{0000-0002-6065-3025}\inst{\ref{aff162},\ref{aff39}}
\and C.~J.~A.~P.~Martins\orcid{0000-0002-4886-9261}\inst{\ref{aff163},\ref{aff164}}
\and L.~Maurin\orcid{0000-0002-8406-0857}\inst{\ref{aff57}}
\and M.~Miluzio\inst{\ref{aff49},\ref{aff165}}
\and P.~Monaco\orcid{0000-0003-2083-7564}\inst{\ref{aff135},\ref{aff8},\ref{aff10},\ref{aff9}}
\and G.~Morgante\inst{\ref{aff2}}
\and S.~Nadathur\orcid{0000-0001-9070-3102}\inst{\ref{aff143}}
\and K.~Naidoo\orcid{0000-0002-9182-1802}\inst{\ref{aff143}}
\and A.~Navarro-Alsina\orcid{0000-0002-3173-2592}\inst{\ref{aff6}}
\and S.~Nesseris\orcid{0000-0002-0567-0324}\inst{\ref{aff125}}
\and L.~Pagano\orcid{0000-0003-1820-5998}\inst{\ref{aff112},\ref{aff113}}
\and F.~Passalacqua\orcid{0000-0002-8606-4093}\inst{\ref{aff21},\ref{aff22}}
\and K.~Paterson\orcid{0000-0001-8340-3486}\inst{\ref{aff71}}
\and L.~Patrizii\inst{\ref{aff29}}
\and A.~Pisani\orcid{0000-0002-6146-4437}\inst{\ref{aff58}}
\and D.~Potter\orcid{0000-0002-0757-5195}\inst{\ref{aff154}}
\and S.~Quai\orcid{0000-0002-0449-8163}\inst{\ref{aff7},\ref{aff2}}
\and M.~Radovich\orcid{0000-0002-3585-866X}\inst{\ref{aff63}}
\and P.~Reimberg\orcid{0000-0003-3410-0280}\inst{\ref{aff86}}
\and P.-F.~Rocci\inst{\ref{aff57}}
\and G.~Rodighiero\orcid{0000-0002-9415-2296}\inst{\ref{aff21},\ref{aff63}}
\and S.~Sacquegna\orcid{0000-0002-8433-6630}\inst{\ref{aff120},\ref{aff121},\ref{aff122}}
\and M.~Sahl\'en\orcid{0000-0003-0973-4804}\inst{\ref{aff166}}
\and D.~B.~Sanders\orcid{0000-0002-1233-9998}\inst{\ref{aff44}}
\and A.~Schneider\orcid{0000-0001-7055-8104}\inst{\ref{aff154}}
\and D.~Sciotti\orcid{0009-0008-4519-2620}\inst{\ref{aff39},\ref{aff40}}
\and E.~Sellentin\inst{\ref{aff167},\ref{aff37}}
\and L.~C.~Smith\orcid{0000-0002-3259-2771}\inst{\ref{aff168}}
\and J.~G.~Sorce\orcid{0000-0002-2307-2432}\inst{\ref{aff169},\ref{aff57}}
\and K.~Tanidis\orcid{0000-0001-9843-5130}\inst{\ref{aff117}}
\and C.~Tao\orcid{0000-0001-7961-8177}\inst{\ref{aff58}}
\and G.~Testera\inst{\ref{aff4}}
\and R.~Teyssier\orcid{0000-0001-7689-0933}\inst{\ref{aff170}}
\and S.~Tosi\orcid{0000-0002-7275-9193}\inst{\ref{aff5},\ref{aff4},\ref{aff3}}
\and A.~Troja\orcid{0000-0003-0239-4595}\inst{\ref{aff21},\ref{aff22}}
\and M.~Tucci\inst{\ref{aff56}}
\and C.~Valieri\inst{\ref{aff29}}
\and A.~Venhola\orcid{0000-0001-6071-4564}\inst{\ref{aff171}}
\and D.~Vergani\orcid{0000-0003-0898-2216}\inst{\ref{aff2}}
\and F.~Vernizzi\orcid{0000-0003-3426-2802}\inst{\ref{aff172}}
\and G.~Verza\orcid{0000-0002-1886-8348}\inst{\ref{aff173}}
\and P.~Vielzeuf\orcid{0000-0003-2035-9339}\inst{\ref{aff58}}
\and N.~A.~Walton\orcid{0000-0003-3983-8778}\inst{\ref{aff168}}}
										   
\institute{Dipartimento di Fisica e Astronomia, Universit\`a di Bologna, Via Gobetti 93/2, 40129 Bologna, Italy\label{aff1}
\and
INAF-Osservatorio di Astrofisica e Scienza dello Spazio di Bologna, Via Piero Gobetti 93/3, 40129 Bologna, Italy\label{aff2}
\and
INAF-Osservatorio Astronomico di Brera, Via Brera 28, 20122 Milano, Italy\label{aff3}
\and
INFN-Sezione di Genova, Via Dodecaneso 33, 16146, Genova, Italy\label{aff4}
\and
Dipartimento di Fisica, Universit\`a di Genova, Via Dodecaneso 33, 16146, Genova, Italy\label{aff5}
\and
Universit\"at Bonn, Argelander-Institut f\"ur Astronomie, Auf dem H\"ugel 71, 53121 Bonn, Germany\label{aff6}
\and
Dipartimento di Fisica e Astronomia "Augusto Righi" - Alma Mater Studiorum Universit\`a di Bologna, via Piero Gobetti 93/2, 40129 Bologna, Italy\label{aff7}
\and
INAF-Osservatorio Astronomico di Trieste, Via G. B. Tiepolo 11, 34143 Trieste, Italy\label{aff8}
\and
IFPU, Institute for Fundamental Physics of the Universe, via Beirut 2, 34151 Trieste, Italy\label{aff9}
\and
INFN, Sezione di Trieste, Via Valerio 2, 34127 Trieste TS, Italy\label{aff10}
\and
Institute of Space Sciences (ICE, CSIC), Campus UAB, Carrer de Can Magrans, s/n, 08193 Barcelona, Spain\label{aff11}
\and
Laboratoire Univers et Th\'eorie, Observatoire de Paris, Universit\'e PSL, Universit\'e Paris Cit\'e, CNRS, 92190 Meudon, France\label{aff12}
\and
Universit\'e Paris-Saclay, Universit\'e Paris Cit\'e, CEA, CNRS, AIM, 91191, Gif-sur-Yvette, France\label{aff13}
\and
SISSA, International School for Advanced Studies, Via Bonomea 265, 34136 Trieste TS, Italy\label{aff14}
\and
Institut d'Estudis Espacials de Catalunya (IEEC),  Edifici RDIT, Campus UPC, 08860 Castelldefels, Barcelona, Spain\label{aff15}
\and
Aix-Marseille Universit\'e, CNRS, CNES, LAM, Marseille, France\label{aff16}
\and
Technion Israel Institute of Technology, Israel\label{aff17}
\and
Aix-Marseille Universit\'e, Universit\'e de Toulon, CNRS, CPT, Marseille, France\label{aff18}
\and
Dipartimento di Fisica "Aldo Pontremoli", Universit\`a degli Studi di Milano, Via Celoria 16, 20133 Milano, Italy\label{aff19}
\and
INFN Gruppo Collegato di Parma, Viale delle Scienze 7/A 43124 Parma, Italy\label{aff20}
\and
Dipartimento di Fisica e Astronomia "G. Galilei", Universit\`a di Padova, Via Marzolo 8, 35131 Padova, Italy\label{aff21}
\and
INFN-Padova, Via Marzolo 8, 35131 Padova, Italy\label{aff22}
\and
Laboratoire d'Annecy-le-Vieux de Physique Theorique, CNRS \& Universite Savoie Mont Blanc, 9 Chemin de Bellevue, BP 110, Annecy-le-Vieux, 74941 ANNECY Cedex, France\label{aff23}
\and
ICSC - Centro Nazionale di Ricerca in High Performance Computing, Big Data e Quantum Computing, Via Magnanelli 2, Bologna, Italy\label{aff24}
\and
INAF - Osservatorio Astronomico di Brera, via Emilio Bianchi 46, 23807 Merate, Italy\label{aff25}
\and
Max Planck Institute for Extraterrestrial Physics, Giessenbachstr. 1, 85748 Garching, Germany\label{aff26}
\and
School of Mathematics and Physics, University of Surrey, Guildford, Surrey, GU2 7XH, UK\label{aff27}
\and
Centre National d'Etudes Spatiales -- Centre spatial de Toulouse, 18 avenue Edouard Belin, 31401 Toulouse Cedex 9, France\label{aff28}
\and
INFN-Sezione di Bologna, Viale Berti Pichat 6/2, 40127 Bologna, Italy\label{aff29}
\and
Department of Physics "E. Pancini", University Federico II, Via Cinthia 6, 80126, Napoli, Italy\label{aff30}
\and
INAF-Osservatorio Astronomico di Capodimonte, Via Moiariello 16, 80131 Napoli, Italy\label{aff31}
\and
Dipartimento di Fisica, Universit\`a degli Studi di Torino, Via P. Giuria 1, 10125 Torino, Italy\label{aff32}
\and
INFN-Sezione di Torino, Via P. Giuria 1, 10125 Torino, Italy\label{aff33}
\and
INAF-Osservatorio Astrofisico di Torino, Via Osservatorio 20, 10025 Pino Torinese (TO), Italy\label{aff34}
\and
European Space Agency/ESTEC, Keplerlaan 1, 2201 AZ Noordwijk, The Netherlands\label{aff35}
\and
Institute Lorentz, Leiden University, Niels Bohrweg 2, 2333 CA Leiden, The Netherlands\label{aff36}
\and
Leiden Observatory, Leiden University, Einsteinweg 55, 2333 CC Leiden, The Netherlands\label{aff37}
\and
INAF-IASF Milano, Via Alfonso Corti 12, 20133 Milano, Italy\label{aff38}
\and
INAF-Osservatorio Astronomico di Roma, Via Frascati 33, 00078 Monteporzio Catone, Italy\label{aff39}
\and
INFN-Sezione di Roma, Piazzale Aldo Moro, 2 - c/o Dipartimento di Fisica, Edificio G. Marconi, 00185 Roma, Italy\label{aff40}
\and
Centro de Investigaciones Energ\'eticas, Medioambientales y Tecnol\'ogicas (CIEMAT), Avenida Complutense 40, 28040 Madrid, Spain\label{aff41}
\and
Port d'Informaci\'{o} Cient\'{i}fica, Campus UAB, C. Albareda s/n, 08193 Bellaterra (Barcelona), Spain\label{aff42}
\and
INFN section of Naples, Via Cinthia 6, 80126, Napoli, Italy\label{aff43}
\and
Institute for Astronomy, University of Hawaii, 2680 Woodlawn Drive, Honolulu, HI 96822, USA\label{aff44}
\and
Dipartimento di Fisica e Astronomia "Augusto Righi" - Alma Mater Studiorum Universit\`a di Bologna, Viale Berti Pichat 6/2, 40127 Bologna, Italy\label{aff45}
\and
Instituto de Astrof\'{\i}sica de Canarias, V\'{\i}a L\'actea, 38205 La Laguna, Tenerife, Spain\label{aff46}
\and
Institute for Astronomy, University of Edinburgh, Royal Observatory, Blackford Hill, Edinburgh EH9 3HJ, UK\label{aff47}
\and
European Space Agency/ESRIN, Largo Galileo Galilei 1, 00044 Frascati, Roma, Italy\label{aff48}
\and
ESAC/ESA, Camino Bajo del Castillo, s/n., Urb. Villafranca del Castillo, 28692 Villanueva de la Ca\~nada, Madrid, Spain\label{aff49}
\and
Universit\'e Claude Bernard Lyon 1, CNRS/IN2P3, IP2I Lyon, UMR 5822, Villeurbanne, F-69100, France\label{aff50}
\and
Institut de Ci\`{e}ncies del Cosmos (ICCUB), Universitat de Barcelona (IEEC-UB), Mart\'{i} i Franqu\`{e}s 1, 08028 Barcelona, Spain\label{aff51}
\and
Instituci\'o Catalana de Recerca i Estudis Avan\c{c}ats (ICREA), Passeig de Llu\'{\i}s Companys 23, 08010 Barcelona, Spain\label{aff52}
\and
UCB Lyon 1, CNRS/IN2P3, IUF, IP2I Lyon, 4 rue Enrico Fermi, 69622 Villeurbanne, France\label{aff53}
\and
Departamento de F\'isica, Faculdade de Ci\^encias, Universidade de Lisboa, Edif\'icio C8, Campo Grande, PT1749-016 Lisboa, Portugal\label{aff54}
\and
Instituto de Astrof\'isica e Ci\^encias do Espa\c{c}o, Faculdade de Ci\^encias, Universidade de Lisboa, Campo Grande, 1749-016 Lisboa, Portugal\label{aff55}
\and
Department of Astronomy, University of Geneva, ch. d'Ecogia 16, 1290 Versoix, Switzerland\label{aff56}
\and
Universit\'e Paris-Saclay, CNRS, Institut d'astrophysique spatiale, 91405, Orsay, France\label{aff57}
\and
Aix-Marseille Universit\'e, CNRS/IN2P3, CPPM, Marseille, France\label{aff58}
\and
INAF-Istituto di Astrofisica e Planetologia Spaziali, via del Fosso del Cavaliere, 100, 00100 Roma, Italy\label{aff59}
\and
Space Science Data Center, Italian Space Agency, via del Politecnico snc, 00133 Roma, Italy\label{aff60}
\and
INFN-Bologna, Via Irnerio 46, 40126 Bologna, Italy\label{aff61}
\and
School of Physics, HH Wills Physics Laboratory, University of Bristol, Tyndall Avenue, Bristol, BS8 1TL, UK\label{aff62}
\and
INAF-Osservatorio Astronomico di Padova, Via dell'Osservatorio 5, 35122 Padova, Italy\label{aff63}
\and
Universit\"ats-Sternwarte M\"unchen, Fakult\"at f\"ur Physik, Ludwig-Maximilians-Universit\"at M\"unchen, Scheinerstrasse 1, 81679 M\"unchen, Germany\label{aff64}
\and
INFN-Sezione di Milano, Via Celoria 16, 20133 Milano, Italy\label{aff65}
\and
Institute of Theoretical Astrophysics, University of Oslo, P.O. Box 1029 Blindern, 0315 Oslo, Norway\label{aff66}
\and
Jet Propulsion Laboratory, California Institute of Technology, 4800 Oak Grove Drive, Pasadena, CA, 91109, USA\label{aff67}
\and
Felix Hormuth Engineering, Goethestr. 17, 69181 Leimen, Germany\label{aff68}
\and
Technical University of Denmark, Elektrovej 327, 2800 Kgs. Lyngby, Denmark\label{aff69}
\and
Cosmic Dawn Center (DAWN), Denmark\label{aff70}
\and
Max-Planck-Institut f\"ur Astronomie, K\"onigstuhl 17, 69117 Heidelberg, Germany\label{aff71}
\and
NASA Goddard Space Flight Center, Greenbelt, MD 20771, USA\label{aff72}
\and
Department of Physics and Astronomy, University College London, Gower Street, London WC1E 6BT, UK\label{aff73}
\and
Department of Physics and Helsinki Institute of Physics, Gustaf H\"allstr\"omin katu 2, 00014 University of Helsinki, Finland\label{aff74}
\and
Universit\'e de Gen\`eve, D\'epartement de Physique Th\'eorique and Centre for Astroparticle Physics, 24 quai Ernest-Ansermet, CH-1211 Gen\`eve 4, Switzerland\label{aff75}
\and
Department of Physics, P.O. Box 64, 00014 University of Helsinki, Finland\label{aff76}
\and
Helsinki Institute of Physics, Gustaf H{\"a}llstr{\"o}min katu 2, University of Helsinki, Helsinki, Finland\label{aff77}
\and
Laboratoire d'etude de l'Univers et des phenomenes eXtremes, Observatoire de Paris, Universit\'e PSL, Sorbonne Universit\'e, CNRS, 92190 Meudon, France\label{aff78}
\and
SKA Observatory, Jodrell Bank, Lower Withington, Macclesfield, Cheshire SK11 9FT, UK\label{aff79}
\and
Centre de Calcul de l'IN2P3/CNRS, 21 avenue Pierre de Coubertin 69627 Villeurbanne Cedex, France\label{aff80}
\and
University of Applied Sciences and Arts of Northwestern Switzerland, School of Computer Science, 5210 Windisch, Switzerland\label{aff81}
\and
Department of Physics, Institute for Computational Cosmology, Durham University, South Road, Durham, DH1 3LE, UK\label{aff82}
\and
Universit\'e Paris Cit\'e, CNRS, Astroparticule et Cosmologie, 75013 Paris, France\label{aff83}
\and
CNRS-UCB International Research Laboratory, Centre Pierre Bin\'etruy, IRL2007, CPB-IN2P3, Berkeley, USA\label{aff84}
\and
University of Applied Sciences and Arts of Northwestern Switzerland, School of Engineering, 5210 Windisch, Switzerland\label{aff85}
\and
Institut d'Astrophysique de Paris, 98bis Boulevard Arago, 75014, Paris, France\label{aff86}
\and
Institut d'Astrophysique de Paris, UMR 7095, CNRS, and Sorbonne Universit\'e, 98 bis boulevard Arago, 75014 Paris, France\label{aff87}
\and
Institute of Physics, Laboratory of Astrophysics, Ecole Polytechnique F\'ed\'erale de Lausanne (EPFL), Observatoire de Sauverny, 1290 Versoix, Switzerland\label{aff88}
\and
Telespazio UK S.L. for European Space Agency (ESA), Camino bajo del Castillo, s/n, Urbanizacion Villafranca del Castillo, Villanueva de la Ca\~nada, 28692 Madrid, Spain\label{aff89}
\and
Institut de F\'{i}sica d'Altes Energies (IFAE), The Barcelona Institute of Science and Technology, Campus UAB, 08193 Bellaterra (Barcelona), Spain\label{aff90}
\and
DARK, Niels Bohr Institute, University of Copenhagen, Jagtvej 155, 2200 Copenhagen, Denmark\label{aff91}
\and
Waterloo Centre for Astrophysics, University of Waterloo, Waterloo, Ontario N2L 3G1, Canada\label{aff92}
\and
Department of Physics and Astronomy, University of Waterloo, Waterloo, Ontario N2L 3G1, Canada\label{aff93}
\and
Perimeter Institute for Theoretical Physics, Waterloo, Ontario N2L 2Y5, Canada\label{aff94}
\and
Institute of Space Science, Str. Atomistilor, nr. 409 M\u{a}gurele, Ilfov, 077125, Romania\label{aff95}
\and
Consejo Superior de Investigaciones Cientificas, Calle Serrano 117, 28006 Madrid, Spain\label{aff96}
\and
Universidad de La Laguna, Departamento de Astrof\'{\i}sica, 38206 La Laguna, Tenerife, Spain\label{aff97}
\and
Institut f\"ur Theoretische Physik, University of Heidelberg, Philosophenweg 16, 69120 Heidelberg, Germany\label{aff98}
\and
Institut de Recherche en Astrophysique et Plan\'etologie (IRAP), Universit\'e de Toulouse, CNRS, UPS, CNES, 14 Av. Edouard Belin, 31400 Toulouse, France\label{aff99}
\and
Universit\'e St Joseph; Faculty of Sciences, Beirut, Lebanon\label{aff100}
\and
Departamento de F\'isica, FCFM, Universidad de Chile, Blanco Encalada 2008, Santiago, Chile\label{aff101}
\and
Universit\"at Innsbruck, Institut f\"ur Astro- und Teilchenphysik, Technikerstr. 25/8, 6020 Innsbruck, Austria\label{aff102}
\and
Satlantis, University Science Park, Sede Bld 48940, Leioa-Bilbao, Spain\label{aff103}
\and
Department of Physics, Royal Holloway, University of London, TW20 0EX, UK\label{aff104}
\and
Instituto de Astrof\'isica e Ci\^encias do Espa\c{c}o, Faculdade de Ci\^encias, Universidade de Lisboa, Tapada da Ajuda, 1349-018 Lisboa, Portugal\label{aff105}
\and
Cosmic Dawn Center (DAWN)\label{aff106}
\and
Niels Bohr Institute, University of Copenhagen, Jagtvej 128, 2200 Copenhagen, Denmark\label{aff107}
\and
Universidad Polit\'ecnica de Cartagena, Departamento de Electr\'onica y Tecnolog\'ia de Computadoras,  Plaza del Hospital 1, 30202 Cartagena, Spain\label{aff108}
\and
Centre for Information Technology, University of Groningen, P.O. Box 11044, 9700 CA Groningen, The Netherlands\label{aff109}
\and
Kapteyn Astronomical Institute, University of Groningen, PO Box 800, 9700 AV Groningen, The Netherlands\label{aff110}
\and
Infrared Processing and Analysis Center, California Institute of Technology, Pasadena, CA 91125, USA\label{aff111}
\and
Dipartimento di Fisica e Scienze della Terra, Universit\`a degli Studi di Ferrara, Via Giuseppe Saragat 1, 44122 Ferrara, Italy\label{aff112}
\and
Istituto Nazionale di Fisica Nucleare, Sezione di Ferrara, Via Giuseppe Saragat 1, 44122 Ferrara, Italy\label{aff113}
\and
INAF, Istituto di Radioastronomia, Via Piero Gobetti 101, 40129 Bologna, Italy\label{aff114}
\and
Astronomical Observatory of the Autonomous Region of the Aosta Valley (OAVdA), Loc. Lignan 39, I-11020, Nus (Aosta Valley), Italy\label{aff115}
\and
Universit\'e C\^{o}te d'Azur, Observatoire de la C\^{o}te d'Azur, CNRS, Laboratoire Lagrange, Bd de l'Observatoire, CS 34229, 06304 Nice cedex 4, France\label{aff116}
\and
Department of Physics, Oxford University, Keble Road, Oxford OX1 3RH, UK\label{aff117}
\and
Aurora Technology for European Space Agency (ESA), Camino bajo del Castillo, s/n, Urbanizacion Villafranca del Castillo, Villanueva de la Ca\~nada, 28692 Madrid, Spain\label{aff118}
\and
Zentrum f\"ur Astronomie, Universit\"at Heidelberg, Philosophenweg 12, 69120 Heidelberg, Germany\label{aff119}
\and
Department of Mathematics and Physics E. De Giorgi, University of Salento, Via per Arnesano, CP-I93, 73100, Lecce, Italy\label{aff120}
\and
INFN, Sezione di Lecce, Via per Arnesano, CP-193, 73100, Lecce, Italy\label{aff121}
\and
INAF-Sezione di Lecce, c/o Dipartimento Matematica e Fisica, Via per Arnesano, 73100, Lecce, Italy\label{aff122}
\and
INAF-Osservatorio Astronomico di Brera, Via Brera 28, 20122 Milano, Italy, and INFN-Sezione di Genova, Via Dodecaneso 33, 16146, Genova, Italy\label{aff123}
\and
ICL, Junia, Universit\'e Catholique de Lille, LITL, 59000 Lille, France\label{aff124}
\and
Instituto de F\'isica Te\'orica UAM-CSIC, Campus de Cantoblanco, 28049 Madrid, Spain\label{aff125}
\and
CERCA/ISO, Department of Physics, Case Western Reserve University, 10900 Euclid Avenue, Cleveland, OH 44106, USA\label{aff126}
\and
Technical University of Munich, TUM School of Natural Sciences, Physics Department, James-Franck-Str.~1, 85748 Garching, Germany\label{aff127}
\and
Max-Planck-Institut f\"ur Astrophysik, Karl-Schwarzschild-Str.~1, 85748 Garching, Germany\label{aff128}
\and
Departamento de F{\'\i}sica Fundamental. Universidad de Salamanca. Plaza de la Merced s/n. 37008 Salamanca, Spain\label{aff129}
\and
Instituto de Astrof\'isica de Canarias (IAC); Departamento de Astrof\'isica, Universidad de La Laguna (ULL), 38200, La Laguna, Tenerife, Spain\label{aff130}
\and
Universit\'e de Strasbourg, CNRS, Observatoire astronomique de Strasbourg, UMR 7550, 67000 Strasbourg, France\label{aff131}
\and
Center for Data-Driven Discovery, Kavli IPMU (WPI), UTIAS, The University of Tokyo, Kashiwa, Chiba 277-8583, Japan\label{aff132}
\and
Ludwig-Maximilians-University, Schellingstrasse 4, 80799 Munich, Germany\label{aff133}
\and
Max-Planck-Institut f\"ur Physik, Boltzmannstr. 8, 85748 Garching, Germany\label{aff134}
\and
Dipartimento di Fisica - Sezione di Astronomia, Universit\`a di Trieste, Via Tiepolo 11, 34131 Trieste, Italy\label{aff135}
\and
Jodrell Bank Centre for Astrophysics, Department of Physics and Astronomy, University of Manchester, Oxford Road, Manchester M13 9PL, UK\label{aff136}
\and
California Institute of Technology, 1200 E California Blvd, Pasadena, CA 91125, USA\label{aff137}
\and
Department of Physics \& Astronomy, University of California Irvine, Irvine CA 92697, USA\label{aff138}
\and
Departamento F\'isica Aplicada, Universidad Polit\'ecnica de Cartagena, Campus Muralla del Mar, 30202 Cartagena, Murcia, Spain\label{aff139}
\and
Instituto de F\'isica de Cantabria, Edificio Juan Jord\'a, Avenida de los Castros, 39005 Santander, Spain\label{aff140}
\and
Observatorio Nacional, Rua General Jose Cristino, 77-Bairro Imperial de Sao Cristovao, Rio de Janeiro, 20921-400, Brazil\label{aff141}
\and
CEA Saclay, DFR/IRFU, Service d'Astrophysique, Bat. 709, 91191 Gif-sur-Yvette, France\label{aff142}
\and
Institute of Cosmology and Gravitation, University of Portsmouth, Portsmouth PO1 3FX, UK\label{aff143}
\and
Department of Computer Science, Aalto University, PO Box 15400, Espoo, FI-00 076, Finland\label{aff144}
\and
Instituto de Astrof\'\i sica de Canarias, c/ Via Lactea s/n, La Laguna 38200, Spain. Departamento de Astrof\'\i sica de la Universidad de La Laguna, Avda. Francisco Sanchez, La Laguna, 38200, Spain\label{aff145}
\and
Ruhr University Bochum, Faculty of Physics and Astronomy, Astronomical Institute (AIRUB), German Centre for Cosmological Lensing (GCCL), 44780 Bochum, Germany\label{aff146}
\and
Department of Physics and Astronomy, Vesilinnantie 5, 20014 University of Turku, Finland\label{aff147}
\and
Serco for European Space Agency (ESA), Camino bajo del Castillo, s/n, Urbanizacion Villafranca del Castillo, Villanueva de la Ca\~nada, 28692 Madrid, Spain\label{aff148}
\and
ARC Centre of Excellence for Dark Matter Particle Physics, Melbourne, Australia\label{aff149}
\and
Centre for Astrophysics \& Supercomputing, Swinburne University of Technology,  Hawthorn, Victoria 3122, Australia\label{aff150}
\and
Department of Physics and Astronomy, University of the Western Cape, Bellville, Cape Town, 7535, South Africa\label{aff151}
\and
DAMTP, Centre for Mathematical Sciences, Wilberforce Road, Cambridge CB3 0WA, UK\label{aff152}
\and
Kavli Institute for Cosmology Cambridge, Madingley Road, Cambridge, CB3 0HA, UK\label{aff153}
\and
Department of Astrophysics, University of Zurich, Winterthurerstrasse 190, 8057 Zurich, Switzerland\label{aff154}
\and
Department of Physics, Centre for Extragalactic Astronomy, Durham University, South Road, Durham, DH1 3LE, UK\label{aff155}
\and
Institute for Theoretical Particle Physics and Cosmology (TTK), RWTH Aachen University, 52056 Aachen, Germany\label{aff156}
\and
IRFU, CEA, Universit\'e Paris-Saclay 91191 Gif-sur-Yvette Cedex, France\label{aff157}
\and
Oskar Klein Centre for Cosmoparticle Physics, Department of Physics, Stockholm University, Stockholm, SE-106 91, Sweden\label{aff158}
\and
Astrophysics Group, Blackett Laboratory, Imperial College London, London SW7 2AZ, UK\label{aff159}
\and
Univ. Grenoble Alpes, CNRS, Grenoble INP, LPSC-IN2P3, 53, Avenue des Martyrs, 38000, Grenoble, France\label{aff160}
\and
INAF-Osservatorio Astrofisico di Arcetri, Largo E. Fermi 5, 50125, Firenze, Italy\label{aff161}
\and
Dipartimento di Fisica, Sapienza Universit\`a di Roma, Piazzale Aldo Moro 2, 00185 Roma, Italy\label{aff162}
\and
Centro de Astrof\'{\i}sica da Universidade do Porto, Rua das Estrelas, 4150-762 Porto, Portugal\label{aff163}
\and
Instituto de Astrof\'isica e Ci\^encias do Espa\c{c}o, Universidade do Porto, CAUP, Rua das Estrelas, PT4150-762 Porto, Portugal\label{aff164}
\and
HE Space for European Space Agency (ESA), Camino bajo del Castillo, s/n, Urbanizacion Villafranca del Castillo, Villanueva de la Ca\~nada, 28692 Madrid, Spain\label{aff165}
\and
Theoretical astrophysics, Department of Physics and Astronomy, Uppsala University, Box 516, 751 37 Uppsala, Sweden\label{aff166}
\and
Mathematical Institute, University of Leiden, Einsteinweg 55, 2333 CA Leiden, The Netherlands\label{aff167}
\and
Institute of Astronomy, University of Cambridge, Madingley Road, Cambridge CB3 0HA, UK\label{aff168}
\and
Univ. Lille, CNRS, Centrale Lille, UMR 9189 CRIStAL, 59000 Lille, France\label{aff169}
\and
Department of Astrophysical Sciences, Peyton Hall, Princeton University, Princeton, NJ 08544, USA\label{aff170}
\and
Space physics and astronomy research unit, University of Oulu, Pentti Kaiteran katu 1, FI-90014 Oulu, Finland\label{aff171}
\and
Institut de Physique Th\'eorique, CEA, CNRS, Universit\'e Paris-Saclay 91191 Gif-sur-Yvette Cedex, France\label{aff172}
\and
Center for Computational Astrophysics, Flatiron Institute, 162 5th Avenue, 10010, New York, NY, USA\label{aff173}}    

%
%
\abstract{We investigate the accuracy and range of validity of the perturbative model for the 2-point (2PCF) and 3-point (3PCF) correlation functions in real space in view of the forthcoming analysis of the \Euclid mission spectroscopic sample. We take advantage of clustering measurements from four snapshots of the Flagship I $N$-body simulations at $z = \left\{0.9, 1.2, 1.5, 1.8\right\}$, which mimic the expected galaxy population in the ideal case of absence of observational effects such as purity and completeness. For the 3PCF we consider all available triangular configurations given a minimal separation $\rmin$. In the first place, we assess the model performance by fixing the cosmological parameters and evaluating the goodness-of-fit provided by the perturbative bias expansion in the joint analysis of the two statistics, finding overall agreement with the data down to separations of $20 \Mpc$. Subsequently, we build on the state-of-the-art and extend the analysis to include the dependence on three cosmological parameters: the amplitude of scalar perturbations $A_{\rm s}$, the matter density $\omega_{\rm cdm}$ and the Hubble parameter $h$.  To achieve this goal, we develop an emulator capable of generating fast and robust modelling predictions for the two summary statistics, allowing efficient sampling of the joint likelihood function. \brown{We therefore present the first joint full-shape analysis of the real-space 2PCF and 3PCF, testing the consistency and constraining power of the perturbative model across both probes, and assessing its performance in a combined likelihood framework.} We explore possible systematic uncertainties induced by the perturbative model at small scales finding an optimal scale cut of $\rmin=30\Mpc$ for the 3PCF, when imposing an additional limitation on nearly isosceles triangular configurations included in the data vector. This work is part of a series of papers validating theoretical models for galaxy clustering measurements in preparation for the \Euclid mission.}
    
%
%
%
%
   \titlerunning{\Euclid preparation. Full-shape 2PCF and 3PCF modelling in real space}
   \authorrunning{Euclid Collaboration: M. Guidi et al.}
   
   \maketitle
%
%
%
%
 
\input{Introduction}

\input{Theory}

\input{Data}

\input{Likelihood}

\input{Results}

\input{Conclusion}

\begin{acknowledgements}

\AckEC

\end{acknowledgements}

\bibliography{cosmology,Euclid}
\input{Appendix}

%

%
%


%

\label{LastPage}
\end{document}

%% file: Introduction.tex
\section{\label{sc:Intro}Introduction}

The \Euclid mission \citep{LaureijsEtal2011} is, along with the Dark Energy Spectroscopic Survey \citep[DESI;][]{AghamousaEtal2016}, \MG{Vera Rubin Observatory \citep{IvezicEtal2009} and the Nancy Grace Roman Telescope \citep{Dore2019}} a Stage IV \MG{ \citep{AlbrechtEtal2006a}} observational campaign expected to provide a major advancement to our understanding of the Universe by mapping cosmological perturbations on an unprecedented volume. 

\Euclid will combine two major probes of the large-scale structure: galaxy clustering and weak lensing. The scientific potential of this combination, along with a description of the instrumentation that will enable it, is described in the recent overview paper \citet{MellierEtal2024}. The spectroscopic galaxy sample, in particular, will cover a redshift range of $0.9 \leq z \leq 1.8$, collecting redshift measurements for millions of H$\alpha$-emitting galaxies across a total area of \num{14000} $\mathrm{deg}^2$. 

Galaxy clustering exploits the statistical properties of the fluctuations in the galaxy distribution at large scale, measuring and analysing, in the standard approach, its correlation functions, starting with the 2-point correlation function (2PCF) or its Fourier-space counterpart, the power spectrum. A major contribution to cosmological studies over the past two decades came from using the baryonic acoustic oscillations present in the 2PCF as a standard ruler to reconstruct the background expansion \citep{SeoEisenstein2003, EisensteinEtal2005B, PercivalEtal2007, WangEtal2017, ZhaoEtal2017, AdameEtal2024DESI_FS}. Additional constraints come as well from the anisotropy induced on two-point correlators by redshift-space distortions \citep{PeacockEtal2001, GuzzoEtal2008, BeutlerEtal2017B, GriebEtal2017, PezzottaEtal2017, HouEtal2018} and, more generally, from the full-shape analysis of their redshift-space multipoles, with the aim of extracting all available information from the clustering measurements \citep{SanchezEtal2013, SanchezEtal2017b, DAmicoEtal2020, IvanovSimonovicZaldarriaga2020, TrosterEtal2020, AdameEtal2024DESI_FS}.  

In recent years, the inclusion in galaxy clustering analyses of higher-order correlation functions such as the 3-point correlation function or (3PCF), in Fourier space, the bispectrum, is becoming more and more part of the standard lore. These statistics quantify the non-Gaussian properties of the galaxy distribution as a random field. Different sources of non-Gaussianity are directly related to nonlinearities in the evolution of matter perturbations \citep{Fry1984}, galaxy bias \citep{FryGaztanaga1993, Fry1994B, FriemanGaztanaga1994}, and redshift-space distortions \citep{HivonEtal1995, ScoccimarroCouchmanFrieman1999}, or possibly to a primordial non-Gaussian component due to inflationary physics \citep{VerdeEtal2000, Scoccimarro2000A, ScoccimarroSefusattiZaldarriaga2004}. The joint analysis of the galaxy power spectrum and bispectrum has been performed, over the last few years, in several studies of the Baryon Oscillations Spectroscopic Survey \citep[BOSS;][]{DawsonEtal2013} data sets, leading to improvements in constraints on cosmological parameters in the context of the standard model and its extensions \citep[see, e.g.,][]{GilMarinEtal2017, DAmicoEtal2020, PhilcoxIvanov2022} or constraints on the initial conditions \citep{DAmicoEtal2022A, CabassEtal2022}.

The full exploitation of the information provided by the 3-point correlation function in configuration space, despite clear advantages over the bispectrum in terms of accounting for survey geometry effects, has been, for a long time, hampered by the computational cost of its estimation in large data sets. Early studies focused on a subset of all potentially measurable configurations, focusing on the galaxy properties and nonlinear bias \citep{JingBorner2004, GaztanagaEtal2005, McBrideEtal2011B, Marin2011, MarinEtal2013, MorescoEtal2017} and the detection of acoustic features \citep{GaztanagaEtal2009, MorescoEtal2021}. The introduction of an estimator based on a spherical harmonic expansion \citep{SlepianEisenstein2015B, SlepianEisenstein2018} has substantially reduced the computational burden, enabling a full exploitation of the information potentially encoded in the 3PCF \citep{SlepianEtal2017, SlepianEtal2018}.

Yet, in a likelihood analysis, an additional disadvantage of the 3PCF compared to the bispectrum is still present at the level of the evaluation of the theoretical model. Indeed, predictions for configuration-space statistics are typically obtained from a first evaluation of their Fourier-space counterpart. For the 2PCF, this mapping is efficiently handled using the \texttt{FFTLog} algorithm \citep{Hamilton2000} in one dimension. In the case of the 3PCF, while some works applied the 1-dimensional Fast-Fourier Transform (FFT) to leading-order, perturbative predictions \citep{SlepianEisenstein2017, SugiyamaEtal2021, VeropalumboEtal2022}, more general methods adopted instead a 2-dimensional FFT \citep{FangEiflerKrause2020} to transform any bispectrum model, when expanded into spherical harmonics, to its configuration-space equivalent \citep{Umeh2021, GuidiEtal2023, FarinaEtal2024a, PugnoEtal2024a}. 

Cosmological perturbation theory (PT) plays a crucial role in our understanding of galaxy clustering at large scales \citep[see][for a classical review]{BernardeauEtal2002}. A significant effort, in the past decade, went into identifying all potential effects that could lead to biased estimates of cosmological parameters if ignored. This includes the impact of small-scale matter fluctuations on large-scale clustering that motivated the effective field theory of large-scale structure \citep[EFTofLSS;][]{BaumannEtal2012, CarrascoHertzbergSenatore2012}, non-local \citep{ChanScoccimarroSheth2012, BaldaufEtal2012}, renormalised \citep{McDonald2006, McDonaldRoy2009, AssassiEtal2014, EggemeierScoccimarroSmith2019}, higher derivative bias \citep{BBKS1986, FujitaEtal2020}, and infrared resummation \citep{BlasEtal2016B}. Here we set aside, for the sake of brevity, redshift-space distortions effects, extremely important in actual data analysis, but not relevant for this work. The state-of-the-art model, which is a perturbative prediction at next-to-leading order for the power spectrum and at leading order for the bispectrum, is at the basis of current full-shape analysis of the galaxy power spectrum, bispectrum, and 2PCF, while a full-shape analysis involving on equal footing the 3PCF is still lacking for the computational burden of its numerical evaluation.

This work is part of a series of papers in preparation for the \Euclid mission, providing an extensive validation of analytical and in some cases numerical models for galaxy clustering statistics, with the first one, already published, focused on the real-space power spectrum \citep{PezzottaEtal2024}. 

Here we explore instead, in the first place, the range of validity of perturbative models in describing real-space measurements of the 2PCF and 3PCF in synthetic catalogues reproducing the density and clustering amplitude expected for the \Euclid spectroscopic sample. In addition, we apply such models to a full-shape, joint analysis of the two statistics. In order to do so, we develop an emulator for both 2PCF and 3PCF prediction as a function of three cosmological parameters, the amplitude of scalar perturbations $A_{\rm s}$, the cold dark matter density parameter $\omega_{\rm cdm}$ and the reduced Hubble parameter $h$. This constitutes an important step toward a complete, configuration-space analysis of galaxy clustering data sets in general and more specifically toward the goals of the \Euclid mission.  

This paper is organised as follows. In Sects. ~\ref{sec:definitions} and ~\ref{sec:theo_model} we introduce the basic definitions of Fourier-space and configuration-space correlators along with a brief overview of their theoretical model in PT. Section ~\ref{sec:data} presents the synthetic galaxy catalogues and the simulation on which they are based, the 2PCF and 3PCF measurements, and the theoretical model for the relative covariance matrices. In Sect.~\ref{sec:likelihood}, we describe the fitting procedure and the performance metrics employed to assess model accuracy and precision. Our results are presented in Sects.~\ref{sec:results_bias} and \ref{sec:results_cosmo}, first fixing the cosmological parameters to assess the goodness of fit of the model, then exploring the full-shape analysis of 2PCF and 3PCF in order to constrain the same parameters. Lastly, in Sect. \ref{sec:conclusion} we present our conclusions. In Appendix \ref{appendix:emulator} we include some more detailed description on the  2PCF and 3PCF emulator and its validation.

%% file: Theory.tex
\section{Definitions and conventions}
\label{sec:definitions}
For convenience, we use the following notation to refer to the integration over the infinite volume of a loop variable $\qv$,
\begin{align}
    \int_{\qv} \vcentcolon = \int \frac{\rm d^3 q}{(2\pi)^3} \ ,
\end{align}
and we also adopt the following convention for the Fourier transform and its inverse
\begin{align}
    \delta(\kv) & \vcentcolon = (2\pi)^3 \int_{\qv} \,\delta(\qv) \, \mathrm{e}^{{\rm i}{\kv \cdot {\qv}}}\,,
\\
\delta(\xv) & \vcentcolon = \int_{\kv} \,\delta(\kv) \, \mathrm{e}^{- {\rm i}\kv \cdot {\xv}}\,.
\end{align}

Under the assumption of statistical homogeneity and isotropy, the power spectrum $P(k)$ of a generic density contrast $\delta(\kv)$ is therefore defined as
\begin{equation}
    \langle \delta(\kv_1)\,\delta(\kv_2) \rangle \vcentcolon = \, (2 \pi)^3  \delta_{\mathrm{D}}(\kv_1 \!+\! \kv_2)\,P(k)\,,
\end{equation}
\MG{where $k \vcentcolon = |\kv_1|$.} The corresponding two-point function $\xi(r) \vcentcolon = \langle \delta(\xv)\delta(\xv+\xr)  \rangle$, \MG{where $r \vcentcolon = |\rv|$}, can be obtained as its Fourier Transform
\begin{equation}
    \xi(r)  = \int_{\kv} P(k)\ \mathrm{e}^{{\rm i} \kv \cdot \rv} = \int \frac{\dd k}{2 \pi^2} k^2 \, j_0(kr) \, P(k) \, ,
    \label{eq:xi_def}
\end{equation}
where $j_n$ represent the spherical Bessel functions of $n$-th order. Similarly, the bispectrum $B(\kv_1,\kv_2,\kv_3)$, i.e. the 3PCF of the Fourier-space density contrast $\delta(\kv)$, is defined as
\begin{equation}
    \langle \delta(\kv_1)  \delta(\kv_2)  \delta(\kv_3)\rangle \vcentcolon = (2 \pi)^3 \, \delta_\mathrm{D}(\kv_1 \!+ \! \kv_2 \!+ \! \kv_3)\,B(k_1, k_2, k_3)\,.
\end{equation}
Here the Dirac delta function forces the three wave vectors $\{\kv_1,\kv_2,\kv_3\}$ to form a closed triangle, so that $k_3$ can be written as a function of $k_1$, $k_2$ and the cosine of the angle between $\kv_1$ and $\kv_2$, denoted as $\mu_{12}\vcentcolon =\kv_1\cdot\kv_2/k_1k_2$. We can therefore consider an expansion of such angle dependence in Legendre polynomials 
\begin{equation}
    B(k_1,k_2,k_3)=\sum_\ell B_\ell(k_1,k_2)\,\mathcal{L}_{\ell}(\mu_{12})\,,
\end{equation}
with coefficients defined as
\begin{align}
    B_\ell(k_1, k_2) = \frac{2 \ell + 1}{2} \int_{-1}^{1} \mathrm{d} \mu_{12} \, B (k_1, k_2, k_3) \,\mathcal{L}_{\ell}(\mu_{12}) \ .
\end{align}

The 3PCF, in configuration space, is defined, adopting the notation $\rv_{ij}\vcentcolon = \rv_i-\rv_j$, as
\begin{equation}
    \zeta(r_{12}, r_{13}, r_{23}) \vcentcolon = \langle \delta(\rv_1)\,\delta(\rv_2)\, \delta(\rv_3)\rangle\,,
\end{equation}
and, in the same way, the dependence on the angle between $\rv_{12}$ and $\rv_{13}$ can be expanded in Legendre polynomials as
\begin{equation}
\label{eq:LegendreExp}
    \zeta(r_{12}, r_{13}, r_{23}) = \sum_{\ell}\,\zeta_\ell(r_{12}, r_{13} ) \mathcal{L}_\ell(\mu_{12,13}) \,,
\end{equation}
where now we define the cosine $\mu_{12,13}\vcentcolon = \rv_{12} \cdot \rv_{13}/ (r_{12} r_{13})$ and the coefficients are given by
\begin{equation}
    \zeta_{\ell}(r_{12}, r_{13}) \vcentcolon = \frac{2 \ell + 1}{2} \!\int_{-1}^{1} \mathrm{d} \mu_{12,13}\, \zeta(r_{12}, r_{13}, r_{23}) \,\mathcal{L}_{\ell}(\mu_{12,13})\,. \label{eq:zeta_ell_def}
\end{equation}
 
The relation between the 3PCF and the bispectrum can be expressed in terms of their respective multipoles $B_\ell$ and $\zeta_\ell$ as
\begin{equation}
    \zeta_{\ell}(r_{12}, r_{13}) = (-1)^{\ell}\int \frac{\mathrm{d}k_1 \, \mathrm{d}k_2}{(2 \pi)^6} B_{\ell}(k_1, k_2)\,j_\ell(k_1r_{12}) \,j_\ell(k_2r_{13})\,. 
    \label{eq:zeta_Bk_ell}
\end{equation}
As we will see the expansion in multipoles will be crucial to define an efficient estimator for the 3PCF, as for a fast evaluation of its theoretical prediction.

\section{\label{sec:theo_model} Theoretical models}

This section provides a brief introduction to the real-space modelling of the 2PCF and 3PCF in cosmological perturbation theory 
\citep[PT,][]{BernardeauEtal2002}. We consider specifically a one-loop expression for the 2PCF  accounting for a general bias expansion \citep{DesjacquesJeongSchmidt2018}, corresponding to the EFTofLSS expression described in a companion paper (Euclid Collaboration: Kärcher et al., in prep.), to which we refer the reader for further details. The prediction for the 3PCF is instead at tree-level in PT.

\subsection{\label{subsec:PT} Eulerian perturbation theory and galaxy bias}

In Eulerian perturbation theory, the equations describing the evolution of the matter density and velocity field, which are continuity and Euler equations, are solved perturbatively assuming the density contrast $\delta(\xv)$ to be small at large scales. The nonlinear solution is given by the sum of the solution to the linearized equations plus higher-order corrections.

In addition, the relation between the galaxy density contrast $\delta_{\rm g}(\xv)$ and the matter field $\delta(\xv)$ is also given in terms of a perturbative expansion. The terms relevant for the one-loop power spectrum and 2PCF models are given by
\begin{align}
        \delta_{\rm g}(\xv)  = \ & b_1\,\delta(\xv) +\frac{b_2}{2}\delta^{\,2}(\xv) + b_{\nabla^2} \nabla^2 \delta(\xv) 
\nonumber \\
& + b_{\mathcal{G}_2}\mathcal{G}_2\left(\Phi_\mathrm{v}\,|\,\xv\right) + b_{\Gamma_3}\Gamma_3(\xv)\,,
   \label{eq:deltag_expansion}
\end{align}
where  $\mathcal{G}_2$ and 
$\Gamma_3$ are non-local operators, defined as 
\begin{align}    
    &\mathcal{G}_2 (\Phi |\textbf{x}) \vcentcolon = \big [\partial_i \partial_j \Phi(\textbf{x}) \big]^2 - [\partial^2 \Phi(\textbf{x})]^2, \\ \notag \\ 
    &\Gamma_3(\textbf{x}) \vcentcolon = \mathcal{G}_2(\Phi |\textbf{x}) - \mathcal{G}_2(\Phi_v |\textbf{x})\,,
\end{align}
and $\Phi(\textbf{x})$ and $\Phi_v(\textbf{v})$ represent the gravitational and velocity potential. The bias relation in Eq.~(\ref{eq:deltag_expansion}) consists of all operators built from second derivatives of the gravitational and velocity potential. It  includes the linear ($b_1$) and quadratic ($b_2$), local bias contributions \citep{Kaiser1984, Szalay1988, ColesEtal1993, FryGaztanaga1993}, omitting the cubic local operator that leads to a correction that can be absorbed in a renormalized linear bias parameter \citep{Szalay1988, McDonald2006, McDonaldRoy2009, AssassiEtal2014, EggemeierScoccimarroSmith2019}. It includes as well non-local contributions ($b_{\mathcal{G}_2}$, $b_{\Gamma_3}$) induced by nonlinear evolution \citep{CatelanEtal1998, ChanScoccimarroSheth2012, BaldaufEtal2012} and higher derivative correction to linear bias \citep[$b_{\nabla^2}$, ][]{BBKS1986, Desjacques2008, FujitaEtal2020}. For a detailed review, see \citet{DesjacquesJeongSchmidt2018}.

The most general and conservative assumption in fitting the galaxy correlation models to their measurements consist in considering all bias parameters as free, independent parameters to be determined by the data along with the cosmological parameters. On the other hand, we do know that the bias parameters are correlated, and large portions of the parameter space are in fact non-physical. We can take advantage of specific relations among the bias parameters to reduce the number of nuisance parameters. We will consider two of such relations. The first, given by
\begin{equation}
    b_{\mathcal{G}_2}^{\rm ex-set}  =  0.524 - 0.547\ b_1 +0.046 \ b_1^2 \ ,
    \label{eq:bg2_bias_rel}
\end{equation}
is a quadratic fit obtained by \citet{EggemeierEtal2020}
to the excursion-set prediction of  \citet{ShethChanScoccimarro2013}. The second relation, given instead by
\begin{equation}    
    b_{\Gamma_3}^{\rm coev}  =  -\frac{1}{6}(b_1 -1) - \frac{5}{2} b_{\mathcal{G}_2} + b_{{\Gamma_3}}^{\mathcal{L}}\,, 
    \label{eq:bg3_bias_rel}
\end{equation}
is derived in \citet{EggemeierScoccimarroSmith2019} assuming the evolution of conserved galaxy number density (hence, ``coevolution'') after formation, with the subscript $\mathcal{L}$ indicates the corresponding Lagrangian quantities at the formation moment. 

The resulting, perturbative expression for the galaxy power spectrum in real space, $\Pgg(k)$, omitting stochastic contributions irrelevant for the 2PCF prediction, then reads
\be
    \Pgg(k)=\Pgg^{\,\rm tree}(k) + \Pgg^{\,\rm1\mbox{-}loop}(k) + \Pgg^{\,\rm ctr}(k) \, .
    \label{eq:Pgg}
\ee
The linear, “tree level”, leading term is simply proportional to the linear matter power spectrum
\be
    \Pgg^{\,\rm tree}(k)=b_1^{\,2} \, \Plin(k)\,,
\label{eq:Pgg_tree}
\ee
while the one-loop correction in standard PT is 
\begin{align}
        \Pgg^{\,\rm 1\mbox{-}loop}(k)  = &  \ P_{{\rm g}, 22}(k)+P_{{\rm g}, 13}(k) \\
          = & \ 2\int_{\qv}K_2^{\,2}\paren{\qv,\kv-\qv} \, \Plin\paren{\abs{\kv-\qv}} \, \Plin(q)  \nonumber \\
        & + 6 \, b_1 \, \Plin(k) \int_{\qv} K_3\paren{\qv,-\qv,\kv} \, \Plin(q) \, ,
    \label{eq:Pgg_oneloop}
\end{align}
where the kernels describing nonlinear matter density evolution and nonlinear bias are
\begin{align}
    K_2(\kv_1, \kv_2)  = &\, b_1 F_2(\kv_1, \kv_2) + \frac{1}{2}\,b_2 + b_{\mathcal{G}_2} S(\kv_1, \kv_2) \ ,
    \\ 
    K_3(\kv_1, \kv_2, \kv_3) = &\, b_1 F_3(\kv_1, \kv_2, \kv_3) + b_2 F_2(\kv_1, \kv_2) 
    \nonumber\\
    & + 2 b_{\mathcal{G}_2}\,S(\kv_1, \kv_{12})\,F_2(\kv_2, \kv_3) 
    \nonumber\\ 
    & + 2b_{\Gamma_3}\,S(\kv_1, \kv_{12})\,\big[F_2(\kv_2, \kv_3) - G_2(\kv_2, \kv_3) \big] \ ,
\end{align}
with
\begin{align}
    S(\kv_1, \kv_{2}) & =  \frac{(\kv_1 \cdot \kv_2)^2}{k^2_1k^2_2} - 1, \\
    F_2(\kv_1, \kv_{2}) & = \frac{5}{7} + \frac{1}{2} \frac{\kv_1 \cdot \kv_2}{k_1k_2}\ \Bigg(\frac{k_2}{k_1} + \frac{k_1}{k_2}\Bigg) + \frac{2}{7}\frac{(\kv_1 \cdot \kv_2)^2}{k^2_1k^2_2}\,, \\
    G_2(\kv_1, \kv_{2}) & = \frac{3}{7} + \frac{1}{2} \frac{\kv_1 \cdot \kv_2}{k_1k_2} \ \Bigg(\frac{k_2}{k_1} + \frac{k_1}{k_2}\Bigg) + \frac{4}{7} \frac{(\kv_1 \cdot \kv_2)^2}{k^2_1k^2_2}\,.
\end{align}
The last term in the power spectrum model accounts for the fully degenerate contributions from the EFT matter counterterms depending on the effective speed of sound $c_s$ \citep{BaumannEtal2012, CarrascoHertzbergSenatore2012} and the higher derivative correction to linear bias, 
\begin{align}
         \Pgg^{\,\rm ctr}(k) &=-2 \, b_1\paren{b_1\,c_{\rm s}^{\,2}+\bdtwod} \, k^2 \, \Plin(k) \\
        &\vcentcolon = -2 \, c_0 \, k^2 \, \Plin(k)\,,
     \label{eq:Pgg_ctr}
\end{align}
with the parameter $c_0\vcentcolon = b_1 c_s^2+\bdtwod$.

Finally, the expression for the galaxy bispectrum at tree-level in perturbation theory is simply
\begin{align}
    B^{\rm tree}_{\rm g}(k_1, k_2, k_3) = 2b_1^2 \, K_2(\kv_1, \kv_2)\,\Plin(k_1)\,\Plin(k_2)  + \mathrm{cyc.}\,,
\end{align}
where, again, we neglect shot-noise contributions.

\subsection{Infrared resummation}
\label{sec:IR-resummation}

The expressions described above do not account for the additional smearing of the baryonic features due to nonlinear evolution \citep{EisensteinSeoWhite2007, SmithScoccimarroSheth2007, CrocceScoccimarro2008, Matsubara2008A, DesjacquesEtal2010, BaldaufEtal2015B, SenatoreZaldarriaga2015}. We model this effect in the power spectrum following the approach of \citet{BlasEtal2016} in terms of a wiggle-no wiggle splitting of the linear power spectrum
\begin{equation}
    \Pl(k)=P_{\rm nw}(k)+P_{\rm w}(k)\,,
\end{equation}
obtained adopting the specific procedure of \citet{VlahEtal2016} -- \citep[see Appendix C in][for a detailed description]{PezzottaEtal2024}. 

The linear galaxy power spectrum in real space is then replaced by a leading-order (LO) contribution 
\be
    P_{\rm g}^{\rm IR,\, LO}(k)=b_1^2\left[P_{\rm nw}(k)+ \mathrm{e}^{-k^2\Sigma^2}P_{\rm w}(k)\right]\,,
\ee
with the constant $\Sigma^2$ representing the anisotropic variance of the relative displacement field \citep{EisensteinSeoWhite2007} 
\begin{align}
    \Sigma^2 = & 
    \frac{1}{6\pi^2}\int_0^{k_S} \dd q\,P_{\rm nw}(q)\,\left[1-j_0\left(\frac{q}{k_{\rm osc}}\right)+2j_2\left(\frac{q}{k_{\rm osc}}\right)\right]\,,
\end{align}
where $k_{\rm osc}=1/\ell_{\rm osc}$ is the wavenumber associated with the BAO scale $\ell_{\rm{osc}}=110\Mpc$, while $k_S=0.2 \, \kMpc$ is an upper limit of integration whose specific choice does not lead to significant differences on the final result.

The one-loop prediction is now included in the next-to-leading-order (NLO) correction as
\begin{align}
\label{eq:PNLO}
    P_{\rm g}^{\rm IR,\, LO+NLO}(k) = & \ b_1^2\,P_{\rm g,\,\rm nw}(k) +\left(1+k^2\Sigma^2\right)\mathrm{e}^{-k^2\Sigma^2} \,b_1^2P_{\rm w}(k) 
    \nonumber\\
    & + P_{\rm g}^{\rm IR,\, 1-loop}(k) \, ,
\end{align}
where the last term denotes the galaxy power spectrum one-loop correction evaluated in terms of the leading-order matter power spectrum, which is, schematically,
\begin{equation}
    P_{\rm g,\, {\rm nw}}^{\rm IR,\,1-loop} (k) \vcentcolon = P_{\rm g}^{\rm 1-loop} \, [P_{\rm nw} (k) +\mathrm{e}^{-k^2\Sigma^2}P_{\rm w}(k)]\,.
\end{equation}

The tree-level bispectrum expression is replaced instead by the leading-order prediction \citep{IvanovSibiryakov2018}
\begin{align}
\label{eq:B_det}
B_{\rm g}^{\rm tree, LO}(k_1, k_2, k_3) = & \ 2\, b_1^2\,K_2(\kv_1, \kv_2)
\nonumber\\ 
& \times \left[P_{\rm nw}(k_1)\,P_{\rm nw}(k_2) \right.
+\mathrm{e}^{-k_1^2\Sigma^2}\,P_{\rm w}(k_1)\,P_{\rm nw}(k_2)
\nonumber\\ & 
\left.+\mathrm{e}^{-k_2^2\Sigma^2}\,P_{\rm w}(k_2)P_{\rm nw}(k_1)\right]
+ {\rm 2~ perm}\,.
\end{align}

In what follows we will retain, for clarity, the notation of Sect. \ref{subsec:PT} for all contributions to each correlation function, while we will assume as implicit the implementation of infrared resummation as described here.

\subsection{\label{subsec:models2PCF} Evaluation of 2PCF and 3PCF}

The 2PCF is computed from the power spectrum of Eq.~\eqref{eq:Pgg} according to the Hankel transform of Eq. \eqref{eq:xi_def}, implemented using the \texttt{FFTlog} approach of \citet{Hamilton2000}. 

The decomposition in Legendre polynomials of the 3PCF defined in Eq.~\eqref{eq:LegendreExp} allows us to apply an analogous procedure to this statistic. We can in fact consider the 2-dimensional extension of the \texttt{FFTlog} algorithm proposed by \citet{FangEiflerKrause2020} for an efficient evaluation of the integral in Eq.~\eqref{eq:zeta_Bk_ell} over the $B_\ell(k_1,k_2)$ functions,  \citep[see][for further details]{Umeh2021, GuidiEtal2023, FarinaEtal2024a, PugnoEtal2024a}.

However, the implementation of the \texttt{2DFFTlog} approach still constitutes a computational burden when considered in the context of Monte Carlo sampling of a likelihood function over a large parameter space. For this reason, we developed an emulator for the 3PCF for a full set of cosmological and nuisance parameters. This allows us to efficiently extend the full-shape analysis of the 2PCF to include the next, higher-order statistic. In fact, our emulator provides all contributions to the 2PCF and 3PCF where a combination of bias parameters can be factorized and describes the cosmological dependence of each term on the scalar amplitude parameter $A_s$, the matter density $\omega_{\rm cdm}$ and the Hubble parameter $h$. A detailed description of the emulator, along with validation tests, can be found in  Appendix \ref{appendix:emulator}.

%% file: Data.tex
\section{\label{sec:data} Data}

We validate our theoretical model against a set of synthetic galaxy catalogues obtained from the Euclid Flagship I N-body simulation. The catalogues adopt a halo occupation distribution (HOD) prescription to describe the population of \Ha galaxies expected for the \Euclid spectroscopic sample. We provide below a concise description of the galaxy catalogues, referring the reader to \citet{PezzottaEtal2024} for further details. 

\subsection{Euclid simulation}

The Flagship I simulation uses the \texttt{PKDGRAV3} code \citep{PotterStadelTeyssier2017}
to track the evolution of two trillion dark matter particles in a box of size $L = 3780 \ h^{-1}  \ \mathrm{Mpc}$, with a mass resolution, $m_{\rm p} \sim 2.398 \times 10^9 \ h^{-1} \ M_\odot$. It assumes a flat, $\Lambda$CDM cosmology with the fiducial parameters given in Table~\ref{tab:simulation_params}. \footnote{For the slight difference w.r.t.  \citet{PotterStadelTeyssier2017}, we refer the reader to the explanation in \citet{PezzottaEtal2024}.}
\begin{table}[t]
    \centering
    \caption{Fiducial parameters of the flat $\Lambda$CDM cosmological model adopted in the Flagship I simulation. The table lists, from left to right, $h$, $\omega_{\mathrm{b}}$ $\omega_{\mathrm{cdm}}$, $A_{\rm s}$, the spectral index $n_{\rm s}$, and the total neutrino mass ($M_{\nu}$).}
    \renewcommand{\arraystretch}{1.5}
    \begin{adjustbox}{width=\columnwidth}
    \begin{tabular}{|c|c|c|c|c|c|}
        \hline
        \rowcolor{blue!15}
        $h$ & $\omega_{\mathrm{b}}$ & $\omega_\mathrm{cdm}$ & $10^9 \times A_s$ & $n_s$ & $M_\nu [\mathrm{eV}]$  \\
        \hline
        0.67 & 0.0219961 & 0.121203 & $2.09427$ & 0.97 &  0.0\\
        \hline
    \end{tabular}
    \end{adjustbox}
    \label{tab:simulation_params}
\end{table}

\begin{table}[t]
    \centering    
    \caption{Summary of key properties for the galaxy synthetic catalogues. The table includes, for each snapshot, the total number of galaxies ($N_{\rm g}$), the number density ($\bar{n}_{\rm g}$) and an estimate of the linear bias parameter ($b_1$) obtained in \citet{PezzottaEtal2024} from measurements of the matter power spectrum. }
    \renewcommand{\arraystretch}{1.5}
    \begin{tabular}{|c|c|c|c|}
         \hline
         \rowcolor{blue!15}
         $z$ & $N_g$ & $\Bar{n}_g\, [ h^{3} \mathrm{Mpc}^{-3} ]$ & $b_1$  \\
         \hline
         $0.9$ & \num{110321755}  & \num{0.0020} & 1.4 \\
         \hline
         $1.2$ & \phantom{0}\num{55563490} & \num{0.0010} & 1.8 \\
         \hline
          $1.5$ & \phantom{0}\num{31613213} & \num{0.0006} & 2.0\\
         \hline
          $1.8$ & \phantom{0}\num{16926864} & \num{0.003}\phantom{0} & 2.5\\
         \hline
     \end{tabular}
    \label{tab:Flagship}
\end{table}

We consider four comoving snapshots at redshift $z=0.9$, 1.2, 1.5, and $1.8$. In each snapshot, a friends-of-friends halo catalogue with a minimum mass corresponding to ten particles, is constructed. Galaxies are then assigned to these halos using a HOD prescription derived from the Flagship I lightcone catalogue, designed to reproduce the Model 3 distribution of \cite{PozzettiEtal2016}. Specifically, the HOD parameters are \MG{are chosen to match the expected selection function of the Euclid spectroscopic sample} with a \Ha flux limit of $f_{{\rm H}\alpha} > 2 \times 10^{-16} \,\text{erg} \,\text{cm}^{-2} \, \text{s}^{-1}$, as outlined by \cite{ScaramellaEtal2022}. Table~\ref{tab:Flagship} shows for each snapshot the total number of galaxies, the number density and a value for the linear bias obtained from measurements of the galaxy and matter power spectra. For a more detailed description of the mock galaxy catalogue we refer the reader, again, to \citet{PezzottaEtal2024}. 


\brown{We fit our model to measurements from the full simulation volume, which is approximately  three to six times larger than the effective volume of the Euclid redshift bins assumed for the forecasts analysis of \citet[][]{BlanchardEtal2020}. The large simulation volume allows for a stringent test of the model ensuring that any residual theory systematic error is well within the expected precision of future measurements.}

The catalogue we consider does not take into account observational systematic effects such as target incompleteness, sample purity, and the impact of the angular footprint or radial selection function. A comprehensive exploration of such effects is left to other works (see Euclid Collaboration: Granett et al., in prep.; Euclid Collaboration: Monaco et al., in prep.; Euclid Collaboration: Risso et al., in prep.; Euclid Collaboration: Lee et al., in prep.). 

\begin{figure*}
    \centering
    \includegraphics[width=1.0\textwidth]{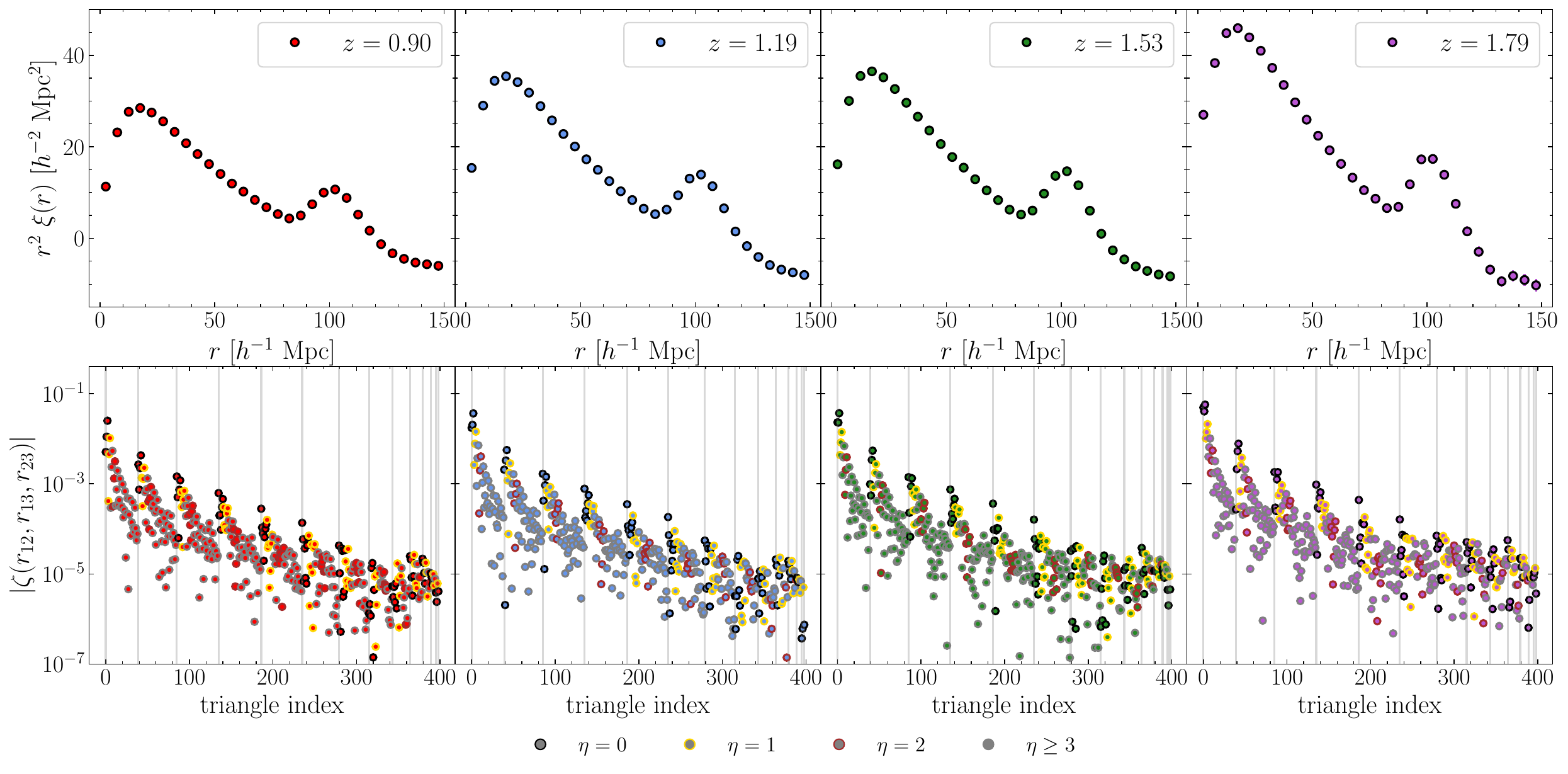}
    \caption{\textit{Top panels}: 2PCF measurements for four comoving snapshots from the Flagship I simulation of the Model 3 HOD. \textit{Bottom panels}: same but for the 3PCF measurements. The ordering of the triangular configurations is defined by increasing values of $r_{12}$, $r_{13}$, and $r_{23}$ under the condition $r_{12} \leq r_{13} \leq r_{23}$, as depicted in Fig. \ref{fig:r_ij}. \MG{Different coulored circles label different choiche of $\eta$, while} the vertical gray lines mark group of triangles sharing the same value of the smaller side $r_{12}$.}
    \label{fig:meas}
\end{figure*}

\subsection{\label{sec:measures} Clustering measurements}



We estimate the 2PCF using the natural estimator \citep{Peebles1973}:
\be
\label{eq:measures:2pcf}
    \hat{\xi}(r) \, \vcentcolon = \, \frac{N_{\rm DD}(r)}{N_{\rm RR}(r)} - 1\,,
\ee
where $N_{\rm DD}(r)$ and $N_{\rm RR}(r)$ represent the pair counts, as a function of separation $r$, in the data and a random distribution with, in this ideal case, constant density. This is equivalent to the usual Landy--Szalay estimator \citep[LS,][]{LandySzalay1993}  in case of a cubic box with periodic boundary conditions (they should both lead to the same variance). We take advantage of this property also to estimate analytically the number of pairs from the random catalogue as
\be
\label{eq:measures:rr_analytic}
    N_{\rm RR}(r) \, = \, \frac{4\pi}{3}\bar{n}^2 V \left[  \left(r + \frac{\Delta r}{2}\right)^3 - \left(r - \frac{\Delta r}{2}\right)^3 \right]\,,
\ee
where we assume a separation bin of size $\Delta r$ centred on $r$, $\bar{n}$ being the mean galaxy number density and $V$ the volume of the box.


We measure the 2PCF in the four snapshots for separations ranging from $\rmin = 0 \Mpc$ to $\rmax = 150 \Mpc$. We use linearly spaced bins with a width of $\Delta r = 5 \Mpc$. The results are displayed in the top panels of Fig. \ref{fig:meas}, where each column represents one of the four redshift snapshots. In the analysis presented in Sects.~\ref{sec:results_bias} and \ref{sec:results_cosmo}, we will restrict the 2PCF data vector to separations $r\leq\rmax = 140 \Mpc$.

The LS estimator for the 2PCF can be easily extended to the analogous estimator for the 3PCF, constructed in terms of triplet counts \citep[see, e.g.,][]{SzapudiSzalay1997}. However, since such an estimator scales with the number of galaxies $N_g$ as $\mathcal{O}(N_g^3)$, its computational cost quickly becomes prohibitive when it comes to very large data sets such as those we are considering here, where $N_g$ is around $10^8$. To resolve this issue, \citet{SlepianEisenstein2015B} introduced an estimator based on the spherical harmonics decomposition (SHD), which improves significantly the computational efficiency. In this approach, the dependence of the 3PCF on the angle between the triangle sides $\vec{r}_{12}$ and $\vec{r}_{13}$ is expanded in terms of Legendre polynomials according to Eq.~\eqref{eq:LegendreExp}. As a result, the full estimator $\hat{\zeta}(r_{12}, r_{13}, r_{23})$ is written as a function of an estimator $\hat{\zeta}_{\ell} (r_{12}, r_{13})$ of the expansion coefficients as 
\begin{equation}
    \label{eq:3pcf_resum}
     \hat{\zeta}(r_{12}, r_{13}, r_{23}) = \sum_{\ell=0}^{\ell_{\rm max}} \hat{\zeta}_{\rm \ell} (r_{12}, r_{13}) \, \mathcal{L}(\mu_{12,13}) \ ,
\end{equation}
thereby reducing the computational complexity to $\mathcal{O}(N_g^2)$. The coefficients are estimated as 
\begin{equation}
    \label{eq:measures:3pcf}
    \hat{\zeta}_\ell (r_{12}, r_{13}) \, = \, \frac{N_{\rm DDD,\,\ell}-3N_{\rm DDR,\,\ell}+3N_{\rm DRR,\,\ell}-N_{\rm RRR,\,\ell}}{N_{\rm RRR,\, 0}}\,,
\end{equation}
where $N_{\rm DDD,\,\ell}$, $N_{\rm DDR,\,\ell}$, $N_{\rm DRR,\,\ell}$ and $N_{\rm DRR,\,\ell}$ are, in turn, the coefficients of the Legendre expansion of the triplets counts from the galaxy catalogue, from the random and the two mixed terms,  directly evaluated in the approach of \citet{SlepianEisenstein2015B}. Again, since we are working with a simulation box with periodic boundary conditions, it is possible to simplify the evaluation of the coefficients $\hat{\zeta}_{\ell} (r_{12}, r_{13})$ and derive an analytical expression for the monopole of the random triplet counts $N_{\rm RRR,\,0}$.

\brown{The SHD estimator is known to be less efficient for isosceles or nearly isosceles triangle configurations (i.e. with $r_{12} \simeq r_{13}$). In these cases, a large value of $\ell_{\max}$ is required to recover results consistent with those of the standard, direct-counting, estimator. Indeed, for isosceles triangles each coefficient of the expansion in Eq.~\eqref{eq:LegendreExp} is the result of an integration over the bispectrum down to small, nonlinear scales, beyond the range of validity of the perturbative model. On the other hand, using a fixed $\ell_{\max}$ ensures internal consistency between measurements and theory. To mitigate this issue, we introduce the quantity \citep[see, e.g.,][]{VeropalumboEtal2022}}
\begin{equation}
    \eta \vcentcolon = \frac{r_{13} - r_{12}}{\Delta r}\,,
    \label{eq:eta}
\end{equation} 
with $\Delta r$ being the separation bin size, to parametrise the proximity of a given configuration to the isosceles case. Fixing a lower limit $\eta_{\rm min}$ amounts to excluding isosceles, or nearly isosceles configurations.

In the bottom panels of Fig.~\ref{fig:meas} we present the 3PCF measurements for the four snapshots. The triangle index is defined by increasing value of $r_{12}$, $r_{13}$, and $r_{23}$ under the condition $r_{12} \leq r_{13} \leq r_{23}$, as depicted in Fig. \ref{fig:r_ij}. The measurements include all configurations with side lengths from $r_{\rm min} = 0 \Mpc$ to $r_{\rm max} = 150 \Mpc$ in bins of width $\Delta r = 5 \Mpc$. We expand the 3PCF up to $\ell_{\rm max} = 10$, which strikes an ideal balance between computational efficiency and information content. 

A more detailed description of the numerical implementation and validation of the SHD estimator for the 3PCF will be given in Euclid Collaboration: Veropalumbo et al. (in prep.), while a similar presentation of the 2PCF estimator will be given in Euclid Collaboration: De La Torre et al. (in prep.).

    \label{fig:StoN}


\subsection{Covariance}
\label{subsec:theo_gauss_cov}

We make use of a covariance matrix obtained from an analytical model for both 2- and 3-point functions. We work in the Gaussian approximation, so that all covariance depends simply on the galaxy power spectrum, and we further assume a linear model for the latter.

For the 2PCF covariance we have \citep{GriebEtal2016, LippichEtal2019}
\begin{equation}
    \mathcal{C}_{\xi}(r, r') = \frac{1}{\pi^2 V} \, \int_0^{\infty} \dd k \, k^2 \, P^2_{\rm tot}(k) \, \bar{j}_{0}(kr) \, \bar{j}_{0}(kr')\,,
\end{equation} 
where $V$ is the box volume, $\bar{j}_0$ are the bin-averaged, zeroth-order Bessel functions \citep[see Eq. A19 in][]{GriebEtal2016}, while 
\be
P_{\rm tot}(k)\vcentcolon = b_1^2 \, P_\mathrm{lin}(k) + \frac{1}{\bar{n}_g}\,,
\ee
is the linear galaxy power spectrum including a Poisson shot-noise contribution, $\bar{n}_g$ being the galaxy number density. For each redshift snapshot, we compute the covariance matrix using the bias and density values listed in Table~\ref{tab:Flagship}. \MG{As noted by \citet{SmithEtal2008}, a more detailed treatment of discrete tracers introduces additional terms in the covariance—particularly relevant at low number density — even under Gaussian assumptions. These include contributions from combinatoric sampling effects, which are not included in our model which is sufficient givent the number density of simulated dataset.}

For the 3PCF covariance, we follow the expression proposed in \citet{SlepianEisenstein2015B}, rigorously validated in \citet{VeropalumboEtal2022}. 
This again takes advantage of the spherical harmonics decomposition writing the covariance between two measurements $\hat{\zeta}(r_{12},r_{13},r_{23})$ and $\hat{\zeta}(r_{12}',r_{13}',r_{23}')$ as
\begin{equation}
    \begin{split}        
    \mathcal{C}_{{\zeta}}(&r_{12},  r_{13}, r_{23};r'_{12}, r'_{13}, r'_{23}) = \\ & =  \sum_{\ell, \ell'}^{\ell_{max}} \mathcal{C}_{{\zeta}, \ell \ell^{'}}(r_{12}, r_{13}; r'_{12}, r'_{13}) \, \mathcal{L}_{\ell}(\mu_{12,13}) \, \mathcal{L}_{\ell'}(\mu_{12,13}), 
    \end{split}
\end{equation}
where $\mathcal{C}_{{\zeta}, \ell \ell^{'}}(r_{12}, r_{13}; r'_{12}, r'_{13})$ represents the covariance between the coefficients  $\hat{\zeta}_\ell(r_{12},r_{13})$ and $\hat{\zeta}_{\ell'}(r_{12}',r_{13}')$. This, in turn, is given by
\begin{align}
    \mathcal{C}_{\zeta,\ell\ell'}&(r_i,  r_j;r'_i, r'_j )= 
    \\ \nonumber
   = &   \frac{(2\ell+1)\, (2\ell'+1)\, (-1)^{\ell+\ell'}}{(2\pi)^6\,V}\!\int\!\dd p_1 \, \dd p_2 \, \dd p_3 \, \delta_{\rm D}(\vec{p}_{123})
   \\ \nonumber & \times 
   P_{\rm tot}(p_1) \, P_{\rm tot}(p_2) \, P_{\rm tot}(p_3) 
   \bar{j}_\ell(p_1r_i) \, \bar{j}_\ell(p_2r_j) \, \Le_\ell(\mu_{12}) 
   \\ \nonumber   & 
   \times \left[\bar{j}_{\ell'}(p_1r_i') \, \bar{j}_{\ell'}(p_2r_j') \, \Le_{\ell'}(\mu_{12})+5~{\rm perm.}\right]\,.
\end{align}
We refer the reader to \citet{SlepianEisenstein2015B} for a detailed description of the practical evaluation of this expression. 

\MG{In the Gaussian approximation, the cross-covariance between the 2- and 3-point functions vanishes, as the only contributions arise from the bispectrum of the density field and the connected 5-point function \citep[see, e.g.,][]{SefusattiEtal2006}. While these terms are expected to introduce non-negligible correlations—potentially at the $\sim 20\%$ level for amplitude-related parameters—we neglect them in this work for consistency with the Gaussian covariance framework. This choice is justified by the scope of our analysis, which focuses on validating the modeling pipeline rather than delivering optimal cosmological constraints.}

%% file: Likelihood.tex
\section{Likelihood analysis and performance metrics}
\label{sec:likelihood}

We define in the section the likelihood function we adopt to fit the parameters of our model to the simulation measurements. We introduce as well the statistical tools we use to determine the range of validity of the model in terms of the goodness-of-fit, the level of theoretical uncertainties introduced by the model and the ability to constrain cosmological parameters. 

\subsection{Likelihood function}

We adopt a Gaussian likelihood function for both the 2PCF and the 3PCF measurements. We have then
\be
-2\ln \mathcal{L}_\xi(\vec{\theta}) = \chi_\xi^2 (\vec{\theta})\,,
\ee
where vector $\vec{\theta}$ denotes the model parameters while the $\chi^2$ for the 2PCF is given by
\begin{equation}
    \chi^2_{\mathrm{\xi}}(\vec{\theta}) = \sum_{i,j=1}^{N_{r}} \big[ \xi_i(\vec{\theta}) - \hat{\xi}_i \big] \ C^{-1}_{\xi, ij} \  \big[ \xi_j(\vec{\theta}) - \hat{\xi}_j \big]\,,
    \label{eq:chi2PCF}
\end{equation}
with $\xi(\vec{\theta})$ and $\hat{\xi}$ representing respectively the 2PCF theoretical prediction and its measurements and with the sum running over the separation bins $r_i$ and $r_j$, $N_r$ being the total number of bins, corresponding to the size of the data-vector. Similarly, for the 3PCF we have
\be
-2\ln \mathcal{L}_\zeta(\vec{\theta}) = \chi_\zeta^2 (\vec{\theta})\,,
\ee
with
\begin{equation}
    \chi^2_{\zeta}(\vec{\theta}) = \sum_{i,j=1}^{N_{\rm t}} \big[ \zeta_i(\vec{\theta}) - \hat{\zeta}_i \big] \ C^{-1}_{\zeta, ij} \ \big[ \zeta_j(\vec{\theta}) - \hat{\zeta}_j \big]\,,
    \label{eq:chi3PCF}
\end{equation}
with the sum now extending over all triangular configuration up to their total number $N_{\rm t}$. 

The likelihood of joint measurements of the 2PCF and 3PCF is simply given by
\be
\mathcal{L}_{\xi+\zeta}(\vec{\theta}) = \mathcal{L}_{\xi}(\vec{\theta}) \, \mathcal{L}_{\zeta}(\vec{\theta})\,,
\ee
and, correspondingly, the joint $\chi^2$ will be
\begin{equation}
    \chi^2_{\mathrm{\xi}+\mathrm{\zeta}}(\vec{\theta}) = \chi^2_{\xi}(\vec{\theta}) + \chi^2_{\zeta}(\vec{\theta}) \, ,
    \label{eq:chijoint}
\end{equation}
since, as mentioned above, we neglect the cross-covariance between the two statistics.

We sample the posterior using the \texttt{emcee} code\footnote{\url{https://emcee.readthedocs.io/}} \citep{ForemanMackeyEtal2013}, implementing an affine-invariant ensemble sampler, particularly well-suited to high-dimensional parameter spaces. We employ 50 walkers in each run and ensure that the chains are terminated only after exceeding 100 integrated autocorrelation times, thereby ensuring full convergence.

The model parameters, along with uniform priors assumed in all likelihood evaluations, are outlined in Table \ref{tab:priors}. The choice of large intervals is meant to ensure uninformative priors.

\subsection{Performance metrics \label{subsec:perf_met}}

Our main goal is the assessment of the range of validity of the model, determining the minimum scale, $r_{\mathrm{min}}$, down to which the model can still accurately describe the measurements without introducing significant systematic uncertainties to the recovered cosmological parameters. As in the companion paper \citep{PezzottaEtal2024}, this task is carried out using three performance metrics: a goodness of fit (GoF), a figure of bias (FoB), and figure of merit (FoM) as defined below.

\subsubsection{Goodness of fit}

For an assessment on the goodness of fit, we simply use the standard $\chi^2$ test. The $\chi^2$ definitions for the 2PCF, 3PCF, and joint at a given point $\vec{\theta}$ in the parameter space are given in Eqs.~\eqref{eq:chi2PCF}, \eqref{eq:chi3PCF}, and \eqref{eq:chijoint}. The posterior-averaged value, $\langle\chi^2\rangle_{\rm post}$, is then compared to the thresholds from the $\chi^2$ distribution at the 68\% and 95\% confidence levels, determined in terms of the appropriate number of degrees of freedom, computed as the difference between the size of the data vector and the number of model parameters.

\subsubsection{Figure of bias}

We quantify the uncertainties in recovering a subset of fiducial, cosmological parameters $\vec{\theta}_{\rm fid}$ due to model systematic uncertainties in terms of a figure of bias (FoB) defined as
\begin{equation}
    {\rm FoB}(\vec{\theta}) = \left[ \left( \langle \vec{\theta} \rangle_{\rm post} - \vec{\theta}_{\rm fid} \right)^{\rm T} S^{-1}(\vec{\theta}) \left( \langle \vec{\theta} \rangle_{\rm post} - \vec{\theta}_{\rm fid} \right) \right]^{1/2},
    \label{eq:fob}
\end{equation}
where, $\langle \vec{\theta} \rangle_{\rm post}$ represents the parameter averaged over the posterior distribution while $S(\theta)$ is the parameters' covariance matrix, evaluated as well from the same posterior distribution.

When $\theta$ consists of a single parameter, the FoB measures the deviation of the posterior mean from the fiducial value in units of the parameter's standard deviation, with FoB values of one and two corresponding to the 68\% and 95\% confidence levels, respectively. However, when dealing with multiple parameters, these thresholds must be recalculated by integrating a multivariate normal distribution over the appropriate number of dimensions. For instance, in the case of three parameters relevant for the results of Sect.~\ref{sec:results_cosmo}, the 68\% and 95\% confidence levels correspond to FoB values of 1.87 and 2.80, respectively.

\begin{table}[t]
\caption{Cosmological and nuisance parameters for the 2PCF and 3PCF model described in Sect.~\ref{sec:theo_model}. The last column provides the interval defining the uniform priors assumed for the analysis, or the bias relation adopted in some runs to reduce the parameters space.}
\renewcommand{\arraystretch}{1.2}
\centering
  \begin{tabular}{|l|l|}
    \hline
    \rowcolor{blue!5}
      Parameter & Prior\\
     \hline
     \rowcolor{blue!5}
    \multicolumn{2}{|l|}{\em Cosmology} \\
    \hline
     $10^{-9}A_s$ & \small{$\mathcal{U}\,[1,3]$}\\
     $h$ & \small{$\mathcal{U}\,[0.50, 0.80]$}\\
     $\omega_\mathrm{cdm}$ & \small{$\mathcal{U}\,[0.09,0.15]$} \\
    \hline
    \rowcolor{blue!5}
    \multicolumn{2}{|l|}{\em Nuisance parameters (bias \& counterterm)} \\
    \hline
     $b_1$ & \small{$\mathcal{U}\,[1,5]$}\\
     $b_2$ & \small{$\mathcal{U}\,[$-$20,20]$}\\
    $b_{\mathcal{G}_2}$ & \small{$\mathcal{U}\,[$-$20,20]$ or fixed to Eq. \eqref{eq:bg2_bias_rel}}\\
     $b_{\Gamma_3}$ & \small{$\mathcal{U}\,[$-$20,20]$ or fixed to Eq. \eqref{eq:bg3_bias_rel}}\\
    \hline
    $c_0\,\big[h^{-2}{\rm  Mpc^2}\big]$ & \small{$\mathcal{U}\,[$-$100,100]$}\\    
    \hline
  \end{tabular}
  \label{tab:priors}
\end{table}

\subsubsection{Figure of merit}

\begin{figure*}[h!]
    \centering
\includegraphics[width=0.89\textwidth]{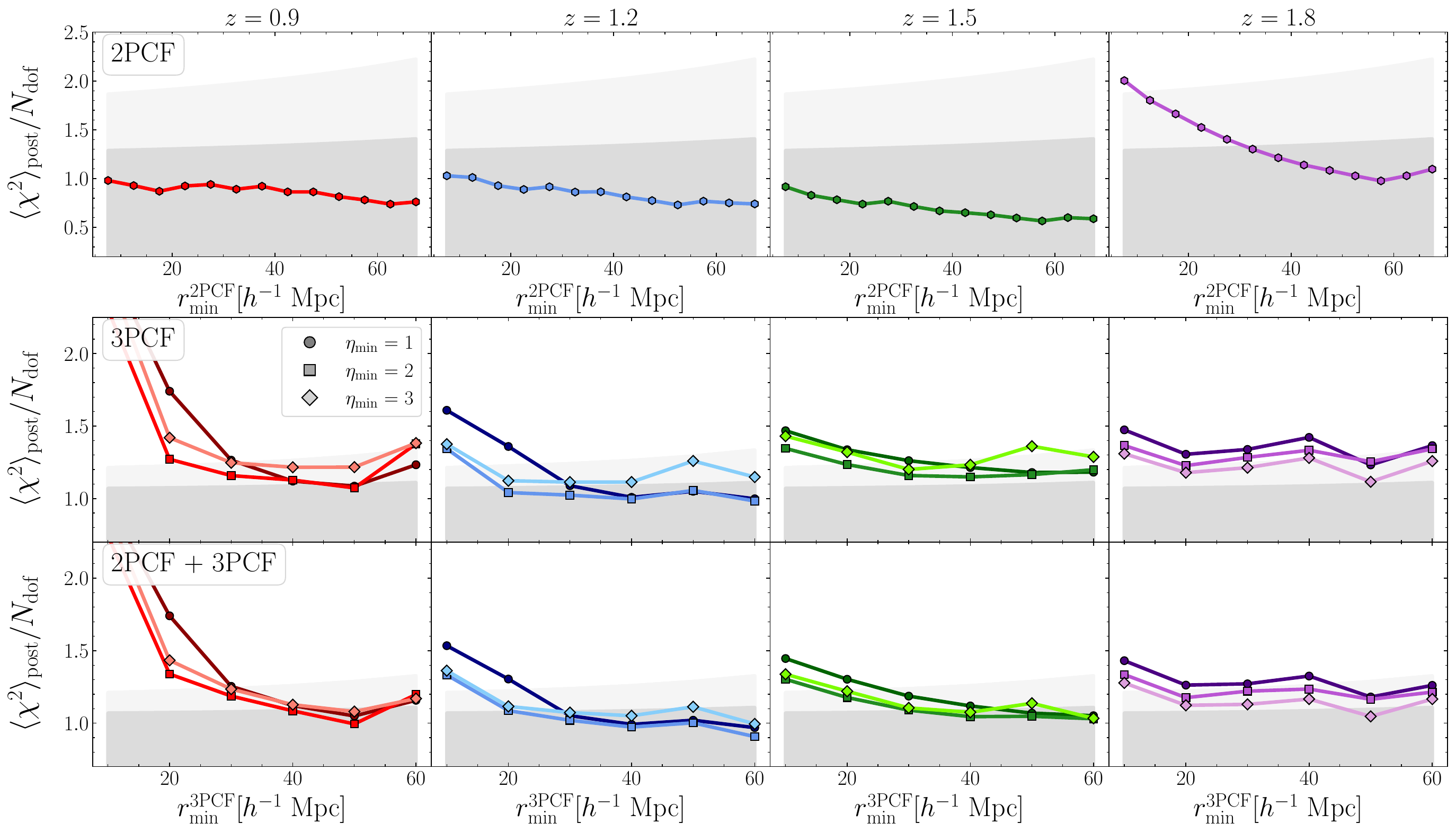}
    \caption{Goodness-of-fit test for the bias 2PCF and 3PCF model with fixed cosmological parameters as a function of the minimal scale, $r_{\rm min}^{\rm 2PCF}$ and $r_{\rm min}^{\rm 3PCF}$, respectively, assumed for the analysis. The posterior-averaged $\chi^2$ is compared to the 68\% and 95\% confidence intervals for the $\chi^2$ distribution, denoted by the gray shaded areas, given the corresponding $N_{\rm dof}$ (see text, \MG{fixing the case $\eta_{\rm min} = 1$ for the 3PCF confidence levels}). For the 3PCF, we consider three different values for the $\eta_{\rm min}$ parameter, Eq.~\eqref{eq:eta}, progressively excluding near-isosceles configurations. In the joint analysis, the minimal scale for the 2PCF is kept fixed at $r_{\rm min}^{\rm 2PCF}=25\Mpc$, with the results shown as a function of $r_{\rm min}^{\rm 3PCF}$ alone. The four columns span the four simulations snapshot at redshift $z=0.9$, 1.2, 1.5, and 1.8. 
    }
    \label{fig:bias_chi2}
\end{figure*}

\begin{figure*}[t]
    \centering
\includegraphics[width=0.494\textwidth]{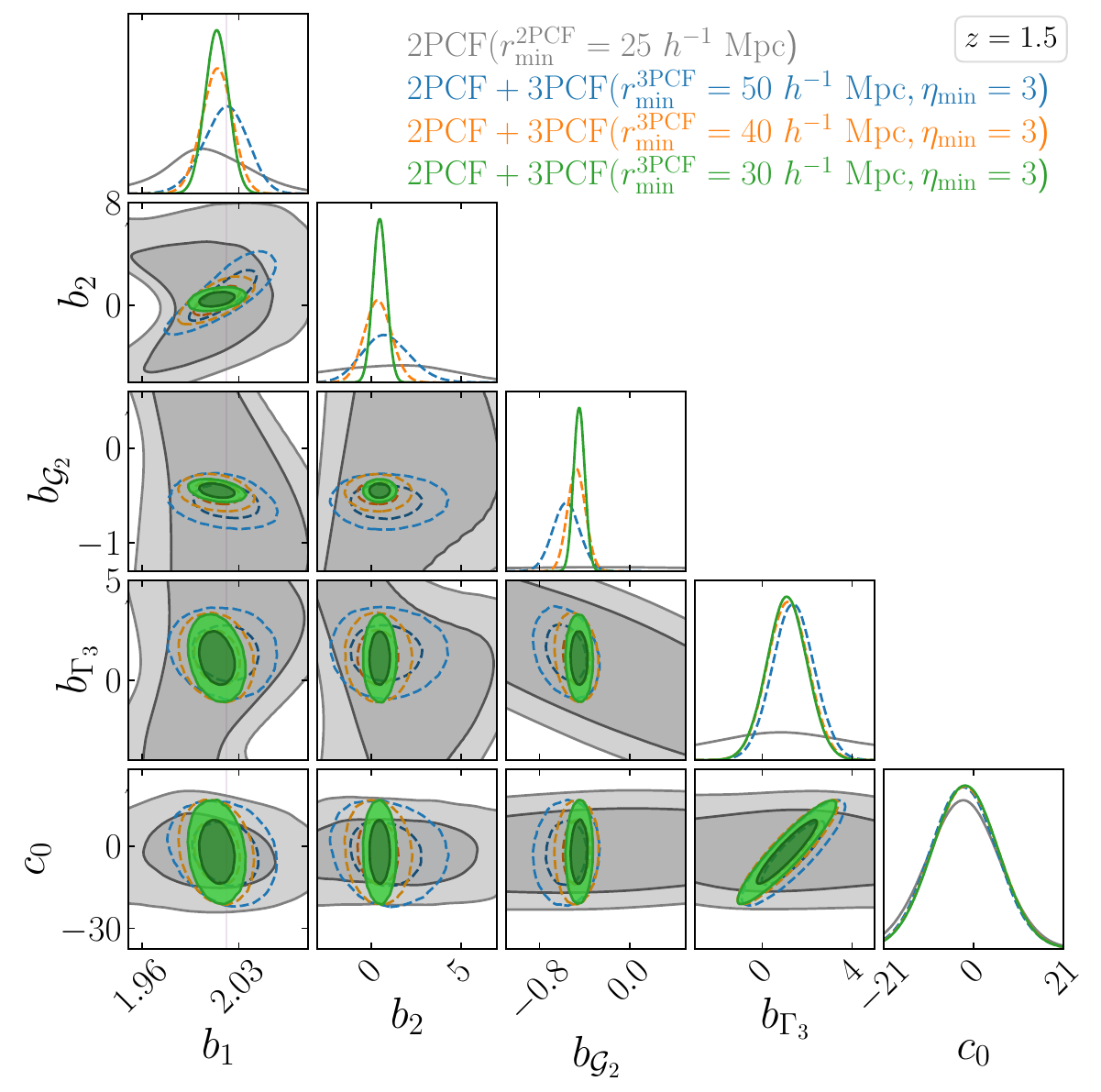}
\includegraphics[width=0.494\textwidth]{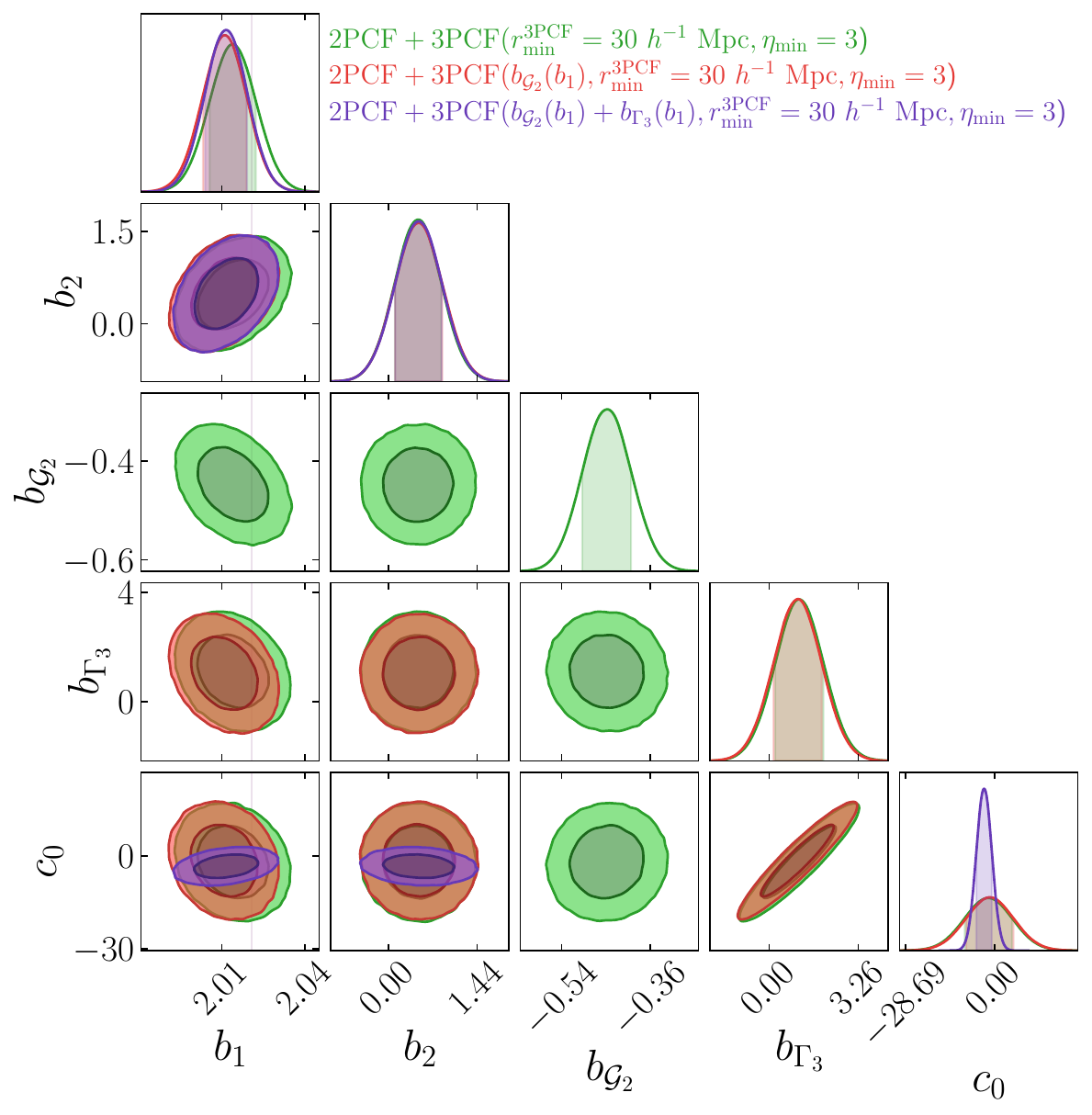}
    \caption{{\em Left panel}: marginalized 2D constraints at confidence intervals of $68.3\%$ and $99.5\%$, on the nuisance parameters (bias and counterterm) from the 2PCF analysis (gray shaded areas) against the joint 2PCF and 3PCF analysis for several values of $\rmin^{\rm \,3PCF}$ at $z=1.5$ (respectively blue dashed lines, orange dashed lines and green shaded areas). All cases assume $\eta=3$, see Eq.~\eqref{eq:eta}. The vertical gray band is an estimate of the linear bias $b_1$ obtained from a comparison of the galaxy power spectrum with measurements of the matter power spectrum in \citet{PezzottaEtal2024}. {\em Right panel}: marginalized 2D constraints at confidence intervals of $68.3\%$ and $99.5\%$, on the nuisance parameters (bias and counterterm) from the joint 2PCF-3PCF analysis at $z=1.5$ for \MG{$\rmin^{\rm \,3PCF}=30\Mpc$} (red contours) compared to the same set-up but with the assumption of the $b_{\mathcal{G}_2}(b_1)$ bias relation alone (\MG{red} contours) and the combination of the two relations $b_{\mathcal{G}_2}(b_1)$ and $b_{\Gamma_3}(b_1)$ (violet contours).}
    \label{fig:bias_3pcf_vs_2pcf}
\end{figure*}

Finally, we define a figure of merit \citep{AlbrechtEtal2006a, Wang2008} to quantify the constraining power of our model over a specific range of scales and under given bias assumptions
\begin{equation}
    {\rm FoM}(\vec{\theta}) = \frac{1}{\sqrt{\det[S(\vec{\theta})]}} \, ,
\end{equation}
where $\det[S(\theta)]$ is the determinant of the covariance matrix of the parameters of interest. This represents the hyper-volume enclosed by the hyper-surface defined by the covariance matrix $S(\vec{\theta})$. Therefore, a larger FoM indicates tighter constraints on the model parameters.

%% file: Results.tex
\section{\label{sec:results_bias} Galaxy bias}

In this section we assess the range of validity of our bias model for both the 2PCF and 3PCF. Our focus will be then on the goodness-of-fit test, keeping fixed the cosmological parameters at their fiducial values. We consider both the case where all bias parameters are free as well as in the case where the bias relations mentioned above is assumed to reduce the parameter space. We also need to make sure that no significant running of the bias parameters is present, \MG{i.e. no signifcant variation in the recovered parameters}, as a function of the minimal scale defining the data vectors.

\subsection{Maximal model}

We start with the maximal model, with all bias parameters free, with the priors given in Table~\ref{tab:priors}.  In Fig.~\ref{fig:bias_chi2} we show the $\chi^2$ averaged over the posterior distribution for the four redshift snapshots and for the 2PCF, 3PCF and the joint 2PCF+3PCF data vector. 

We notice in the first place that while the one-loop 2PCF model provides a good fit for all scales considered, the tree-level 3PCF prediction fails for separations below $20\Mpc$ at $z=0.9$, with a better performance at higher redshift. Under the assumption of $\eta_{\mathrm{min}}=1$, Eq.~\eqref{eq:eta}, including configurations closer to the isosceles case, the validity of the model is further restricted to $r_{\rm min}>20\Mpc$ at $z=0.9$ and $r_{\rm min}>10\Mpc$ for $z=1.2$ and above.

In Fig. ~\ref{fig:bias_3pcf_vs_2pcf} we show the marginalized 2D constraints on the nuisance parameters (bias and counterterm) from the 2PCF analysis against the joint 2PCF plus 3PCF analysis for several values of $\rmin^{\rm \,3PCF}$. This shows in the first place the peculiar and significant constraining power of higher-order statistics on linear and nonlinear bias. This is particularly evident in the breaking of the bimodality of the constraints on $b_2$ and in the reduction of the posterior uncertainty on the non-local parameters $b_{\mathcal{G}_2}$ and $b_{\Gamma_3}$. It further shows that down to \MG{$\rmin^{\rm \,3PCF}=30\Mpc$} no running of the bias parameter is evident, with all analyses at different $r_{\rm min}$ consistently reducing the posterior uncertainty.

The plots include, as a vertical gray band, an estimate of the value of the linear bias $b_1$ obtained from a comparison of the galaxy power spectrum with measurements of the matter power spectrum in \citet{PezzottaEtal2024}. \MG{We find that the inclusion of the 3PCF consistently yields posterior distributions for $b_1$ that are visually closer to the reference value across all the minimum scale cuts explored. While the 2PCF-only results remain statistically consistent, they exhibit broader posteriors, reflecting a stronger degeneracy with other model parameters. We will revisit this aspect in the context of cosmological parameter constraints in the following section.}


\subsection{Bias relations}

The right panel of Fig. ~\ref{fig:bias_3pcf_vs_2pcf} shows the same results for the joint analysis with \MG{$\rmin=30\Mpc$ and $\eta_{\mathrm{min}}=3$} (green contours), here compared to the same set-up but with the additional assumptions of two bias relations. We consider, in particular, the $b_{\mathcal{G}_2}(b_1)$ relation alone, \MG{Eq. \eqref{eq:bg2_bias_rel}, (red contours) and its combination with the relation for $b_{{\Gamma}_3}$ of Eq. \eqref{eq:bg3_bias_rel}}. We display only the results for the $z=1.5$ snapshots, noticing that for the measurements at other redshifts we obtain qualitatively similar results. 

\begin{figure*}[h!]
    \centering
    \includegraphics[width=1.\textwidth]{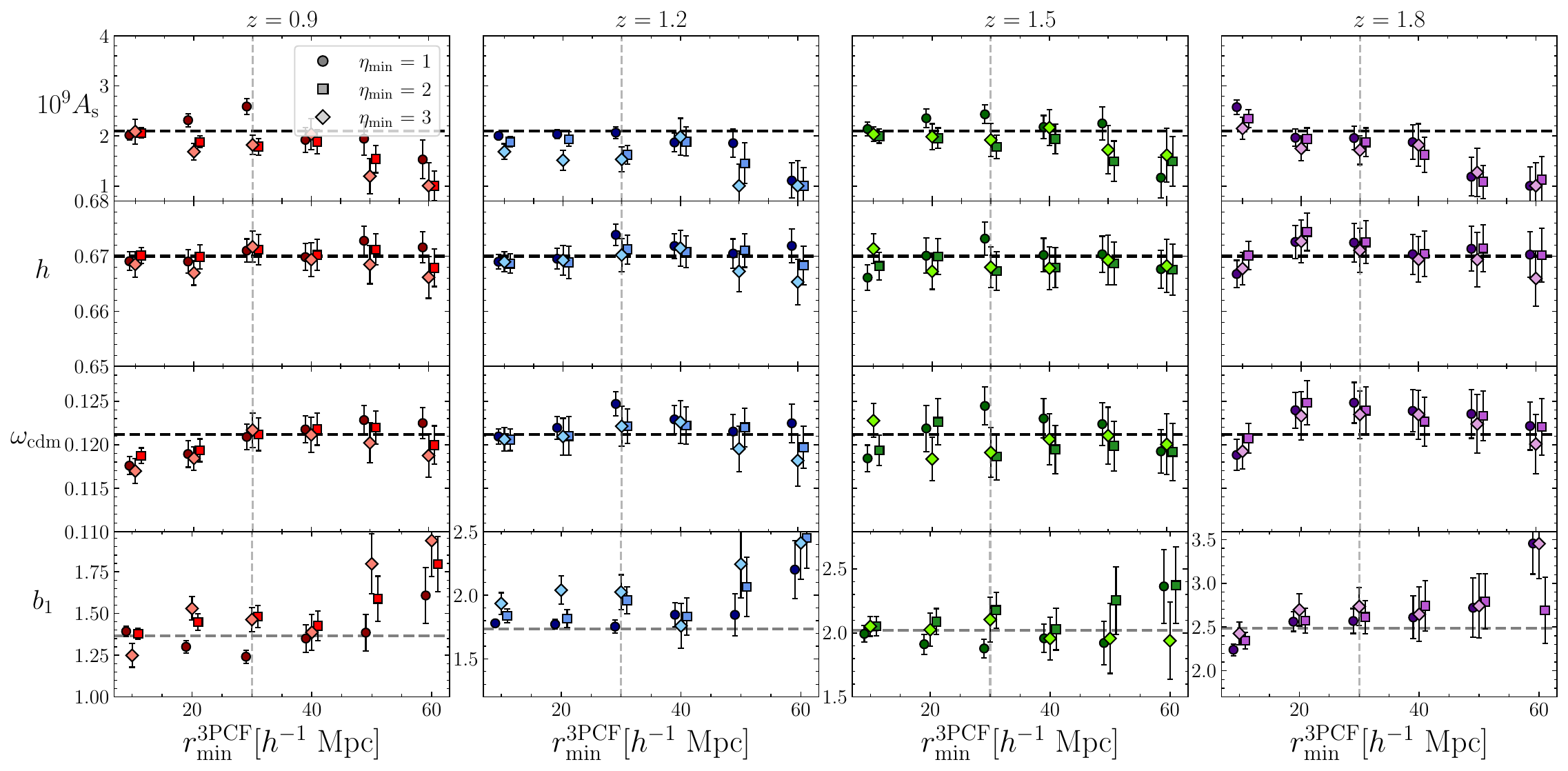}
    \caption{Marginalised constraints on the cosmological parameters and linear bias from a joint 2PCF and 3PCF analysis. The posterior mean along with its uncertainties are displayed as a function of the minimal scale $\rmin^{\rm \,3PCF}$ for different choice of $\eta_{\rm min}$ and fixing $\rmin^{\rm \,2PCF}=25\Mpc$ and $\rmax = 140 \Mpc$ for both statistics. The black dashed line refers to the fiducial values assumed in the simulation, while the dashed grey line refers to the fiducial $b_1$ adopted in \cite{PezzottaEtal2024}, also shown in Fig. \ref{fig:bias_3pcf_vs_2pcf}.}
    \label{fig:joint_best_fit_cosmo}
\end{figure*}

\begin{figure*}
    \centering
    \includegraphics[width=1.0002\textwidth]{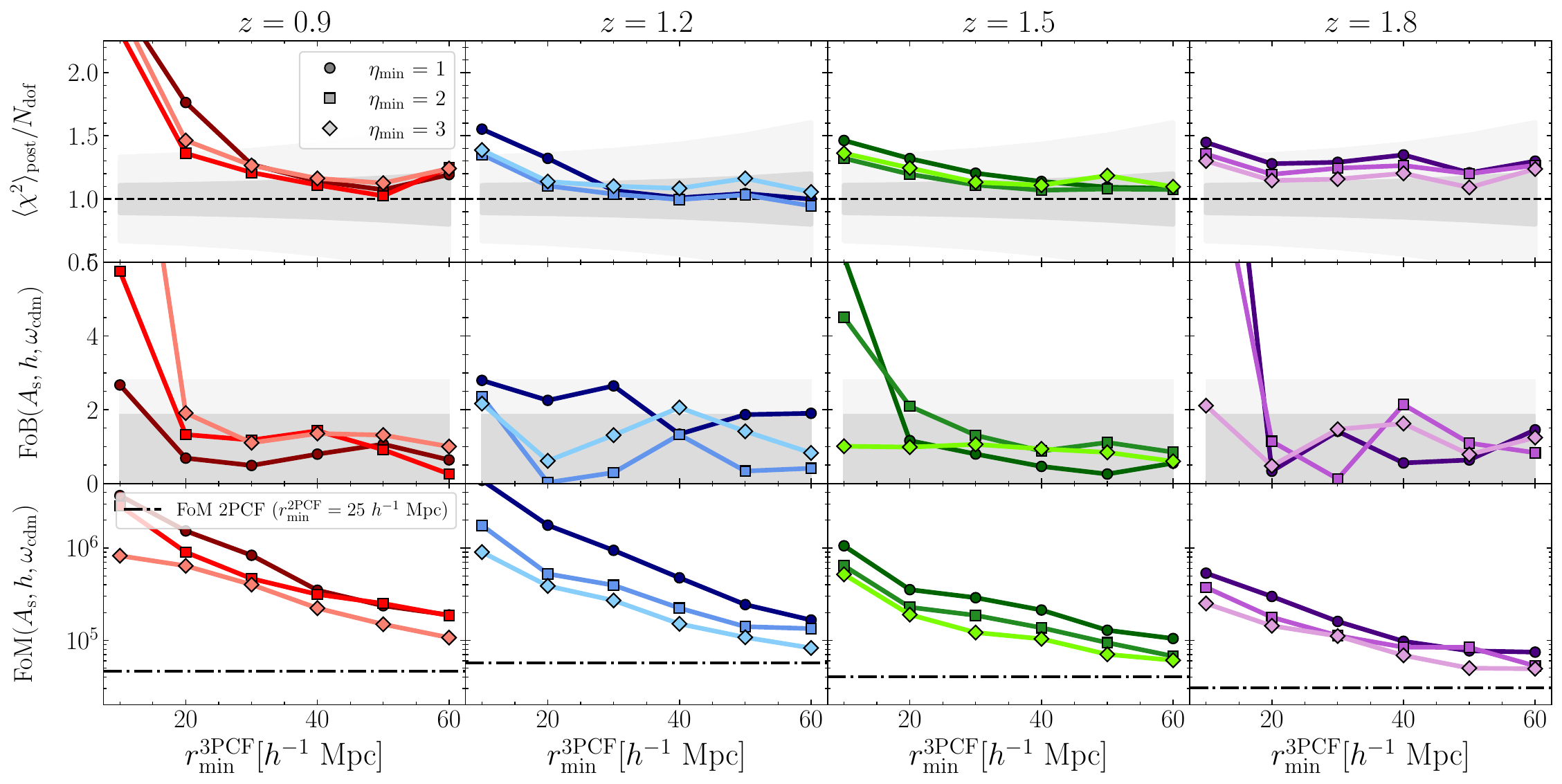}
    \caption{Performance metrics for the full-shape, joint analysis of the 2PCF and the 3PCF. All quantities are shown as a function of the minimal scales $\rmin^{\rm \,3PCF}$, with $r^{\mathrm{2PCF}}_{\mathrm{min}}$ fixed at $25 \ h^{-1} \ \mathrm{Mpc}$, and for the same three different values of $\eta_{\rm min}$ in different colour shades. From top to bottom we show the posterior-averaged $\langle\chi^2\rangle_{\rm post}$ for the GoF, the FoB and the FoM, these last two both defined in terms of the three cosmological parameters  $A_{\rm s}$, $\omega_{\rm cdm}$, and $h$. The grey bands denote the 68\% and 99.7\% confidence levels for the $\chi^2$ distribution and the FoB.}
    \label{fig:joint_perf_met_cosmo}
\end{figure*}

We notice that in both cases, the bias relations provide a reduction of the parameter space by one and two parameters, respectively, without affecting the $b_1, b_2$ constraints. On this specific contour in fact we do not observe any induced systematic shift as we do not notice any appreciable reduction in the uncertainty. This is noticeable instead on the $c_0$ parameter, describing the combined effect of the EFT counterterm and higher derivative bias. These results provide some evidence that the bias relations can be used to speed up the likelihood evaluation without compromising the main results. We will return to this point when discussing cosmological parameters constraints.

We close the section by noticing that the adoption of one or two bias relations does not lead to any worsening of the goodness of fit, in terms of the $\chi^2$ test. We avoid a dedicated figure since the difference with the maximal model case would be barely visible.

\section{\label{sec:results_cosmo} Cosmological parameters from the joint, full-shape analysis}

In this section we present the results for the full-shape, joint analysis of 2PCF and 3PCF aiming at constraining three cosmological parameters:  the amplitude of scalar fluctuations, $A_{\rm s}$, the cold dark matter density $\omega_\mathrm{cdm}$, and the Hubble constant, $h$. To the best of our knowledge, this is the first attempt at estimating the improvement due to adding the 3PCF to the full-shape analysis of the 2PCF, although still limited to real space. \brown{Until now, 3PCF analyses have been restricted to template fitting, i.e. using fixed-shape models to extract cosmological information from the BAO scale position or from the anisotropy induced by redshift-space distortions, rather than performing full-shape parameter inference}, due to the considerable computational demands of a full cosmology-dependent 3PCF model. 

A key ingredient, to achieve this goal, is the emulator described in the Appendix \ref{appendix:emulator}. Constructed using a \texttt{PyTorch}\footnote{\url{https://pytorch.org/docs/stable/index.html}} framework, the emulator has undergone extensive testing and has been fully integrated into the same sampler.  The emulator extends to the 2PCF prediction, further enhancing the overall efficiency of our analysis.

Figure ~\ref{fig:joint_best_fit_cosmo} shows marginalized constraints on cosmological parameters $A_{\rm s}$, $\omega_{\mathrm{cdm}}$, $h$, and on the linear bias $b_1$ derived from the joint 2PCF and 3PCF analysis as a function of the minimal scale $\rmin$ that characterizes the 3PCF data vector, with the minimal separation in the 2PCF measurements fixed at $\rmin^{\rm \,2PCF}=25\Mpc$. The four columns span the different redshifts. As in Fig.~\ref{fig:bias_chi2}, varying shades of these colours correspond to different choices for the parameter $\eta_{\rm min}$. The black dashed lines denote the fiducial values assumed for the simulations, together with the gray line that shows the same fiducial values for $b_1$ considered in Fig. \ref{fig:bias_3pcf_vs_2pcf}. Overall, the results show good agreement between the emulated predictions and the expected values. For large values of $\rmin$, the estimates for $A_{\rm s}$ and $b_1$ tend to deviate further from the expected values, largely due to projection effects caused by a large degeneracy between these parameters. On the other hand, moving to the small-scale regime, a possible failure of the model to recover the fiducial values is only evident for $\omega_{\rm cdm}$ at the lowest redshift of $z=0.9$.  Below $\rmin = 40 \Mpc$ the estimates for $A_{\rm s}$ and $b_1(z)$ exhibit a slight dependence on the choice of $\eta_{\rm min}$, with the case of $\eta_{\rm min}=1$  leading to more biased values for these parameters, particularly at low redshift. 

In Fig.~\ref{fig:joint_perf_met_cosmo}, we present the performance metrics corresponding to the results of Fig.~\ref{fig:joint_best_fit_cosmo}, all as a function of the minimal scales $\rmin^{\rm \,3PCF}$, with $r^{\mathrm{2PCF}}_{\mathrm{min}}$ fixed at $25 \ h^{-1} \ \mathrm{Mpc}$, and for the same three different values of $\eta_{\rm min}$. The top panels show the posterior-averaged $\langle\chi^2\rangle_{\rm post}$. The grey bands denote the 68\% and 99.7\% confidence levels for the $\chi^2$ distribution. We find a good fit for $\rmin^{\rm \,3PCF}\gtrsim 20\Mpc$ except for the lowest redshift where $\rmin^{\rm \,3PCF}\gtrsim 30\Mpc$ ensures a safer result. This is naturally consistent with the results at fixed cosmology. The middle panels show the figure of bias, defined in terms of the parameters $A_{\rm s}$, $\omega_{\rm cdm}$, and $h$. The uncertainties are shown again in two shades of grey, corresponding to 68\% and 95\% confidence levels. The model provides overall unbiased results except for the first separation bin at $\rmin=10\Mpc$, for all redshifts and for all values of $\eta_{\rm min}$. Finally, the bottom panels of Fig.~\ref{fig:joint_perf_met_cosmo} show the figure of merit, defined again in terms of the three cosmological parameters. The FoM increases significantly across the whole separation range (notice the log-scale on the $y$-axis) and presents also larger values in the lowest redshift snapshots. In addition, the choice of a low value for $\eta_{\rm min}$ can also make a large difference, since this increase the size of the data vector and relative $\rm S/N$.

\begin{figure*}[t]
    \centering
    \includegraphics[width=0.48\textwidth]{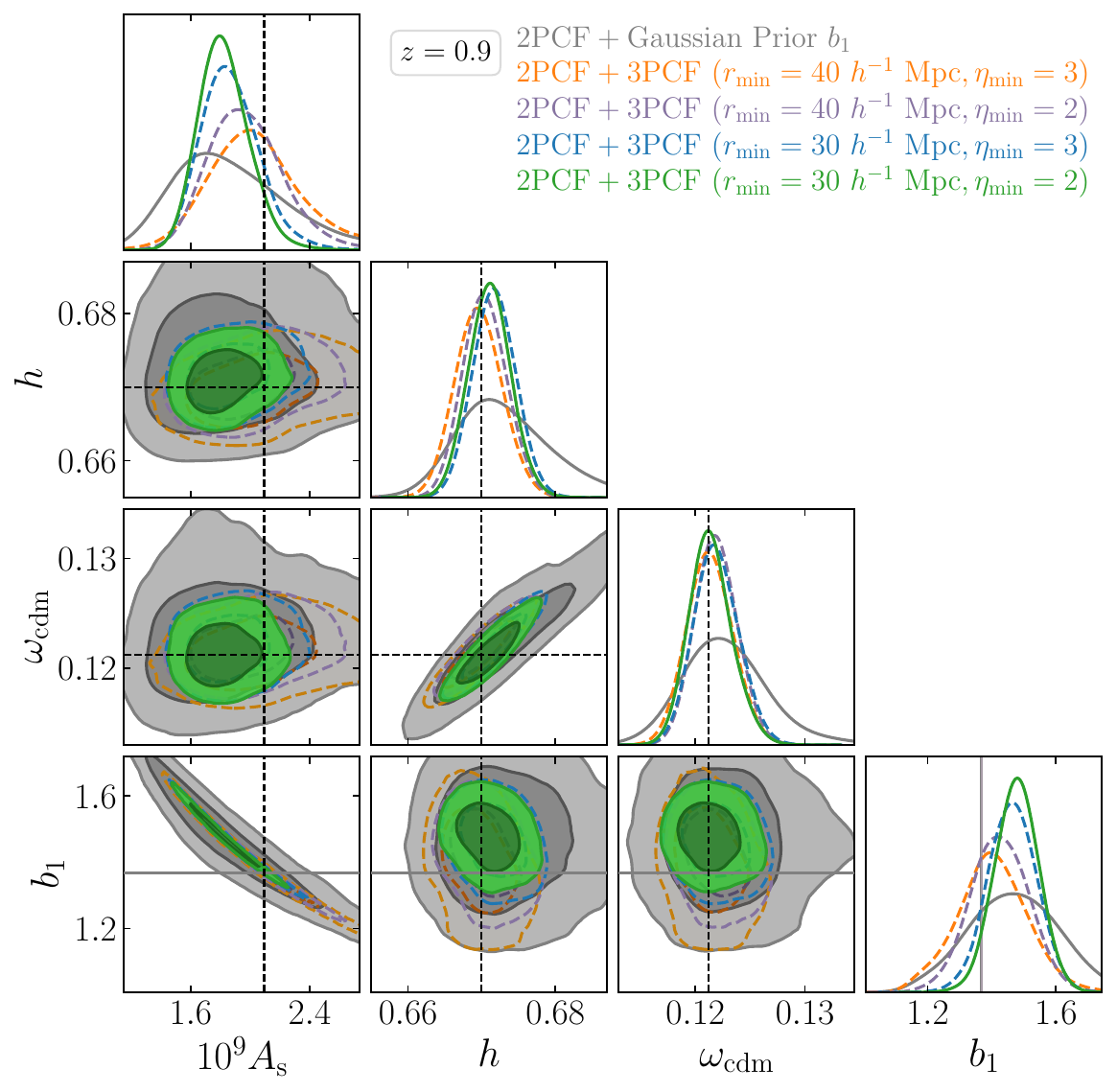}
    \includegraphics[width=0.48\textwidth]{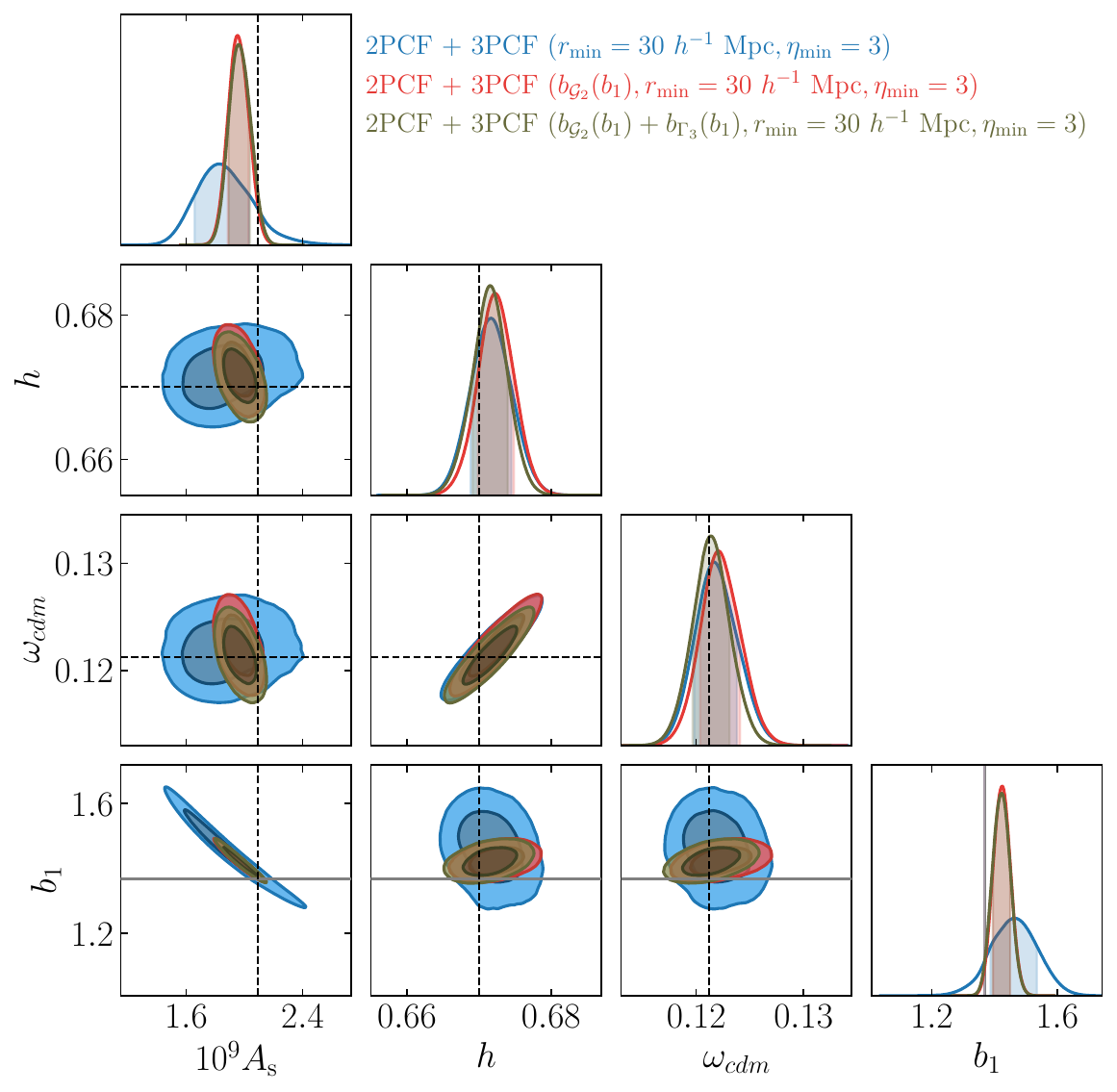}
    \caption{\textit{Left panel}:Marginalized 2D and 1D posterior distributions \MG{at $z = 0.9$} from the joint 2PCF and 3PCF analysis of the cosmological parameters $h$, $\omega_{\rm cdm}$ and $A_{\rm s}$ and the linear bias parameter $b_1$ for various choices for the 3PCF scale cut $\rminthree$ and the $\eta_{\rm min}$ parameter defined in Eq.~\eqref{eq:eta}, fixing $\rmintwo = 25 \Mpc$. Gray contours correspond to the 2PCF-only analysis assuming a Gaussian prior on $b_1$ given by a 10\% uncertainties around the best-fit value obtained from the joint analysis. Each panel shows the results for a different snapshot. Vertical and horizontal lines denote the fiducial values of the cosmological parameters. \textit{Right panel}: Marginalized 1D and 2D posterior distributions of the same subset of model parameters of the left panel, for the combined analysis of 2PCF and 3PCF at $z=0.9$ fixing $\rmintwo = 25 \Mpc$. The reference case of the maximal model with $\rminthree = 30 \Mpc$, and $\etamin = 3$ is compared to the results obtained adopting the bias relations of Eq.~\eqref{eq:bg2_bias_rel}, in red,  and its combination with Eq.~\eqref{eq:bg3_bias_rel}, in brown.}
    \label{fig:contours_cosmo_2_3_joint_main}
\end{figure*}

\MG{The left plot of Fig.}~\ref{fig:contours_cosmo_2_3_joint_main} shows the constraints on the cosmological parameters \MG{at $z = 0.9$} plus the linear bias obtained from the combined statistics for different choices of $\rminthree$ and $\eta_{\rm min}$, compared to the 2PCF-only constraints where we include an additional, informative Gaussian prior on $b_1$ \MG{for illustrative purposes} defined by a 10\% error on the best-fit value obtained from the joint analysis. This is required by the strong degeneracy between $A_{\rm s}$ and $b_1$ in the 2PCF likelihood. Such degeneracy is still present in the joint analysis, although significantly reduced \MG{without imposing any prior on $b_1$}. We obtain a discrepancy with the fiducial value of $A_{\rm s}$ at confidence interval of $68.3\%$ for the configuration with $\rminthree=30\Mpc$ and $\eta_{\rm min}=2$, \MG{while the discrepancy is smaller for the other parameters}, despite the additional degeneracy between $h$ and $\omega_{\rm cdm}$. Interestingly, no tension is visible on the $h$ and $\omega_{\mathrm{cdm}}$ constraints. 
 
The same results are found with no relevant difference for all snapshots, \MG{as described in Fig. \ref{fig:contours_cosmo_2_3_joint} in Appendix \ref{sect:appendixcosmo}}. We remind the reader that the posterior distributions are not statistically independent since each realisation is a different snapshot of the same evolving dark matter perturbations, sharing the same initial seeds. We observe that at lower redshift the constraints are tighter due to the larger density of the galaxy distribution and that, in general, both $h$ and $\omega_{\rm cdm}$ exhibit accurate estimates of the fiducial values. The largest discrepancy, at the limit of the confidence interval of $99.5\%$ contour, is obtained for $A_{\rm s}$ at $z = 1.2$. We notice that now also the linear bias parameter is correctly recovered at all redshifts.

\MG{In the right plot of Fig.} ~\ref{fig:contours_cosmo_2_3_joint_main}, we compare the maximal model with all free parameters, assuming $\rminthree = 30 \Mpc$ and $\etamin = 3$, to the results obtained assuming the bias relation of Eq.~\eqref{eq:bg2_bias_rel} for $b_{{\mathcal G}_2}$, and its combination with the relation for $b_{{\Gamma}_3}$ of Eq. \eqref{eq:bg3_bias_rel}, \MG{showing the same set of parameters as in the right plot}. Overall, we notice how the adoption of the bias relations enhances the precision of the cosmological parameter constraints. Notably, the 2D posterior distribution of $A_{\rm s}$ and $b_1$, affected by a strong degeneracy, benefits the most from this additional information, although, again, we notice a slightly biased estimate for $A_{\rm s}$. \MG{As presented in Fig. \ref{fig:contour_full_bias_045}, showing the full set of parameters, } as in the bias-only case, the $b_{\Gamma_3}$ relation, greatly reduces the uncertainty on the higher derivatives/counterterm parameter $c_0$.

%% file: Conclusion.tex
\section{\label{sec:conclusion} Conclusions}

In this work, we provided a validation of perturbation theory models for the 2PCF and 3PCF in real-space across a redshift range and against a synthetic galaxy catalogue representative of the \Euclid spectroscopic galaxy survey \citep[see, e.g.,][]{MellierEtal2024}.  

We take advantage of galaxy clustering measurements obtained from four redshift snapshots of the Flagship I simulation populated with the HOD following the Model 3 prescription of \citet{PozzettiEtal2016}. In particular, HOD parameters are determined by selecting galaxies with a \Ha flux limit of $f_{{\rm H}\alpha} = 2 \times 10^{-16} \ \text{erg} \ \text{cm}^{-2} \ \text{s}^{-1}$. The 3PCF measurements, presented in Sect.~\ref{sec:data}, in particular, are based on the spherical harmonics decomposition introduced by \citet{SlepianEisenstein2015B}, see also Euclid Collaboration: Veropalumbo et al. (in prep). This estimator allows us to measure all 3PCF configurations while keeping a manageable computational cost.

The galaxy 2PCF is modelled at next-to-leading order within the framework of EFTofLSS, involving up to five nuisance parameters: $b_1$, $b_2$, $b_{\mathcal{G}_{2}}$, $b_{\Gamma_3}$, and the parameter $c_0$ describing an EFT counterterm and higher-derivative bias. The 3PCF, on the other hand, is modelled at leading order in perturbation theory, with the tree-level expression depending only on the bias parameters $b_1$, $b_2$, and $b_{\mathcal{G}_2}$.

We first assessed, in Sect.~\ref{sec:results_bias}, the goodness of fit provided by the model in terms of a $\chi^2$ test and explored its range of validity as a function of the minimal scale included in the data vector and of the additional parameter $\eta_{\rm min}$, Eq.~\eqref{eq:eta}, progressively excluding nearly isosceles triangles from the analysis. The predictions for these configurations are, in fact, affected by systematic uncertainties when the Legendre expansion of Eq.~\eqref{eq:LegendreExp} is limited to a relatively small number of multipoles, as in any practical application. We find that the tree-level model for the 3PCF fails below separations of $20\Mpc$ when $\eta_{\min}\ge 2$, while a larger minimum scale should be considered for when $\eta_{\min}=1$. Commonly adopted relations among bias parameters such as those for $b_{{\mathcal{G}_2}}(b_1)$ and $b_{\Gamma_3}(b_1)$ can help to reduce the parameter space without affecting the determination of the other bias parameter, except for the $c_0$ parameter.

The main results of the paper, presented in Sect.~\ref{sec:results_cosmo}, consist however in the joint, full-shape analysis of 2- and 3-point correlation functions, aiming at constraining three cosmological parameters: $A_s$, $\omega_{\rm cdm}$, and $h$. Indeed, cosmological analyses involving the 3PCF have so far been based on template fitting \citep[see, e.g.,][]{SlepianEtal2017, SlepianEisenstein2018, VeropalumboEtal2021, SugiyamaEtal2021, FarinaEtal2024a}, as direct evaluation of the 3PCF model is computationally too expensive. To overcome this hurdle, in this work we developed an emulator able to provide fast and accurate full predictions for both 2PCF and 3PCF, thereby enabling a complete joint analysis of the two statistics.

In this case, in addition to the $\chi^2$ test, we adopt a figure of bias and a figure of merit to quantify the model systematic uncertainties along with its ability to constrain cosmological parameters. Focusing only on the results for the joint analysis, since only this configuration allows us to properly constrain all three parameters without the need for informative priors, we find that a conservative choice for the scale cuts is given by $\rminthree = 40 \Mpc$ and $\eta_{\rm min} = 3$. A more aggressive configuration with $\rminthree = 30 \Mpc$ and $\eta_{\rm min} = 2$ is still acceptable, although a tension at the confidence interval of $68.3\%$ with the fiducial value $A_s$ begins to show. This finding aligns with conclusions drawn from previous methodological studies on the galaxy bias model \citep[see, e.g.,][for further details]{VeropalumboEtal2022}. On the other hand, selecting $\eta_{\rm min} = 1$ led to more unstable results, particularly at lower redshift. 

The adoption of bias relations, while it  provides a limited advantage in a bias-only analysis, when cosmological parameters are included can lead to tighter constraints, most notably on $A_s$, particularly in the case of Eq.~\eqref{eq:bg2_bias_rel}.

In conclusion, we confirmed that, as expected, the combination of 2PCF and 3PCF in real space provides greater constraining power w.r.t. the 2PCF alone,  effectively breaking the degeneracy between $A_s$ and $b_1(z)$. We expect as well the improvement to be less pronounced in redshift space, where the 2PCF multipoles typically show reduced degeneracies among cosmological and bias parameters. Yet, this work represents a first step toward a full-shape joint analysis in redshift space that, out of necessity, will require tools as the emulator developed here.  In this perspective, this study paves the way for the application of emulated predictions in comparisons to real data, setting the stage for a still lacking, complete analysis of galaxy clustering data in configuration space, on the same footing as power spectrum and bispectrum joint analyses.

This work is part of a series of preparation papers aiming at validating the theoretical models for the analysis of galaxy clustering statistics from the \Euclid spectroscopic sample. A first assessment of the real-space model for the galaxy power spectrum has been presented in \citet{PezzottaEtal2024}. The cases of the redshift space power spectrum and 2PCF will be presented respectively in Euclid Collaboration: Camacho et al. (in prep.) and Euclid Collaboration: K{\"a}rcher et al. (in prep.). In addition, the redshift-space, joint power spectrum and bispectrum, and 2PCF-3PCF cases will appear in Euclid Collaboration: Pardede et al. (in prep.) and Euclid Collaboration: Pugno et al. (in prep.). Finally, another study will explore the BAO signal in the redshift-space 3PCF (Euclid Collaboration: Moresco et al., in prep).

%% file: Appendix.tex
\begin{appendix}

\section{Dataset}
\begin{figure}[ht]
    \centering
    \includegraphics[width=0.5\textwidth]{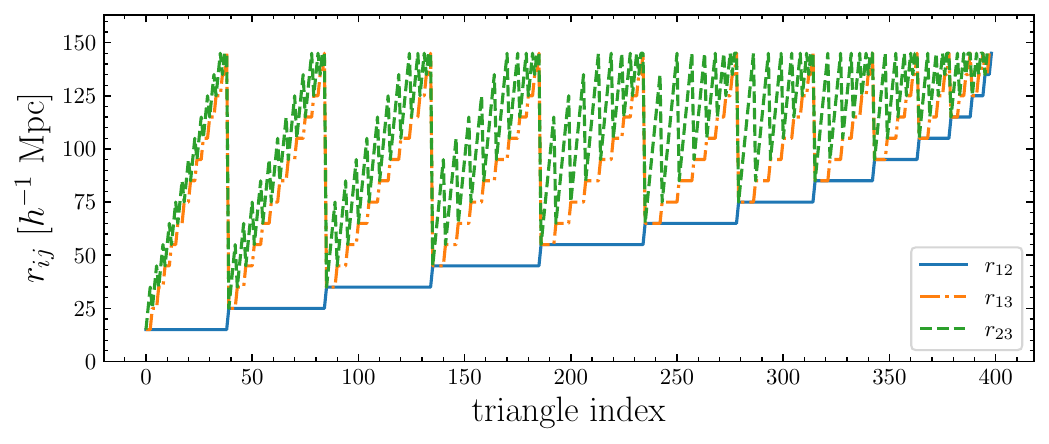}
    \caption{Triangle sides $r_{ij}$ as a function of the Triangle Index, following the condition $r_{12} \leq r_{13} \leq r_{23}$.}
    \label{fig:r_ij}
\end{figure}

\MG{In Fig. \ref{fig:r_ij}, we illustrate how the triangle sides, $r_{ij}$, vary as a function of the triangle index, with each side shown with dashed lines of different colors.}

\begin{figure*}[ht]
    \centering
    \includegraphics[width=.98\textwidth]{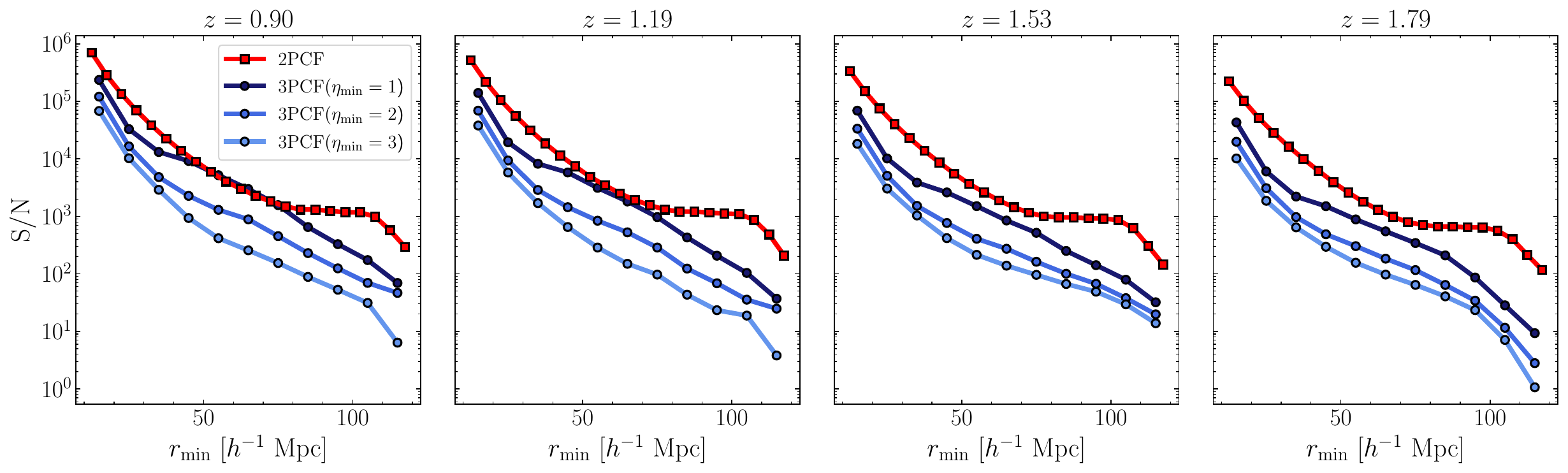}
    \caption{Cumulative $\rm S/N$ for the 2PCF and 3PCF measurements computed with the predicted Gaussian covariance according to Eqs.~\eqref{eq:StoNxi} and ~\eqref{eq:StoNzeta}, as a function of the minimal scale included, $\rmin$.}
    \label{fig:StoN}
\end{figure*}

\MG{In Fig. \ref{fig:StoN} we show the cumulative signal-to-noise ratio ($\rm S/N$) for the two statistics as a function of the minimal separation $\rmin$ computed using the expressions above for the covariance and the $\hat{\xi}$ and $\hat{\zeta}$ measurements as
\be
\label{eq:StoNxi}
\left(\rm \frac{S}{N}\right)_\xi^2\vcentcolon =\sum_{r,r'\ge\rmin}\hat{\xi}(r)\,C_\xi^{-1}(r,r')\,\hat{\xi}(r')\,,
\ee
and 
\be
\label{eq:StoNzeta}
\left(\rm \frac{S}{N}\right)_\zeta^2\vcentcolon =\sum_{r_{\rm ij},r_{\rm ij}'\ge\rmin}\hat{\zeta}(t)\,C_\zeta^{-1}(t,t')\,\hat{\zeta}(t')\,,
\ee
with $t$ and $t'$ corresponding respectively to the triplets $\left\{r_{12}, r_{13}, r_{23}\right\}$ and $\left\{r'_{12}, r'_{13}, r'_{23}\right\}$. The $\rm S/N$ for the 2PCF is, as might be expected, higher than the 3PCF, but only slightly, and less so at lower redshift. The two indeed become comparable at mildly nonlinear scales. We notice as well that the total signal for the subset of 3PCF configurations defined by $\eta_{\rm min}=2$ can be larger by a factor of a few w.r.t. the subset defined by $\eta_{\rm min}=2$, and, again, it increases at low $z$.}

\section{2PCF and 3PCF emulators}
\label{appendix:emulator}

\begin{figure*}[t]
\includegraphics[width=0.5\textwidth]{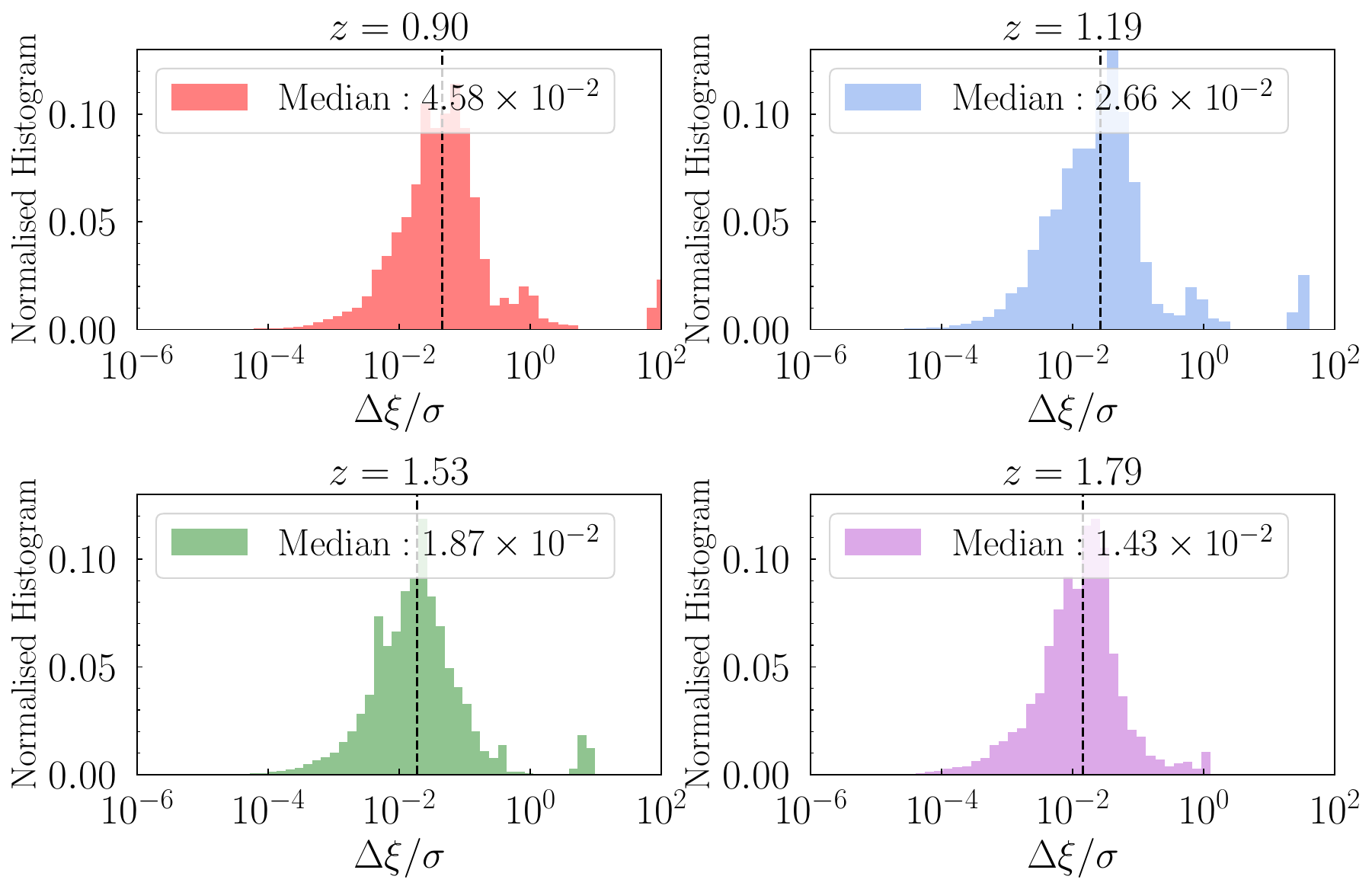}
\includegraphics[width=0.5\textwidth]{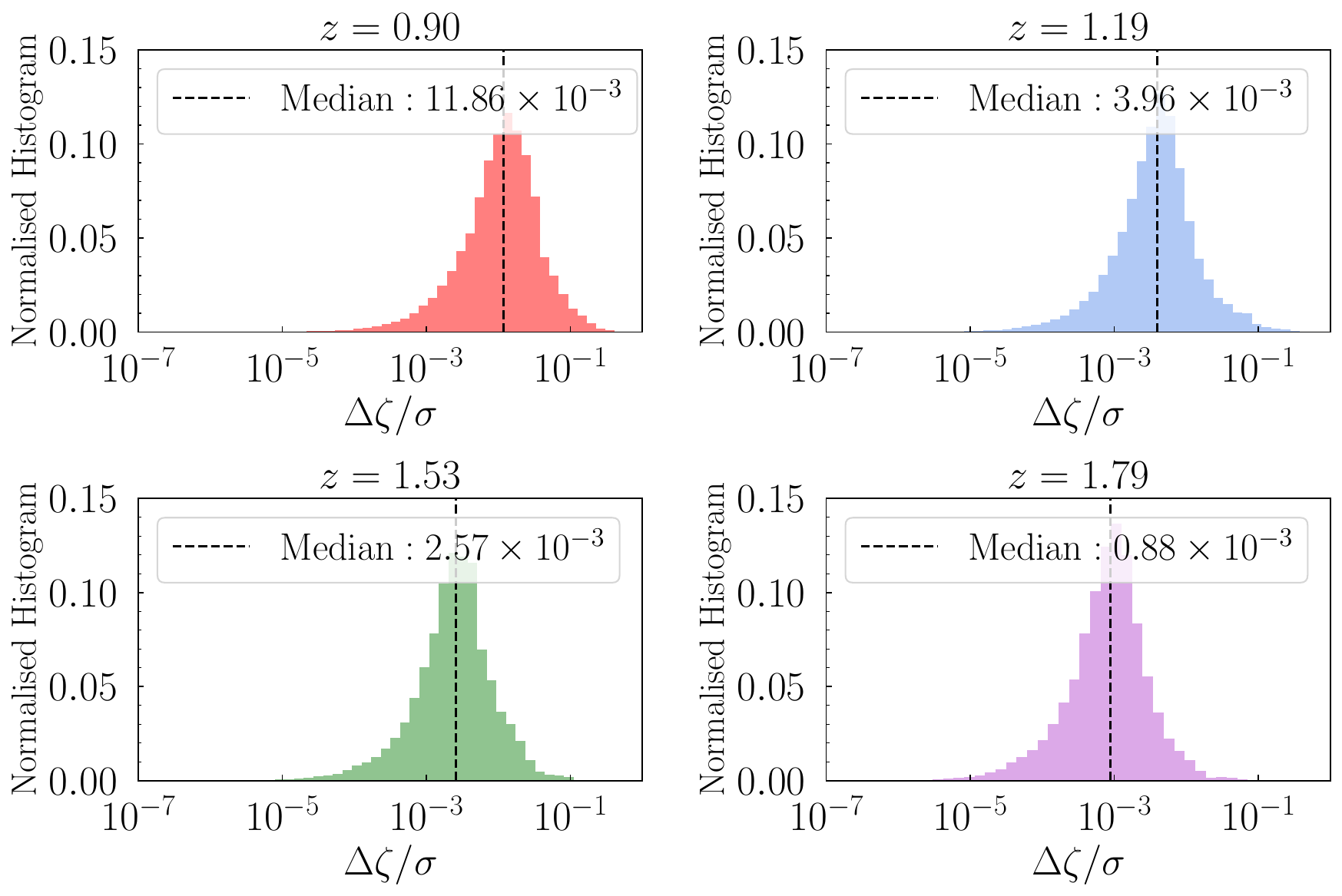}
    \caption{Normalised histogram of the ratio between the difference between emulated and exact modelling 2PCF predictions and the corresponding uncertainty, $\Delta_\xi/\sigma_\xi$ (top four panels). Different colours refer to different redshift snapshot, while the dashed black line represent the median of the distribution. The same quantity for the 3PCF, $\Delta_\zeta/\sigma_\zeta$, is shown in the bottom four panels, for the different redshifts.}
    \label{fig:emu_error}
\end{figure*}

As detailed in Sect.~\ref{sec:likelihood}, to efficiently sample the likelihood function as across the three cosmological parameters, we developed an emulator to obtain model predictions rapidly and accurately, in the first place, for the 3PCF, but also extending it to include the 2PCF. Indeed, the computational cost of the predictions for the coefficients $\zeta_\ell$ via Eq.~\eqref{eq:zeta_Bk_ell} as well as the loop integrals coupled to the Hankel transform of Eq.~\eqref{eq:xi_def} for the 2PCF makes it impractical to explore a broad range of cosmological parameters directly. 

\begin{table}[t]
    \centering
    \caption{The neural network architecture of the emulator.}
    \renewcommand{\arraystretch}{1.5}
    \begin{tabular}{|l|c|}
         \hline
         \rowcolor{blue!15}
         Hyperparameter & Value \\
         \hline
         Number of layers & \phantom{0}\phantom{0}3 \\
         \hline
         Number of neurons & \phantom{0}64 \\ 
         \hline
         Batch size & 256 \\
         \hline
         Learning rate & $10^{-3}$ \\ 
         \hline
         Patience value & \phantom{0}30 \\
         \hline
         Max epochs & 300 \\
         \hline
         Validation split & \phantom{0}20\% \\
         \hline
         Optimizer & Adam \\
         \hline
     \end{tabular}
    \label{table:NN_architecture}
\end{table}

To address this issue, we constructed an emulator implemented in \texttt{PyTorch}\footnote{\url{https://pytorch.org/docs/stable/index.html}}. The neural network, whose architecture is described in Table ~\ref{table:NN_architecture}, was trained using input samples drawn from a Latin hypercube sampling of the 3-dimensional, cosmological parameter space. This spans the ranges $10^{9}A_{\rm s} \in [1.045, 3.135]$, $h \in [0.63, 0.73]$ and $\omega_{\rm cdm} \in [0.09, 0.15]$. The full models of both 2PCF and 3PCF can be written as sums of contributions each factorising the dependence on bias and nuisance parameters. This means that we can keep the analytical dependence on these parameters emulating the cosmology-dependence of each individual contribution. The emulator's predictions were subsequently validated against a test set drawn from an independent sample to ensure their accuracy and reliability. 

The metric chosen for assessing the emulator's performance compares the systematic error $\Delta\xi$ on each separation bin for the 2PCF and $\Delta\zeta$ on each triangular configuration  for the 3PCF to their statistical error estimated from the covariance computed in Sect.~\ref{sec:data} and with the binning choice adopted there. The statistical error is therefore relative to an observed volume of about $54 \ h^{-3}\,{\rm Gpc}^3$, larger than any redshift bin expected for the \Euclid spectroscopic sample \citep{BlanchardEtal2020}.  In particular, we require that the ratios $\Delta{\xi}/\sigma_\xi$ and $\Delta{\zeta}/\sigma_\zeta$ are kept, overall, below the 10\% level. The distribution of this metric depends on both the separation $r$ and the triangle configuration under consideration, the bias contributions, as well as on the sampled values of the cosmological parameters. This is illustrated by the normalised histograms shown in Fig. \ref{fig:emu_error}. In this test, involving the full model predictions, we assume a unitary linear bias $b_1$ and $c_0$, the bias relations presented in Eqs. ~\eqref{eq:bg2_bias_rel} and \eqref{eq:bg3_bias_rel}, and the fitting function for $b_2(b_1)$ obtained from numerical simulations in \citet{LazeyrasEtal2016}. All distributions, for both 2PCf and 3PCF at all four redshifts, are characterised by a median value for $\Delta{\xi}/\sigma_\xi$ or $\Delta{\zeta}/\sigma_\zeta$ well below $10\%$, decreasing slightly with redshift.

This assessment ensures that the emulation process for both the 2PCF and the 3PCF is sufficiently accurate for the analysis presented in this work, but, in practice, also for a direct application to \Euclid data or measurements from Stage IV surveys. We estimate a possible effect of systematic errors in the emulated prediction by adding in quadrature the mean value of $\Delta \xi$ and $\Delta \zeta$ to the diagonal of the two covariance matrices finding no appreciable difference.

Finally we notice that the emulator reduces the computational time required for the 3PCF evaluation from approximately 5 minutes per cosmological model to the order of $10^{-3}$ seconds.

\section{Cosmological constraints}

\MG{As a complement to the left panel of Fig. \ref{fig:contours_cosmo_2_3_joint_main}, we present here the corresponding results at higher redshifts, namely $z = 1.2, 1.5, 1.8$. We find that the same qualitative conclusions drawn at $z = 0.9$ continue to hold across redshift, with a general improvement in the precision of parameter inference at lower redshift. A mild bias in the inferred values of $A_s$ and $b_1$ is observed, though still well within statistical compatibility with the expected values. We interpret this as a residual effect of projection, and note that all snapshots share the same seed.}

\begin{figure*}[t]
    \centering
    \includegraphics[width=0.49\textwidth]{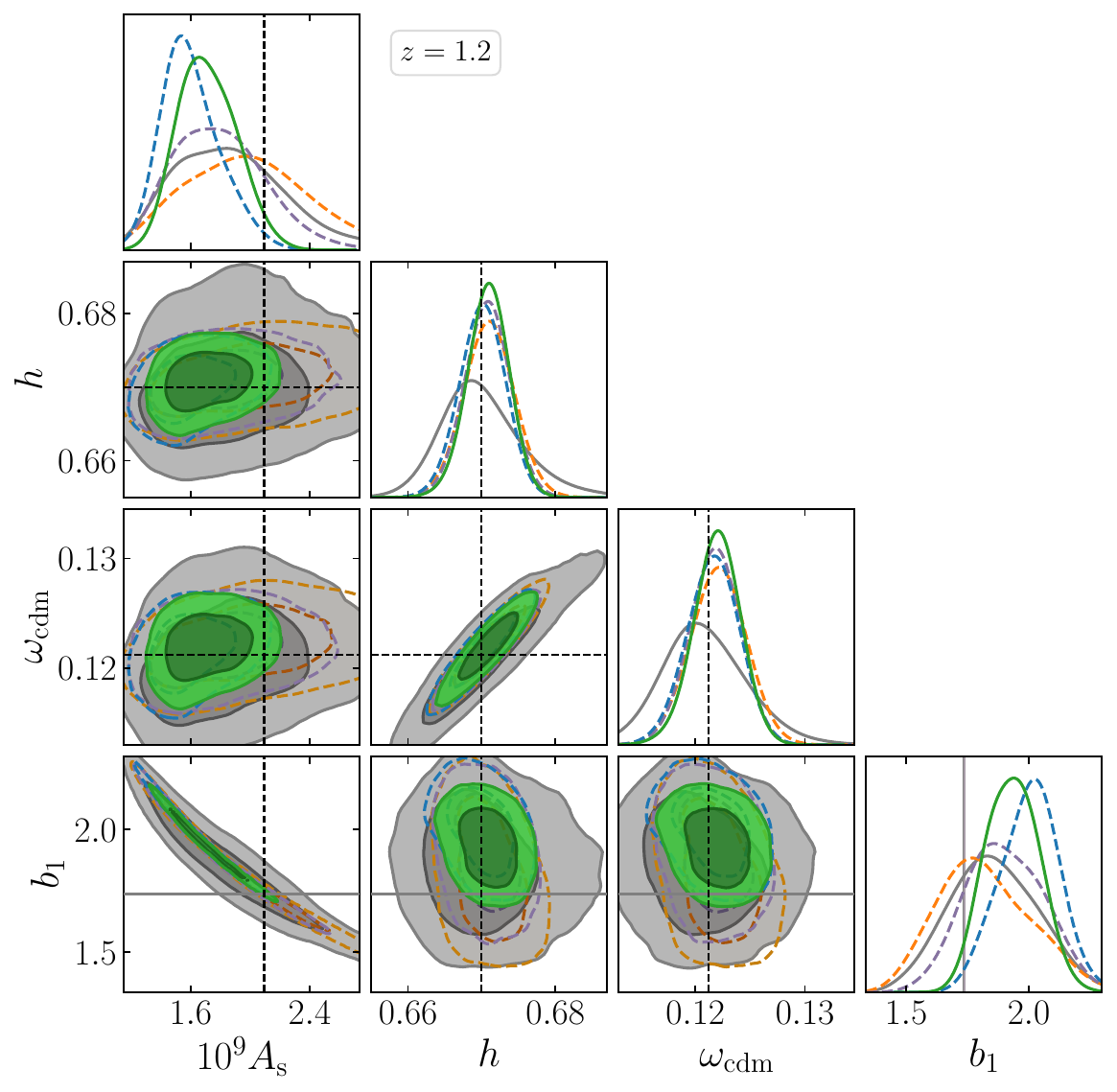}    
    \includegraphics[width=0.49\textwidth]{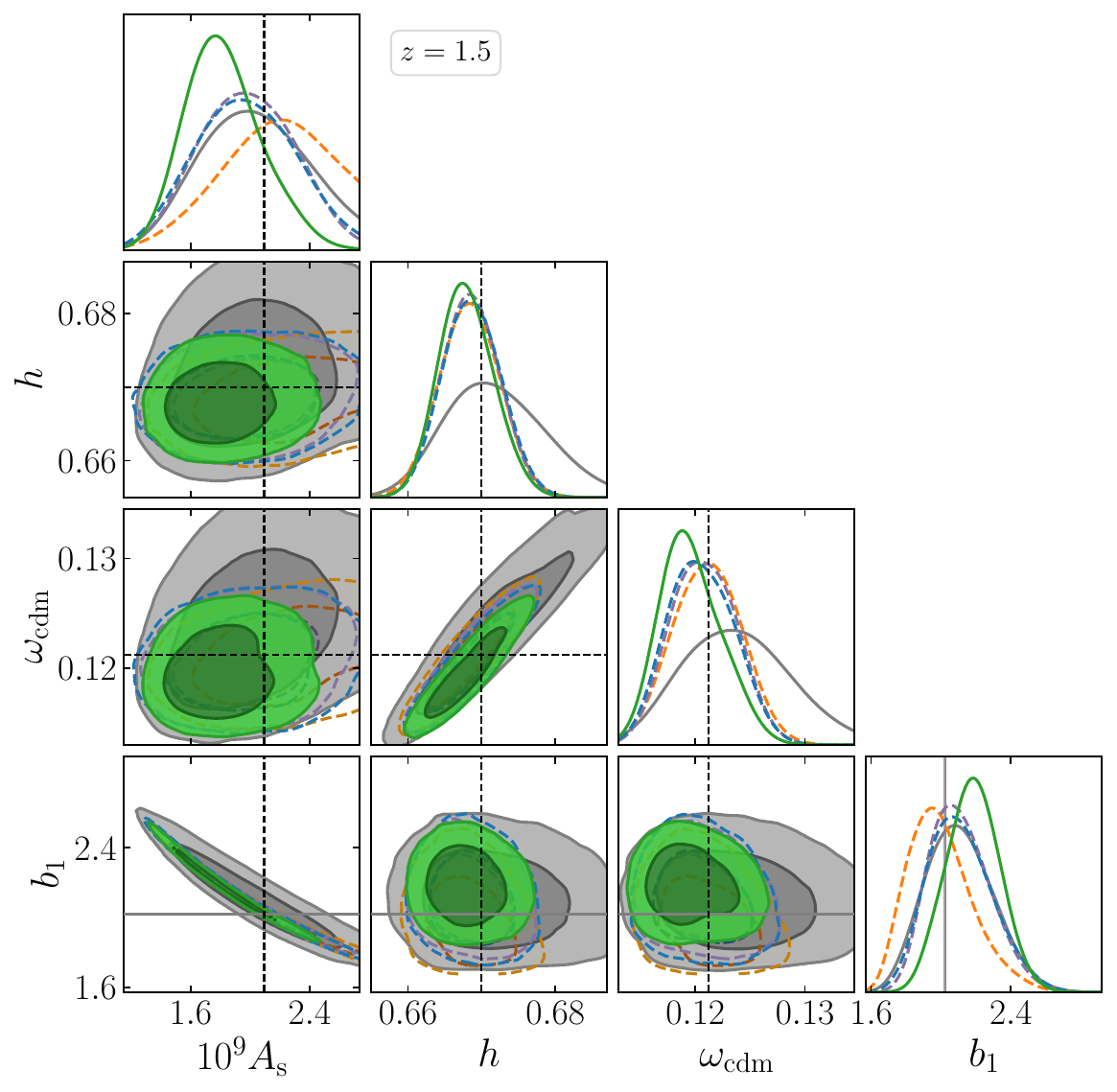}
    \includegraphics[width=0.49\textwidth]{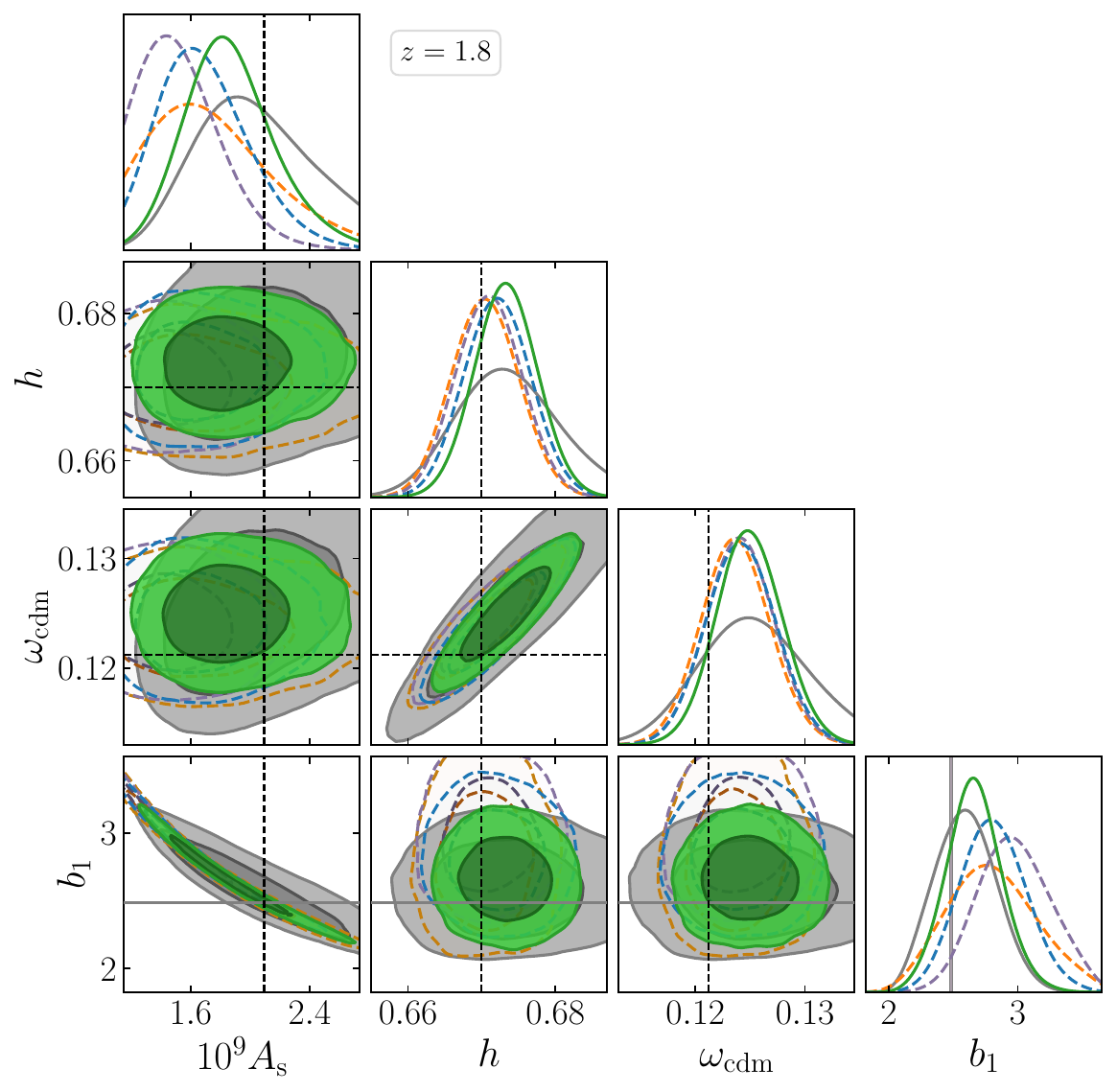}    
    \caption{\MG{As in the left panel of Fig. \ref{fig:contours_cosmo_2_3_joint_main}, considering other redshift snapshots from $z=1.2$ to $z=1.8$.}}
    \label{fig:contours_cosmo_2_3_joint}
\end{figure*}

\MG{In addition, Fig. \ref{fig:contours_cosmo_2_3_joint} shows the full extension of the right panel of Fig. \ref{fig:contours_cosmo_2_3_joint_main}, including all model parameters and, in particular, the nonlinear galaxy bias terms, $b_2, b_{\mathcal{G_2}},b_{\Gamma_3}$ and the EFT parameters $c_0$}.
\label{sect:appendixcosmo}

\begin{figure*}[htbp]
    \centering
        \centering
        \includegraphics[width=0.8\textwidth]{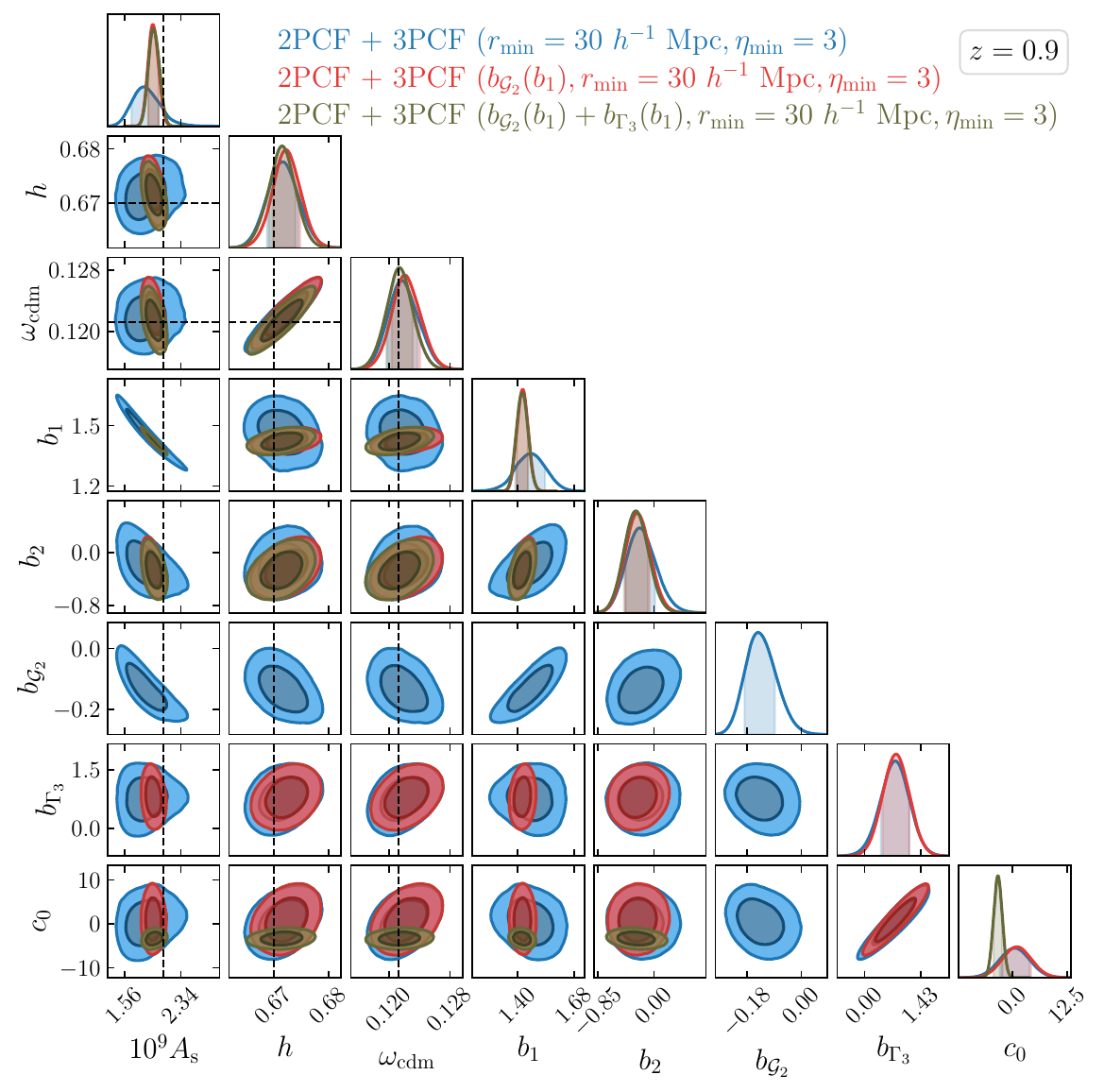}
    \caption{\MG{As in the right panel of Fig. \ref{fig:contours_cosmo_2_3_joint_main}, showing the full set of model parameters}}
    \label{fig:contour_full_bias_045}
\end{figure*}
\end{appendix}

%% file: Main.bbl
\begin{thebibliography}{104}
\expandafter\ifx\csname natexlab\endcsname\relax\def\natexlab#1{#1}\fi

\bibitem[{{Adame} {et~al.}(2025){Adame}, {Aguilar}, {Ahlen}, {Alam},
  {Alexander}, {Alvarez}, {Alves}, {Anand}, {Andrade}, {Armengaud}, {Avila},
  {Aviles}, {Awan}, {Bahr-Kalus}, {Bailey}, {Baltay}, {Bault}, {Behera},
  {BenZvi}, {Bera}, {Beutler}, {Bianchi}, {Blake}, {Blum}, {Brieden},
  {Brodzeller}, {Brooks}, {Buckley-Geer}, {Burtin}, {Calderon}, {Canning},
  {Carnero Rosell}, {Cereskaite}, {Cervantes-Cota}, {Chabanier}, {Chaussidon},
  {Chaves-Montero}, {Chen}, {Chen}, {Claybaugh}, {Cole}, {Cuceu}, {Davis},
  {Dawson}, {de la Macorra}, {de Mattia}, {Deiosso}, {Dey}, {Dey}, {Ding},
  {Doel}, {Edelstein}, {Eftekharzadeh}, {Eisenstein}, {Elliott}, {Fagrelius},
  {Fanning}, {Ferraro}, {Ereza}, {Findlay}, {Flaugher}, {Font-Ribera},
  {Forero-S{\'a}nchez}, {Forero-Romero}, {Frenk}, {Garcia-Quintero},
  {Gazta{\~n}aga}, {Gil-Mar{\'\i}n}, {Gontcho a Gontcho}, {Gonzalez-Morales},
  {Gonzalez-Perez}, {Gordon}, {Green}, {Gruen}, {Gsponer}, {Gutierrez}, {Guy},
  {Hadzhiyska}, {Hahn}, {Hanif}, {Herrera-Alcantar}, {Honscheid}, {Howlett},
  {Huterer}, {Ir{\v{s}}i{\v{c}}}, {Ishak}, {Juneau}, {Kara{\c{c}}ayl{\i}},
  {Kehoe}, {Kent}, {Kirkby}, {Kremin}, {Krolewski}, {Lai}, {Lan}, {Landriau},
  {Lang}, {Lasker}, {Le Goff}, {Le Guillou}, {Leauthaud}, {Levi}, {Li},
  {Linder}, {Lodha}, {Magneville}, {Manera}, {Margala}, {Martini}, {Maus},
  {McDonald}, {Medina-Varela}, {Meisner}, {Mena-Fern{\'a}ndez}, {Miquel},
  {Moon}, {Moore}, {Moustakas}, {Mueller}, {Mu{\~n}oz-Guti{\'e}rrez}, {Myers},
  {Nadathur}, {Napolitano}, {Neveux}, {Newman}, {Nguyen}, {Nie}, {Niz},
  {Noriega}, {Padmanabhan}, {Paillas}, {Palanque-Delabrouille}, {Pan},
  {Penmetsa}, {Percival}, {Pieri}, {Pinon}, {Poppett}, {Porredon}, {Prada},
  {P{\'e}rez-Fern{\'a}ndez}, {P{\'e}rez-R{\`a}fols}, {Rabinowitz}, {Raichoor},
  {Ram{\'\i}rez-P{\'e}rez}, {Ramirez-Solano}, {Rashkovetskyi}, {Ravoux},
  {Rezaie}, {Rich}, {Rocher}, {Rockosi}, {Roe}, {Rosado-Marin}, {Ross},
  {Rossi}, {Ruggeri}, {Ruhlmann-Kleider}, {Samushia}, {Sanchez}, {Saulder},
  {Schlafly}, {Schlegel}, {Schubnell}, {Seo}, {Shafieloo}, {Sharples},
  {Silber}, {Slosar}, {Smith}, {Sprayberry}, {Tan}, {Tarl{\'e}}, {Taylor},
  {Trusov}, {Ure{\~n}a-L{\'o}pez}, {Vaisakh}, {Valcin}, {Valdes},
  {Vargas-Maga{\~n}a}, {Verde}, {Walther}, {Wang}, {Wang}, {Weaver},
  {Weaverdyck}, {Wechsler}, {Weinberg}, {White}, {Yu}, {Yu}, {Yuan},
  {Y{\`e}che}, {Zaborowski}, {Zarrouk}, {Zhang}, {Zhao}, {Zhao}, {Zhou}, \&
  {Zhuang}}]{AdameEtal2024DESI_FS}
{Adame}, A.~G., {Aguilar}, J., {Ahlen}, S., {et~al.} 2025, JCAP, 2025, 021

\bibitem[{Albrecht {et~al.}(2006)Albrecht, Bernstein, Cahn, \&
  et~al.}]{AlbrechtEtal2006a}
Albrecht, A., Bernstein, G., Cahn, R., \& et~al. 2006, arXiv:astro-ph/0609591

\bibitem[{{Assassi} {et~al.}(2014){Assassi}, {Baumann}, {Green}, \&
  {Zaldarriaga}}]{AssassiEtal2014}
{Assassi}, V., {Baumann}, D., {Green}, D., \& {Zaldarriaga}, M. 2014, JCAP, 08,
  056

\bibitem[{{Baldauf} {et~al.}(2015){Baldauf}, {Mirbabayi}, {Simonovi{\'c}}, \&
  {Zaldarriaga}}]{BaldaufEtal2015B}
{Baldauf}, T., {Mirbabayi}, M., {Simonovi{\'c}}, M., \& {Zaldarriaga}, M. 2015,
  \prd, 92, 043514

\bibitem[{{Baldauf} {et~al.}(2012){Baldauf}, {Seljak}, {Desjacques}, \&
  {McDonald}}]{BaldaufEtal2012}
{Baldauf}, T., {Seljak}, U., {Desjacques}, V., \& {McDonald}, P. 2012, \prd,
  86, 083540

\bibitem[{Bardeen {et~al.}(1986)Bardeen, Bond, Kaiser, \& Szalay}]{BBKS1986}
Bardeen, J.~M., Bond, J.~R., Kaiser, N., \& Szalay, A.~S. 1986, \apj, 304, 15

\bibitem[{{Baumann} {et~al.}(2012){Baumann}, {Nicolis}, {Senatore}, \&
  {Zaldarriaga}}]{BaumannEtal2012}
{Baumann}, D., {Nicolis}, A., {Senatore}, L., \& {Zaldarriaga}, M. 2012, JCAP,
  07, 051

\bibitem[{Bernardeau {et~al.}(2002)Bernardeau, Colombi, Gazta{\~n}aga, \&
  Scoccimarro}]{BernardeauEtal2002}
Bernardeau, F., Colombi, S., Gazta{\~n}aga, E., \& Scoccimarro, R. 2002,
  \physrep, 367, 1

\bibitem[{{Beutler} {et~al.}(2017){Beutler}, {Seo}, {Saito}, {Chuang},
  {Cuesta}, {Eisenstein}, {Gil-Mar{\'{\i}}n}, {Grieb}, {Hand}, {Kitaura},
  {Modi}, {Nichol}, {Olmstead}, {Percival}, {Prada}, {S{\'a}nchez},
  {Rodriguez-Torres}, {Ross}, {Ross}, {Schneider}, {Tinker}, {Tojeiro}, \&
  {Vargas-Maga{\~n}a}}]{BeutlerEtal2017B}
{Beutler}, F., {Seo}, H.-J., {Saito}, S., {et~al.} 2017, \mnras, 466, 2242

\bibitem[{{Blas} {et~al.}(2016{\natexlab{a}}){Blas}, {Garny}, {Ivanov}, \&
  {Sibiryakov}}]{BlasEtal2016}
{Blas}, D., {Garny}, M., {Ivanov}, M.~M., \& {Sibiryakov}, S.
  2016{\natexlab{a}}, JCAP, 07, 052

\bibitem[{{Blas} {et~al.}(2016{\natexlab{b}}){Blas}, {Garny}, {Ivanov}, \&
  {Sibiryakov}}]{BlasEtal2016B}
{Blas}, D., {Garny}, M., {Ivanov}, M.~M., \& {Sibiryakov}, S.
  2016{\natexlab{b}}, JCAP, 07, 028

\bibitem[{{Cabass} {et~al.}(2022){Cabass}, {Ivanov}, {Philcox},
  {Simonovi{\'c}}, \& {Zaldarriaga}}]{CabassEtal2022}
{Cabass}, G., {Ivanov}, M.~M., {Philcox}, O. H.~E., {Simonovi{\'c}}, M., \&
  {Zaldarriaga}, M. 2022, \prl, 129, 021301

\bibitem[{{Carrasco} {et~al.}(2012){Carrasco}, {Hertzberg}, \&
  {Senatore}}]{CarrascoHertzbergSenatore2012}
{Carrasco}, J.~J.~M., {Hertzberg}, M.~P., \& {Senatore}, L. 2012, Journal of
  High Energy Physics, 9, 82

\bibitem[{Catelan {et~al.}(1998)Catelan, Lucchin, Matarrese, \&
  Porciani}]{CatelanEtal1998}
Catelan, P., Lucchin, F., Matarrese, S., \& Porciani, C. 1998, \mnras, 297, 692

\bibitem[{Chan {et~al.}(2012)Chan, Scoccimarro, \&
  Sheth}]{ChanScoccimarroSheth2012}
Chan, K.~C., Scoccimarro, R., \& Sheth, R.~K. 2012, \prd, 85, 083509

\bibitem[{Coles {et~al.}(1993)Coles, Moscardini, Lucchin, Matarrese, \&
  Messina}]{ColesEtal1993}
Coles, P., Moscardini, L., Lucchin, F., Matarrese, S., \& Messina, A. 1993,
  \mnras, 264, 749

\bibitem[{Crocce \& Scoccimarro(2008)}]{CrocceScoccimarro2008}
Crocce, M. \& Scoccimarro, R. 2008, \prd, 77, 023533

\bibitem[{{D'Amico} {et~al.}(2020){D'Amico}, {Gleyzes}, {Kokron}, {Markovic},
  {Senatore}, {Zhang}, {Beutler}, \& {Gil-Mar{\'\i}n}}]{DAmicoEtal2020}
{D'Amico}, G., {Gleyzes}, J., {Kokron}, N., {et~al.} 2020, JCAP, 05, 005

\bibitem[{{D'Amico} {et~al.}(2025){D'Amico}, {Lewandowski}, {Senatore}, \&
  {Zhang}}]{DAmicoEtal2022A}
{D'Amico}, G., {Lewandowski}, M., {Senatore}, L., \& {Zhang}, P. 2025, \prd,
  111, 063514

\bibitem[{{Dawson} {et~al.}(2013){Dawson}, {Schlegel}, {Ahn}, {Anderson},
  {Aubourg}, {Bailey}, {Barkhouser}, {Bautista}, {Beifiori}, {Berlind},
  {Bhardwaj}, {Bizyaev}, {Blake}, {Blanton}, {Blomqvist}, {Bolton}, {Borde},
  {Bovy}, {Brandt}, {Brewington}, {Brinkmann}, {Brown}, {Brownstein}, {Bundy},
  {Busca}, {Carithers}, {Carnero}, {Carr}, {Chen}, {Comparat}, {Connolly},
  {Cope}, {Croft}, {Cuesta}, {da Costa}, {Davenport}, {Delubac}, {de Putter},
  {Dhital}, {Ealet}, {Ebelke}, {Eisenstein}, {Escoffier}, {Fan}, {Filiz Ak},
  {Finley}, {Font-Ribera}, {G{\'e}nova-Santos}, {Gunn}, {Guo}, {Haggard},
  {Hall}, {Hamilton}, {Harris}, {Harris}, {Ho}, {Hogg}, {Holder}, {Honscheid},
  {Huehnerhoff}, {Jordan}, {Jordan}, {Kauffmann}, {Kazin}, {Kirkby}, {Klaene},
  {Kneib}, {Le Goff}, {Lee}, {Long}, {Loomis}, {Lundgren}, {Lupton}, {Maia},
  {Makler}, {Malanushenko}, {Malanushenko}, {Mandelbaum}, {Manera}, {Maraston},
  {Margala}, {Masters}, {McBride}, {McDonald}, {McGreer}, {McMahon}, {Mena},
  {Miralda-Escud{\'e}}, {Montero-Dorta}, {Montesano}, {Muna}, {Myers},
  {Naugle}, {Nichol}, {Noterdaeme}, {Nuza}, {Olmstead}, {Oravetz}, {Oravetz},
  {Owen}, {Padmanabhan}, {Palanque-Delabrouille}, {Pan}, {Parejko},
  {P{\^a}ris}, {Percival}, {P{\'e}rez-Fournon}, {P{\'e}rez-R{\`a}fols},
  {Petitjean}, {Pfaffenberger}, {Pforr}, {Pieri}, {Prada}, {Price-Whelan},
  {Raddick}, {Rebolo}, {Rich}, {Richards}, {Rockosi}, {Roe}, {Ross}, {Ross},
  {Rossi}, {Rubi{\~n}o-Martin}, {Samushia}, {S{\'a}nchez}, {Sayres}, {Schmidt},
  {Schneider}, {Sc{\'o}ccola}, {Seo}, {Shelden}, {Sheldon}, {Shen}, {Shu},
  {Slosar}, {Smee}, {Snedden}, {Stauffer}, {Steele}, {Strauss}, {Streblyanska},
  {Suzuki}, {Swanson}, {Tal}, {Tanaka}, {Thomas}, {Tinker}, {Tojeiro},
  {Tremonti}, {Vargas Maga{\~n}a}, {Verde}, {Viel}, {Wake}, {Watson}, {Weaver},
  {Weinberg}, {Weiner}, {West}, {White}, {Wood-Vasey}, {Yeche}, {Zehavi},
  {Zhao}, \& {Zheng}}]{DawsonEtal2013}
{Dawson}, K.~S., {Schlegel}, D.~J., {Ahn}, C.~P., {et~al.} 2013, \aj, 145, 10

\bibitem[{{DESI Collaboration} {et~al.}(2016){DESI Collaboration}, {Aghamousa},
  {Aguilar}, {Ahlen}, {Alam}, {Allen}, {Allende Prieto}, {Annis}, {Bailey},
  {Balland}, {Ballester}, {Baltay}, {Beaufore}, {Bebek}, {Beers}, {Bell},
  {Bernal}, {Besuner}, {Beutler}, {Blake}, {Bleuler}, {Blomqvist}, {Blum},
  {Bolton}, {Briceno}, {Brooks}, {Brownstein}, {Buckley-Geer}, {Burden},
  {Burtin}, {Busca}, {Cahn}, {Cai}, {Cardiel-Sas}, {Carlberg}, {Carton},
  {Casas}, {Castander}, {Cervantes-Cota}, {Claybaugh}, {Close}, {Coker},
  {Cole}, {Comparat}, {Cooper}, {Cousinou}, {Crocce}, {Cuby}, {Cunningham},
  {Davis}, {Dawson}, {de la Macorra}, {De Vicente}, {Delubac}, {Derwent},
  {Dey}, {Dhungana}, {Ding}, {Doel}, {Duan}, {Ealet}, {Edelstein},
  {Eftekharzadeh}, {Eisenstein}, {Elliott}, {Escoffier}, {Evatt}, {Fagrelius},
  {Fan}, {Fanning}, {Farahi}, {Farihi}, {Favole}, {Feng}, {Fernandez},
  {Findlay}, {Finkbeiner}, {Fitzpatrick}, {Flaugher}, {Flender}, {Font-Ribera},
  {Forero-Romero}, {Fosalba}, {Frenk}, {Fumagalli}, {Gaensicke}, {Gallo},
  {Garcia-Bellido}, {Gaztanaga}, {Pietro Gentile Fusillo}, {Gerard},
  {Gershkovich}, {Giannantonio}, {Gillet}, {Gonzalez-de-Rivera},
  {Gonzalez-Perez}, {Gott}, {Graur}, {Gutierrez}, {Guy}, {Habib}, {Heetderks},
  {Heetderks}, {Heitmann}, {Hellwing}, {Herrera}, {Ho}, {Holland}, {Honscheid},
  {Huff}, {Hutchinson}, {Huterer}, {Hwang}, {Illa Laguna}, {Ishikawa},
  {Jacobs}, {Jeffrey}, {Jelinsky}, {Jennings}, {Jiang}, {Jimenez}, {Johnson},
  {Joyce}, {Jullo}, {Juneau}, {Kama}, {Karcher}, {Karkar}, {Kehoe}, {Kennamer},
  {Kent}, {Kilbinger}, {Kim}, {Kirkby}, {Kisner}, {Kitanidis}, {Kneib},
  {Koposov}, {Kovacs}, {Koyama}, {Kremin}, {Kron}, {Kronig}, {Kueter-Young},
  {Lacey}, {Lafever}, {Lahav}, {Lambert}, {Lampton}, {Landriau}, {Lang},
  {Lauer}, {Le Goff}, {Le Guillou}, {Le Van Suu}, {Lee}, {Lee}, {Leitner},
  {Lesser}, {Levi}, {L'Huillier}, {Li}, {Liang}, {Lin}, {Linder}, {Loebman},
  {Luki{\'c}}, {Ma}, {MacCrann}, {Magneville}, {Makarem}, {Manera}, {Manser},
  {Marshall}, {Martini}, {Massey}, {Matheson}, {McCauley}, {McDonald},
  {McGreer}, {Meisner}, {Metcalfe}, {Miller}, {Miquel}, {Moustakas}, {Myers},
  {Naik}, {Newman}, {Nichol}, {Nicola}, {Nicolati da Costa}, {Nie}, {Niz},
  {Norberg}, {Nord}, {Norman}, {Nugent}, {O'Brien}, {Oh}, \&
  {Olsen}}]{AghamousaEtal2016}
{DESI Collaboration}, {Aghamousa}, A., {Aguilar}, J., {et~al.} 2016,
  arXiv:1611.00036

\bibitem[{Desjacques(2008)}]{Desjacques2008}
Desjacques, V. 2008, \prd, 78, 103503

\bibitem[{Desjacques {et~al.}(2010)Desjacques, Crocce, Scoccimarro, \&
  Sheth}]{DesjacquesEtal2010}
Desjacques, V., Crocce, M., Scoccimarro, R., \& Sheth, R.~K. 2010, \prd, 82,
  103529

\bibitem[{{Desjacques} {et~al.}(2018){Desjacques}, {Jeong}, \&
  {Schmidt}}]{DesjacquesJeongSchmidt2018}
{Desjacques}, V., {Jeong}, D., \& {Schmidt}, F. 2018, \physrep, 733, 1

\bibitem[{{Dore} {et~al.}(2019){Dore}, {Hirata}, {Wang}, {Weinberg}, {Eifler},
  {Foley}, {Heinrich}, {Krause}, {Perlmutter}, {Pisani}, {Scolnic}, {Spergel},
  {Suntzeff}, {Aldering}, {Baltay}, {Capak}, {Choi}, {Dvorkin}, {Fall}, {Fang},
  {Fruchter}, {Galbany}, {Ho}, {Hounsell}, {Izard}, {Jain}, {Koekemoer},
  {Kruk}, {Leauthaud}, {Malhotra}, {Mandelbaum}, {Massara}, {Masters},
  {Miyatake}, {Plazas}, {Rhoads}, {Rhodes}, {Rose}, {Rubin}, {Sako},
  {Samushia}, {Shirasaki}, {Simet}, {Takada}, {Troxel}, {Wu}, {Yoshida}, \&
  {Zhai}}]{Dore2019}
{Dore}, O., {Hirata}, C., {Wang}, Y., {et~al.} 2019, \baas, 51, 341

\bibitem[{{Eggemeier} {et~al.}(2020){Eggemeier}, {Scoccimarro}, {Crocce},
  {Pezzotta}, \& {S{\'a}nchez}}]{EggemeierEtal2020}
{Eggemeier}, A., {Scoccimarro}, R., {Crocce}, M., {Pezzotta}, A., \&
  {S{\'a}nchez}, A.~G. 2020, \prd, 102, 103530

\bibitem[{{Eggemeier} {et~al.}(2019){Eggemeier}, {Scoccimarro}, \&
  {Smith}}]{EggemeierScoccimarroSmith2019}
{Eggemeier}, A., {Scoccimarro}, R., \& {Smith}, R.~E. 2019, \prd, 99, 123514

\bibitem[{{Eisenstein} {et~al.}(2007){Eisenstein}, {Seo}, \&
  {White}}]{EisensteinSeoWhite2007}
{Eisenstein}, D.~J., {Seo}, H.-J., \& {White}, M. 2007, \apj, 664, 660

\bibitem[{Eisenstein {et~al.}(2005)Eisenstein, Zehavi, Hogg, Scoccimarro,
  Blanton, Nichol, Scranton, Seo, Tegmark, Zheng, Anderson, Annis, Bahcall,
  Brinkmann, {Burles}, {Castander}, {Connolly}, {Csabai}, {Doi}, {Fukugita},
  {Frieman}, {Glazebrook}, {Gunn}, {Hendry}, {Hennessy}, {Ivezi{\'c}}, {Kent},
  {Knapp}, {Lin}, {Loh}, {Lupton}, {Margon}, {McKay}, {Meiksin}, {Munn},
  {Pope}, {Richmond}, {Schlegel}, {Schneider}, {Shimasaku}, {Stoughton},
  {Strauss}, {SubbaRao}, {Szalay}, {Szapudi}, {Tucker}, {Yanny}, \&
  {York}}]{EisensteinEtal2005B}
Eisenstein, D.~J., Zehavi, I., Hogg, D.~W., {et~al.} 2005, \apj, 633, 560

\bibitem[{Euclid Collaboration:~{Blanchard} {et~al.}(2020)Euclid
  Collaboration:~{Blanchard}, {Camera}, {Carbone}, {Cardone}, {Casas},
  {Clesse}, {Ili{\'c}}, {Kilbinger}, {Kitching}, {Kunz}, {Lacasa}, {Linder},
  {Majerotto}, {Markovi{\v{c}}}, {Martinelli}, {Pettorino}, {Pourtsidou},
  {Sakr}, {S{\'a}nchez}, {Sapone}, {Tutusaus}, {Yahia-Cherif}, {Yankelevich},
  {Andreon}, {Aussel}, {Balaguera-Antol{\'\i}nez}, {Baldi}, {Bardelli},
  {Bender}, {Biviano}, {Bonino}, {Boucaud}, {Bozzo}, {Branchini}, {Brau-Nogue},
  {Brescia}, {Brinchmann}, {Burigana}, {Cabanac}, {Capobianco}, {Cappi},
  {Carretero}, {Carvalho}, {Casas}, {Castander}, {Castellano}, {Cavuoti},
  {Cimatti}, {Cledassou}, {Colodro-Conde}, {Congedo}, {Conselice}, {Conversi},
  {Copin}, {Corcione}, {Coupon}, {Courtois}, {Cropper}, {Da Silva}, {de la
  Torre}, {Di Ferdinando}, {Dubath}, {Ducret}, {Duncan}, {Dupac}, {Dusini},
  {Fabbian}, {Fabricius}, {Farrens}, {Fosalba}, {Fotopoulou}, {Fourmanoit},
  {Frailis}, {Franceschi}, {Franzetti}, {Fumana}, {Galeotta}, {Gillard},
  {Gillis}, {Giocoli}, {G{\'o}mez-Alvarez}, {Graci{\'a}-Carpio}, {Grupp},
  {Guzzo}, {Hoekstra}, {Hormuth}, {Israel}, {Jahnke}, {Keihanen}, {Kermiche},
  {Kirkpatrick}, {Kohley}, {Kubik}, {Kurki-Suonio}, {Ligori}, {Lilje}, {Lloro},
  {Maino}, {Maiorano}, {Marggraf}, {Martinet}, {Marulli}, {Massey},
  {Medinaceli}, {Mei}, {Mellier}, {Metcalf}, {Metge}, {Meylan}, {Moresco},
  {Moscardini}, {Munari}, {Nichol}, {Niemi}, {Nucita}, {Padilla}, {Paltani},
  {Pasian}, {Percival}, {Pires}, {Polenta}, {Poncet}, {Pozzetti}, {Racca},
  {Raison}, {Renzi}, {Rhodes}, {Romelli}, {Roncarelli}, {Rossetti}, {Saglia},
  {Schneider}, {Scottez}, {Secroun}, {Sirri}, {Stanco}, {Starck}, {Sureau},
  {Tallada-Cresp{\'\i}}, {Tavagnacco}, {Taylor}, {Tenti}, {Tereno},
  {Toledo-Moreo}, {Torradeflot}, {Valenziano}, {Vassallo}, {Verdoes Kleijn},
  {Viel}, {Wang}, {Zacchei}, {Zoubian}, \& {Zucca}}]{BlanchardEtal2020}
Euclid Collaboration:~{Blanchard}, A., {Camera}, S., {Carbone}, C., {et~al.}
  2020, \aap, 642, A191

\bibitem[{{Euclid Collaboration: Mellier} {et~al.}(2024){Euclid Collaboration:
  Mellier}, {Abdurro'uf}, {Acevedo~Barroso}, {et~al.}}]{MellierEtal2024}
{Euclid Collaboration: Mellier}, Y., {Abdurro'uf}, {Acevedo~Barroso}, J.,
  {et~al.} 2024, \aap, accepted, arXiv:2405.13491

\bibitem[{Euclid Collaboration:~{Pezzotta} {et~al.}(2024)Euclid
  Collaboration:~{Pezzotta}, {Moretti}, {Zennaro}, {Moradinezhad Dizgah},
  {Crocce}, {Sefusatti}, {Ferrero}, {Pardede}, {Eggemeier}, {Barreira},
  {Angulo}, {Marinucci}, {Camacho Quevedo}, {de la Torre}, {Alkhanishvili},
  {Biagetti}, {Breton}, {Castorina}, {D'Amico}, {Desjacques}, {Guidi},
  {K{\"a}rcher}, {Oddo}, {Pellejero Ibanez}, {Porciani}, {Pugno},
  {Salvalaggio}, {Sarpa}, {Veropalumbo}, {Vlah}, {Amara}, {Andreon},
  {Auricchio}, {Baldi}, {Bardelli}, {Bender}, {Bodendorf}, {Bonino},
  {Branchini}, {Brescia}, {Brinchmann}, {Camera}, {Capobianco}, {Carbone},
  {Cardone}, {Carretero}, {Casas}, {Castander}, {Castellano}, {Cavuoti},
  {Cimatti}, {Congedo}, {Conselice}, {Conversi}, {Copin}, {Corcione},
  {Courbin}, {Courtois}, {Da Silva}, {Degaudenzi}, {Di Giorgio}, {Dinis},
  {Dupac}, {Dusini}, {Ealet}, {Farina}, {Farrens}, {Fosalba}, {Frailis},
  {Franceschi}, {Galeotta}, {Gillis}, {Giocoli}, {Granett}, {Grazian}, {Grupp},
  {Guzzo}, {Haugan}, {Hormuth}, {Hornstrup}, {Jahnke}, {Joachimi},
  {Keih{\"a}nen}, {Kermiche}, {Kiessling}, {Kilbinger}, {Kitching}, {Kubik},
  {Kunz}, {Kurki-Suonio}, {Ligori}, {Lilje}, {Lindholm}, {Lloro}, {Maiorano},
  {Mansutti}, {Marggraf}, {Markovic}, {Martinet}, {Marulli}, {Massey},
  {Medinaceli}, {Mellier}, {Meneghetti}, {Merlin}, {Meylan}, {Moresco},
  {Moscardini}, {Munari}, {Niemi}, {Padilla}, {Paltani}, {Pasian}, {Pedersen},
  {Percival}, {Pettorino}, {Pires}, {Polenta}, {Pollack}, {Poncet}, {Popa},
  {Pozzetti}, {Raison}, {Renzi}, {Rhodes}, {Riccio}, {Romelli}, {Roncarelli},
  {Rossetti}, {Saglia}, {Sapone}, {Sartoris}, {Schneider}, {Schrabback},
  {Secroun}, {Seidel}, {Seiffert}, {Serrano}, {Sirignano}, {Sirri}, {Stanco},
  {Surace}, {Tallada-Cresp{\'\i}}, {Taylor}, {Tereno}, {Toledo-Moreo},
  {Torradeflot}, {Tutusaus}, {Valentijn}, {Valenziano}, {Vassallo}, {Wang},
  {Weller}, {Zamorani}, {Zoubian}, {Zucca}, {Biviano}, {Bozzo}, {Burigana},
  {Colodro-Conde}, {Di Ferdinando}, {Mainetti}, {Martinelli}, {Mauri}, {Sakr},
  {Scottez}, {Tenti}, {Viel}, {Wiesmann}, {Akrami}, {Allevato}, {Anselmi},
  {Baccigalupi}, {Ballardini}, {Bernardeau}, {Blanchard}, {Borgani}, {Bruton},
  {Cabanac}, {Cappi}, {Carvalho}, {Castignani}, {Castro}, {Ca{\~n}as-Herrera},
  {Chambers}, {Contarini}, {Cooray}, {Coupon}, {Davini}, {De Lucia}, {Desprez},
  {Di Domizio}, {Dole}, {D{\'\i}az-S{\'a}nchez}, {Escartin Vigo}, {Escoffier},
  {Ferreira}, {Finelli}, {Gabarra}, {Ganga}, {Garc{\'\i}a-Bellido},
  {Giacomini}, {Gozaliasl}, {Hall}, {Ili{\'c}}, {Joudaki}, {Kajava}, {Kansal},
  {Kirkpatrick}, {Legrand}, {Loureiro}, {Macias-Perez}, {Magliocchetti},
  {Mannucci}, {Maoli}, {Martins}, {Matthew}, {Maurin}, {Metcalf}, {Migliaccio},
  {Monaco}, {Morgante}, {Nadathur}, {Walton}, {Patrizii}, {Popa}, {Potter},
  {Pourtsidou}, {P{\"o}ntinen}, {Risso}, {Rocci}, {Sahl{\'e}n}, {S{\'a}nchez},
  {Schneider}, {Sereno}, {Simon}, {Spurio Mancini}, {Steinwagner}, {Testera},
  {Teyssier}, {Toft}, {Tosi}, {Troja}, {Tucci}, {Valiviita}, {Vergani},
  {Verza}, \& {Vielzeuf}}]{PezzottaEtal2024}
Euclid Collaboration:~{Pezzotta}, A., {Moretti}, C., {Zennaro}, M., {et~al.}
  2024, \aap, 687, A216

\bibitem[{Euclid Collaboration:~{Scaramella} {et~al.}(2022)Euclid
  Collaboration:~{Scaramella}, {Amiaux}, {Mellier}, {Burigana}, {Carvalho},
  {Cuillandre}, {Da Silva}, {Derosa}, {Dinis}, {Maiorano}, {Maris}, {Tereno},
  {Laureijs}, {Boenke}, {Buenadicha}, {Dupac}, {Gaspar Venancio},
  {G{\'o}mez-{\'A}lvarez}, {Hoar}, {Lorenzo Alvarez}, {Racca},
  {Saavedra-Criado}, {Schwartz}, {Vavrek}, {Schirmer}, {Aussel}, {Azzollini},
  {Cardone}, {Cropper}, {Ealet}, {Garilli}, {Gillard}, {Granett}, {Guzzo},
  {Hoekstra}, {Jahnke}, {Kitching}, {Maciaszek}, {Meneghetti}, {Miller},
  {Nakajima}, {Niemi}, {Pasian}, {Percival}, {Pottinger}, {Sauvage},
  {Scodeggio}, {Wachter}, {Zacchei}, {Aghanim}, {Amara}, {Auphan}, {Auricchio},
  {Awan}, {Balestra}, {Bender}, {Bodendorf}, {Bonino}, {Branchini},
  {Brau-Nogue}, {Brescia}, {Candini}, {Capobianco}, {Carbone}, {Carlberg},
  {Carretero}, {Casas}, {Castander}, {Castellano}, {Cavuoti}, {Cimatti},
  {Cledassou}, {Congedo}, {Conselice}, {Conversi}, {Copin}, {Corcione},
  {Costille}, {Courbin}, {Degaudenzi}, {Douspis}, {Dubath}, {Duncan}, {Dusini},
  {Farrens}, {Ferriol}, {Fosalba}, {Fourmanoit}, {Frailis}, {Franceschi},
  {Franzetti}, {Fumana}, {Gillis}, {Giocoli}, {Grazian}, {Grupp}, {Haugan},
  {Holmes}, {Hormuth}, {Hudelot}, {Kermiche}, {Kiessling}, {Kilbinger},
  {Kohley}, {Kubik}, {K{\"u}mmel}, {Kunz}, {Kurki-Suonio}, {Lahav}, {Ligori},
  {Lilje}, {Lloro}, {Mansutti}, {Marggraf}, {Markovic}, {Marulli}, {Massey},
  {Maurogordato}, {Melchior}, {Merlin}, {Meylan}, {Mohr}, {Moresco}, {Morin},
  {Moscardini}, {Munari}, {Nichol}, {Padilla}, {Paltani}, {Peacock},
  {Pedersen}, {Pettorino}, {Pires}, {Poncet}, {Popa}, {Pozzetti}, {Raison},
  {Rebolo}, {Rhodes}, {Rix}, {Roncarelli}, {Rossetti}, {Saglia}, {Schneider},
  {Schrabback}, {Secroun}, {Seidel}, {Serrano}, {Sirignano}, {Sirri},
  {Skottfelt}, {Stanco}, {Starck}, {Tallada-Cresp{\'\i}}, {Tavagnacco},
  {Taylor}, {Teplitz}, {Toledo-Moreo}, {Torradeflot}, {Trifoglio}, {Valentijn},
  {Valenziano}, {Verdoes Kleijn}, {Wang}, {Welikala}, {Weller}, {Wetzstein},
  {Zamorani}, {Zoubian}, {Andreon}, {Baldi}, {Bardelli}, {Boucaud}, {Camera},
  {Di Ferdinando}, {Fabbian}, {Farinelli}, {Galeotta}, {Graci{\'a}-Carpio},
  {Maino}, {Medinaceli}, {Mei}, {Neissner}, {Polenta}, {Renzi}, {Romelli},
  {Rosset}, {Sureau}, {Tenti}, {Vassallo}, {Zucca}, {Baccigalupi},
  {Balaguera-Antol{\'\i}nez}, {Battaglia}, {Biviano}, {Borgani}, {Bozzo},
  {Cabanac}, {Cappi}, {Casas}, {Castignani}, {Colodro-Conde}, {Coupon},
  {Courtois}, {Cuby}, {de la Torre}, {Desai}, {Dole}, {Fabricius}, {Farina},
  {Ferreira}, {Finelli}, {Flose-Reimberg}, {Fotopoulou}, {Ganga}, {Gozaliasl},
  {Hook}, {Keihanen}, {Kirkpatrick}, {Liebing}, {Lindholm}, {Mainetti},
  {Martinelli}, {Martinet}, {Maturi}, {McCracken}, {Metcalf}, {Morgante},
  {Nightingale}, {Nucita}, {Patrizii}, {Potter}, {Riccio}, {S{\'a}nchez},
  {Sapone}, {Schewtschenko}, {Schultheis}, {Scottez}, {Teyssier}, {Tutusaus},
  {Valiviita}, {Viel}, {Vriend}, \& {Whittaker}}]{ScaramellaEtal2022}
Euclid Collaboration:~{Scaramella}, R., {Amiaux}, J., {Mellier}, Y., {et~al.}
  2022, \aap, 662, A112

\bibitem[{{Fang} {et~al.}(2020){Fang}, {Eifler}, \&
  {Krause}}]{FangEiflerKrause2020}
{Fang}, X., {Eifler}, T., \& {Krause}, E. 2020, \mnras, 497, 2699

\bibitem[{{Farina} {et~al.}(2024){Farina}, {Veropalumbo}, {Branchini}, \&
  {Guidi}}]{FarinaEtal2024a}
{Farina}, A., {Veropalumbo}, A., {Branchini}, E., \& {Guidi}, M. 2024,
  arXiv:2408.03036

\bibitem[{{Foreman-Mackey} {et~al.}(2013){Foreman-Mackey}, {Hogg}, {Lang}, \&
  {Goodman}}]{ForemanMackeyEtal2013}
{Foreman-Mackey}, D., {Hogg}, D.~W., {Lang}, D., \& {Goodman}, J. 2013, \pasp,
  125, 306

\bibitem[{{Frieman} \& {Gaztanaga}(1994)}]{FriemanGaztanaga1994}
{Frieman}, J.~A. \& {Gaztanaga}, E. 1994, \apj, 425, 392

\bibitem[{Fry(1984)}]{Fry1984}
Fry, J.~N. 1984, \apj, 279, 499

\bibitem[{Fry(1994)}]{Fry1994B}
Fry, J.~N. 1994, Physical Review Letters, 73, 215

\bibitem[{Fry \& Gazta{\~n}aga(1993)}]{FryGaztanaga1993}
Fry, J.~N. \& Gazta{\~n}aga, E. 1993, \apj, 413, 447

\bibitem[{{Fujita} {et~al.}(2020){Fujita}, {Mauerhofer}, {Senatore}, {Vlah}, \&
  {Angulo}}]{FujitaEtal2020}
{Fujita}, T., {Mauerhofer}, V., {Senatore}, L., {Vlah}, Z., \& {Angulo}, R.
  2020, JCAP, 01, 009

\bibitem[{{Gazta{\~n}aga} {et~al.}(2009){Gazta{\~n}aga}, {Cabr{\'e}},
  {Castander}, {Crocce}, \& {Fosalba}}]{GaztanagaEtal2009}
{Gazta{\~n}aga}, E., {Cabr{\'e}}, A., {Castander}, F., {Crocce}, M., \&
  {Fosalba}, P. 2009, \mnras, 399, 801

\bibitem[{{Gazta{\~n}aga} {et~al.}(2005){Gazta{\~n}aga}, {Norberg}, {Baugh}, \&
  {Croton}}]{GaztanagaEtal2005}
{Gazta{\~n}aga}, E., {Norberg}, P., {Baugh}, C.~M., \& {Croton}, D.~J. 2005,
  \mnras, 364, 620

\bibitem[{{Gil-Mar{\'{\i}}n} {et~al.}(2017){Gil-Mar{\'{\i}}n}, {Percival},
  {Verde}, {Brownstein}, {Chuang}, {Kitaura}, {Rodr{\'{\i}}guez-Torres}, \&
  {Olmstead}}]{GilMarinEtal2017}
{Gil-Mar{\'{\i}}n}, H., {Percival}, W.~J., {Verde}, L., {et~al.} 2017, \mnras,
  465, 1757

\bibitem[{Grieb {et~al.}(2016)Grieb, S{\'a}nchez, Salazar-Albornoz, \&
  Dalla~Vecchia}]{GriebEtal2016}
Grieb, J.~N., S{\'a}nchez, A.~G., Salazar-Albornoz, S., \& Dalla~Vecchia, C.
  2016, \mnras, 457, 1577

\bibitem[{{Grieb} {et~al.}(2017){Grieb}, {S{\'a}nchez}, {Salazar-Albornoz},
  {Scoccimarro}, {Crocce}, {Dalla Vecchia}, {Montesano}, {Gil-Mar{\'{\i}}n},
  {Ross}, {Beutler}, {Rodr{\'{\i}}guez-Torres}, {Chuang}, {Prada}, {Kitaura},
  {Cuesta}, {Eisenstein}, {Percival}, {Vargas-Maga{\~n}a}, {Tinker}, {Tojeiro},
  {Brownstein}, {Maraston}, {Nichol}, {Olmstead}, {Samushia}, {Seo},
  {Streblyanska}, \& {Zhao}}]{GriebEtal2017}
{Grieb}, J.~N., {S{\'a}nchez}, A.~G., {Salazar-Albornoz}, S., {et~al.} 2017,
  \mnras, 467, 2085

\bibitem[{{Guidi} {et~al.}(2023){Guidi}, {Veropalumbo}, {Branchini},
  {Eggemeier}, \& {Carbone}}]{GuidiEtal2023}
{Guidi}, M., {Veropalumbo}, A., {Branchini}, E., {Eggemeier}, A., \& {Carbone},
  C. 2023, JCAP, 08, 066

\bibitem[{Guzzo {et~al.}(2008)Guzzo, {Pierleoni}, {Meneux}, {Branchini}, {Le
  F{\`e}vre}, {Marinoni}, {Garilli}, {Blaizot}, {De Lucia}, {Pollo},
  {McCracken}, {Bottini}, {Le Brun}, {Maccagni}, {Picat}, {Scaramella},
  {Scodeggio}, {Tresse}, {Vettolani}, {Zanichelli}, {Adami}, {Arnouts},
  {Bardelli}, {Bolzonella}, {Bongiorno}, {Cappi}, {Charlot}, {Ciliegi},
  {Contini}, {Cucciati}, {de la Torre}, {Dolag}, {Foucaud}, {Franzetti},
  {Gavignaud}, {Ilbert}, {Iovino}, {Lamareille}, {Marano}, {Mazure}, {Memeo},
  {Merighi}, {Moscardini}, {Paltani}, {Pell{\`o}}, {Perez-Montero}, {Pozzetti},
  {Radovich}, {Vergani}, {Zamorani}, \& {Zucca}}]{GuzzoEtal2008}
Guzzo, L., {Pierleoni}, M., {Meneux}, B., {et~al.} 2008, \nat, 451, 541

\bibitem[{{Hamilton}(2000)}]{Hamilton2000}
{Hamilton}, A.~J.~S. 2000, \mnras, 312, 257

\bibitem[{{Hivon} {et~al.}(1995){Hivon}, {Bouchet}, {Colombi}, \&
  {Juszkiewicz}}]{HivonEtal1995}
{Hivon}, E., {Bouchet}, F.~R., {Colombi}, S., \& {Juszkiewicz}, R. 1995, \aap,
  298, 643

\bibitem[{{Hou} {et~al.}(2018){Hou}, {S{\'a}nchez}, {Scoccimarro},
  {Salazar-Albornoz}, {Burtin}, {Gil-Mar{\'{\i}}n}, {Percival}, {Ruggeri},
  {Zarrouk}, {Zhao}, {Bautista}, {Brinkmann}, {Brownstein}, {Dawson}, {Devi},
  {Myers}, {Habib}, {Heitmann}, {Tojeiro}, {Rossi}, {Schneider}, {Seo}, \&
  {Wang}}]{HouEtal2018}
{Hou}, J., {S{\'a}nchez}, A.~G., {Scoccimarro}, R., {et~al.} 2018, \mnras, 480,
  2521

\bibitem[{{Ivanov} \& {Sibiryakov}(2018)}]{IvanovSibiryakov2018}
{Ivanov}, M.~M. \& {Sibiryakov}, S. 2018, JCAP, 07, 053

\bibitem[{{Ivanov} {et~al.}(2020){Ivanov}, {Simonovi{\'c}}, \&
  {Zaldarriaga}}]{IvanovSimonovicZaldarriaga2020}
{Ivanov}, M.~M., {Simonovi{\'c}}, M., \& {Zaldarriaga}, M. 2020, JCAP, 05, 042

\bibitem[{{Ivezic} {et~al.}(2009){Ivezic}, {Tyson}, {Axelrod}, {Burke},
  {Claver}, {Cook}, {Kahn}, {Lupton}, {Monet}, {Pinto}, {Strauss}, {Stubbs},
  {Jones}, {Saha}, {Scranton}, {Smith}, \& {LSST
  Collaboration}}]{IvezicEtal2009}
{Ivezic}, Z., {Tyson}, J.~A., {Axelrod}, T., {et~al.} 2009, in American
  Astronomical Society Meeting Abstracts, Vol. 213, 460.03

\bibitem[{{Jing} \& {B{\"o}rner}(2004)}]{JingBorner2004}
{Jing}, Y.~P. \& {B{\"o}rner}, G. 2004, \apj, 607, 140

\bibitem[{Kaiser(1984)}]{Kaiser1984}
Kaiser, N. 1984, \apjl, 284, L9

\bibitem[{Landy \& Szalay(1993)}]{LandySzalay1993}
Landy, S.~D. \& Szalay, A.~S. 1993, \apj, 412, 64

\bibitem[{{Laureijs} {et~al.}(2011){Laureijs}, {Amiaux}, {Arduini},
  {Augu{\`e}res}, {Brinchmann}, {Cole}, {Cropper}, {Dabin}, {Duvet}, {Ealet},
  {Garilli}, {Gondoin}, {Guzzo}, {Hoar}, {Hoekstra}, {Holmes}, {Kitching},
  {Maciaszek}, {Mellier}, {Pasian}, {Percival}, {Rhodes}, {Saavedra Criado},
  {Sauvage}, {Scaramella}, {Valenziano}, {Warren}, {Bender}, {Castander},
  {Cimatti}, {Le F{\`e}vre}, {Kurki-Suonio}, {Levi}, {Lilje}, {Meylan},
  {Nichol}, {Pedersen}, {Popa}, {Rebolo Lopez}, {Rix}, {Rottgering},
  {Zeilinger}, {Grupp}, {Hudelot}, {Massey}, {Meneghetti}, {Miller}, {Paltani},
  {Paulin-Henriksson}, {Pires}, {Saxton}, {Schrabback}, {Seidel}, {Walsh},
  {Aghanim}, {Amendola}, {Bartlett}, {Baccigalupi}, {Beaulieu}, {Benabed},
  {Cuby}, {Elbaz}, {Fosalba}, {Gavazzi}, {Helmi}, {Hook}, {Irwin}, {Kneib},
  {Kunz}, {Mannucci}, {Moscardini}, {Tao}, {Teyssier}, {Weller}, {Zamorani},
  {Zapatero Osorio}, {Boulade}, {Foumond}, {Di Giorgio}, {Guttridge}, {James},
  {Kemp}, {Martignac}, {Spencer}, {Walton}, {Bl{\"u}mchen}, {Bonoli},
  {Bortoletto}, {Cerna}, {Corcione}, {Fabron}, {Jahnke}, {Ligori}, {Madrid},
  {Martin}, {Morgante}, {Pamplona}, {Prieto}, {Riva}, {Toledo}, {Trifoglio},
  {Zerbi}, {Abdalla}, {Douspis}, {Grenet}, {Borgani}, {Bouwens}, {Courbin},
  {Delouis}, {Dubath}, {Fontana}, {Frailis}, {Grazian}, {Koppenh{\"o}fer},
  {Mansutti}, {Melchior}, {Mignoli}, {Mohr}, {Neissner}, {Noddle}, {Poncet},
  {Scodeggio}, {Serrano}, {Shane}, {Starck}, {Surace}, {Taylor},
  {Verdoes-Kleijn}, {Vuerli}, {Williams}, {Zacchei}, {Altieri}, {Escudero
  Sanz}, {Kohley}, {Oosterbroek}, {Astier}, {Bacon}, {Bardelli}, {Baugh},
  {Bellagamba}, {Benoist}, {Bianchi}, {Biviano}, {Branchini}, {Carbone},
  {Cardone}, {Clements}, {Colombi}, {Conselice}, {Cresci}, {Deacon}, {Dunlop},
  {Fedeli}, {Fontanot}, {Franzetti}, {Giocoli}, {Garcia-Bellido}, {Gow},
  {Heavens}, {Hewett}, {Heymans}, {Holland}, {Huang}, {Ilbert}, {Joachimi},
  {Jennins}, {Kerins}, {Kiessling}, {Kirk}, {Kotak}, {Krause}, {Lahav}, {van
  Leeuwen}, {Lesgourgues}, {Lombardi}, {Magliocchetti}, {Maguire}, {Majerotto},
  {Maoli}, {Marulli}, {Maurogordato}, {McCracken}, {McLure}, {Melchiorri},
  {Merson}, {Moresco}, {Nonino}, {Norberg}, {Peacock}, {Pello}, {Penny},
  {Pettorino}, {Di Porto}, {Pozzetti}, {Quercellini}, {Radovich}, {Rassat},
  {Roche}, {Ronayette}, \& {Rossetti}}]{LaureijsEtal2011}
{Laureijs}, R., {Amiaux}, J., {Arduini}, S., {et~al.} 2011, arXiv:1110.3193

\bibitem[{{Lazeyras} {et~al.}(2016){Lazeyras}, Wagner, Baldauf, \&
  Schmidt}]{LazeyrasEtal2016}
{Lazeyras}, T., Wagner, C., Baldauf, T., \& Schmidt, F. 2016, JCAP, 02, 018

\bibitem[{{Lippich} {et~al.}(2019){Lippich}, {S{\'a}nchez}, {Colavincenzo},
  {Sefusatti}, {Monaco}, {Blot}, {Crocce}, {Alvarez}, {Agrawal}, {Avila},
  {Balaguera-Antol{\'{\i}}nez}, {Bond}, {Codis}, {Dalla Vecchia}, {Dorta},
  {Fosalba}, {Izard}, {Kitaura}, {Pellejero-Ibanez}, {Stein}, {Vakili}, \&
  {Yepes}}]{LippichEtal2019}
{Lippich}, M., {S{\'a}nchez}, A.~G., {Colavincenzo}, M., {et~al.} 2019, \mnras,
  482, 1786

\bibitem[{{Mar{\'{\i}}n}(2011)}]{Marin2011}
{Mar{\'{\i}}n}, F. 2011, \apj, 737, 97

\bibitem[{{Mar{\'\i}n} {et~al.}(2013){Mar{\'\i}n}, {Blake}, {Poole}, {McBride},
  {Brough}, {Colless}, {Contreras}, {Couch}, {Croton}, {Croom}, {Davis},
  {Drinkwater}, {Forster}, {Gilbank}, {Gladders}, {Glazebrook}, {Jelliffe},
  {Jurek}, {Li}, {Madore}, {Martin}, {Pimbblet}, {Pracy}, {Sharp}, {Wisnioski},
  {Woods}, {Wyder}, \& {Yee}}]{MarinEtal2013}
{Mar{\'\i}n}, F.~A., {Blake}, C., {Poole}, G.~B., {et~al.} 2013, \mnras, 432,
  2654

\bibitem[{Matsubara(2008)}]{Matsubara2008A}
Matsubara, T. 2008, \prd, 77, 063530

\bibitem[{{McBride} {et~al.}(2011){McBride}, {Connolly}, {Gardner}, {Scranton},
  {Scoccimarro}, {Berlind}, {Mar{\'{\i}}n}, \& {Schneider}}]{McBrideEtal2011B}
{McBride}, C.~K., {Connolly}, A.~J., {Gardner}, J.~P., {et~al.} 2011, \apj,
  739, 85

\bibitem[{McDonald(2006)}]{McDonald2006}
McDonald, P. 2006, \prd, 74, 103512

\bibitem[{{McDonald} \& {Roy}(2009)}]{McDonaldRoy2009}
{McDonald}, P. \& {Roy}, A. 2009, JCAP, 08, 020

\bibitem[{{Moresco} {et~al.}(2017){Moresco}, {Marulli}, {Moscardini},
  {Branchini}, {Cappi}, {Davidzon}, {Granett}, {de la Torre}, {Guzzo}, {Abbas},
  {Adami}, {Arnouts}, {Bel}, {Bolzonella}, {Bottini}, {Carbone}, {Coupon},
  {Cucciati}, {De Lucia}, {Franzetti}, {Fritz}, {Fumana}, {Garilli}, {Ilbert},
  {Iovino}, {Krywult}, {Le Brun}, {Le F{\`e}vre}, {Ma{\l}ek}, {McCracken},
  {Polletta}, {Pollo}, {Scodeggio}, {Tasca}, {Tojeiro}, {Vergani}, \&
  {Zanichelli}}]{MorescoEtal2017}
{Moresco}, M., {Marulli}, F., {Moscardini}, L., {et~al.} 2017, \aap, 604, A133

\bibitem[{{Moresco} {et~al.}(2021){Moresco}, {Veropalumbo}, {Marulli},
  {Moscardini}, \& {Cimatti}}]{MorescoEtal2021}
{Moresco}, M., {Veropalumbo}, A., {Marulli}, F., {Moscardini}, L., \&
  {Cimatti}, A. 2021, \apj, 919, 144

\bibitem[{{Peacock} {et~al.}(2001){Peacock}, {Cole}, {Norberg}, {Baugh},
  {Bland-Hawthorn}, {Bridges}, {Cannon}, {Colless}, {Collins}, {Couch},
  {Dalton}, {Deeley}, {De Propris}, {Driver}, {Efstathiou}, {Ellis}, {Frenk},
  {Glazebrook}, {Jackson}, {Lahav}, {Lewis}, {Lumsden}, {Maddox}, {Percival},
  {Peterson}, {Price}, {Sutherland}, \& {Taylor}}]{PeacockEtal2001}
{Peacock}, J.~A., {Cole}, S., {Norberg}, P., {et~al.} 2001, \nat, 410, 169

\bibitem[{Peebles(1973)}]{Peebles1973}
Peebles, P. J.~E. 1973, \apj, 185, 413

\bibitem[{Percival {et~al.}(2007)Percival, Cole, Eisenstein, Nichol, Peacock,
  Pope, \& Szalay}]{PercivalEtal2007}
Percival, W.~J., Cole, S., Eisenstein, D.~J., {et~al.} 2007, \mnras, 381, 1053

\bibitem[{{Pezzotta} {et~al.}(2017){Pezzotta}, {de la Torre}, {Bel}, {Granett},
  {Guzzo}, {Peacock}, {Garilli}, {Scodeggio}, {Bolzonella}, {Abbas}, {Adami},
  {Bottini}, {Cappi}, {Cucciati}, {Davidzon}, {Franzetti}, {Fritz}, {Iovino},
  {Krywult}, {Le Brun}, {Le F{\`e}vre}, {Maccagni}, {Ma{\l}ek}, {Marulli},
  {Polletta}, {Pollo}, {Tasca}, {Tojeiro}, {Vergani}, {Zanichelli}, {Arnouts},
  {Branchini}, {Coupon}, {De Lucia}, {Koda}, {Ilbert}, {Mohammad}, {Moutard},
  \& {Moscardini}}]{PezzottaEtal2017}
{Pezzotta}, A., {de la Torre}, S., {Bel}, J., {et~al.} 2017, \aap, 604, A33

\bibitem[{{Philcox} \& {Ivanov}(2022)}]{PhilcoxIvanov2022}
{Philcox}, O. H.~E. \& {Ivanov}, M.~M. 2022, \prd, 105, 043517

\bibitem[{{Potter} {et~al.}(2017){Potter}, {Stadel}, \&
  {Teyssier}}]{PotterStadelTeyssier2017}
{Potter}, D., {Stadel}, J., \& {Teyssier}, R. 2017, Computational Astrophysics
  and Cosmology, 4, 2

\bibitem[{{Pozzetti} {et~al.}(2016){Pozzetti}, {Hirata}, {Geach}, {Cimatti},
  {Baugh}, {Cucciati}, {Merson}, {Norberg}, \& {Shi}}]{PozzettiEtal2016}
{Pozzetti}, L., {Hirata}, C.~M., {Geach}, J.~E., {et~al.} 2016, \aap, 590, A3

\bibitem[{{Pugno} {et~al.}(2025){Pugno}, {Eggemeier}, {Porciani}, \&
  {Kuruvilla}}]{PugnoEtal2024a}
{Pugno}, A., {Eggemeier}, A., {Porciani}, C., \& {Kuruvilla}, J. 2025, JCAP,
  01, 075

\bibitem[{{S{\'a}nchez} {et~al.}(2013){S{\'a}nchez}, {Kazin}, {Beutler},
  {Chuang}, {Cuesta}, {Eisenstein}, {Manera}, {Montesano}, {Nichol},
  {Padmanabhan}, {Percival}, {Prada}, {Ross}, {Schlegel}, {Tinker}, {Tojeiro},
  {Weinberg}, {Xu}, {Brinkmann}, {Brownstein}, {Schneider}, \&
  {Thomas}}]{SanchezEtal2013}
{S{\'a}nchez}, A.~G., {Kazin}, E.~A., {Beutler}, F., {et~al.} 2013, \mnras,
  433, 1202

\bibitem[{{S{\'a}nchez} {et~al.}(2017){S{\'a}nchez}, {Scoccimarro}, {Crocce},
  {Grieb}, {Salazar-Albornoz}, {Dalla Vecchia}, {Lippich}, {Beutler},
  {Brownstein}, {Chuang}, {Eisenstein}, {Kitaura}, {Olmstead}, {Percival},
  {Prada}, {Rodr{\'{\i}}guez-Torres}, {Ross}, {Samushia}, {Seo}, {Tinker},
  {Tojeiro}, {Vargas-Maga{\~n}a}, {Wang}, \& {Zhao}}]{SanchezEtal2017b}
{S{\'a}nchez}, A.~G., {Scoccimarro}, R., {Crocce}, M., {et~al.} 2017, \mnras,
  464, 1640

\bibitem[{Scoccimarro(2000)}]{Scoccimarro2000A}
Scoccimarro, R. 2000, \apj, 542, 1

\bibitem[{Scoccimarro {et~al.}(1999)Scoccimarro, Couchman, \&
  Frieman}]{ScoccimarroCouchmanFrieman1999}
Scoccimarro, R., Couchman, H. M.~P., \& Frieman, J.~A. 1999, \apj, 517, 531

\bibitem[{Scoccimarro {et~al.}(2004)Scoccimarro, Sefusatti, \&
  Zaldarriaga}]{ScoccimarroSefusattiZaldarriaga2004}
Scoccimarro, R., Sefusatti, E., \& Zaldarriaga, M. 2004, \prd, 69, 103513

\bibitem[{Sefusatti {et~al.}(2006)Sefusatti, Crocce, Pueblas, \&
  Scoccimarro}]{SefusattiEtal2006}
Sefusatti, E., Crocce, M., Pueblas, S., \& Scoccimarro, R. 2006, \prd, 74,
  023522

\bibitem[{Senatore \& Zaldarriaga(2015)}]{SenatoreZaldarriaga2015}
Senatore, L. \& Zaldarriaga, M. 2015, JCAP, 02, 013

\bibitem[{{Seo} \& {Eisenstein}(2003)}]{SeoEisenstein2003}
{Seo}, H.-J. \& {Eisenstein}, D.~J. 2003, \apj, 598, 720

\bibitem[{{Sheth} {et~al.}(2013){Sheth}, {Chan}, \&
  {Scoccimarro}}]{ShethChanScoccimarro2013}
{Sheth}, R.~K., {Chan}, K.~C., \& {Scoccimarro}, R. 2013, \prd, 87, 083002

\bibitem[{{Slepian} \& {Eisenstein}(2015)}]{SlepianEisenstein2015B}
{Slepian}, Z. \& {Eisenstein}, D.~J. 2015, \mnras, 454, 4142

\bibitem[{{Slepian} \& {Eisenstein}(2017)}]{SlepianEisenstein2017}
{Slepian}, Z. \& {Eisenstein}, D.~J. 2017, \mnras, 469, 2059

\bibitem[{{Slepian} \& {Eisenstein}(2018)}]{SlepianEisenstein2018}
{Slepian}, Z. \& {Eisenstein}, D.~J. 2018, \mnras, 478, 1468

\bibitem[{{Slepian} {et~al.}(2018){Slepian}, {Eisenstein}, {Blazek},
  {Brownstein}, {Chuang}, {Gil-Mar{\'\i}n}, {Ho}, {Kitaura}, {McEwen},
  {Percival}, {Ross}, {Rossi}, {Seo}, {Slosar}, \&
  {Vargas-Maga{\~n}a}}]{SlepianEtal2018}
{Slepian}, Z., {Eisenstein}, D.~J., {Blazek}, J.~A., {et~al.} 2018, \mnras,
  474, 2109

\bibitem[{{Slepian} {et~al.}(2017){Slepian}, {Eisenstein}, {Brownstein},
  {Chuang}, {Gil-Mar{\'{\i}}n}, {Ho}, {Kitaura}, {Percival}, {Ross}, {Rossi},
  {Seo}, {Slosar}, \& {Vargas-Maga{\~n}a}}]{SlepianEtal2017}
{Slepian}, Z., {Eisenstein}, D.~J., {Brownstein}, J.~R., {et~al.} 2017, \mnras,
  469, 1738

\bibitem[{{Smith}(2009)}]{SmithEtal2008}
{Smith}, R.~E. 2009, \mnras, 400, 851

\bibitem[{Smith {et~al.}(2007)Smith, Scoccimarro, \&
  Sheth}]{SmithScoccimarroSheth2007}
Smith, R.~E., Scoccimarro, R., \& Sheth, R.~K. 2007, \prd, 75, 063512

\bibitem[{{Sugiyama} {et~al.}(2021){Sugiyama}, {Saito}, {Beutler}, \&
  {Seo}}]{SugiyamaEtal2021}
{Sugiyama}, N.~S., {Saito}, S., {Beutler}, F., \& {Seo}, H.-J. 2021, \mnras,
  501, 2862

\bibitem[{{Szalay}(1988)}]{Szalay1988}
{Szalay}, A.~S. 1988, \apj, 333, 21

\bibitem[{Szapudi \& Szalay(1998)}]{SzapudiSzalay1997}
Szapudi, I. \& Szalay, A.~S. 1998, ApJ, 494, L41

\bibitem[{{Tr{\"o}ster} {et~al.}(2020){Tr{\"o}ster}, {S{\'a}nchez}, {Asgari},
  {Blake}, {Crocce}, {Heymans}, {Hildebrandt}, {Joachimi}, {Joudaki},
  {Kannawadi}, {Lin}, \& {Wright}}]{TrosterEtal2020}
{Tr{\"o}ster}, T., {S{\'a}nchez}, A.~G., {Asgari}, M., {et~al.} 2020, \aap,
  633, L10

\bibitem[{{Umeh}(2021)}]{Umeh2021}
{Umeh}, O. 2021, JCAP, 05, 035

\bibitem[{Verde {et~al.}(2000)Verde, Wang, Heavens, \&
  Kamionkowski}]{VerdeEtal2000}
Verde, L., Wang, L., Heavens, A.~F., \& Kamionkowski, M. 2000, \mnras, 313, 141

\bibitem[{{Veropalumbo} {et~al.}(2022){Veropalumbo}, {Binetti}, {Branchini},
  {Moresco}, {Monaco}, {Oddo}, {S{\'a}nchez}, \&
  {Sefusatti}}]{VeropalumboEtal2022}
{Veropalumbo}, A., {Binetti}, A., {Branchini}, E., {et~al.} 2022, JCAP, 09, 033

\bibitem[{{Veropalumbo} {et~al.}(2021){Veropalumbo}, {S{\'a}ez Casares},
  {Branchini}, {Granett}, {Guzzo}, {Marulli}, {Moresco}, {Moscardini},
  {Pezzotta}, \& {de la Torre}}]{VeropalumboEtal2021}
{Veropalumbo}, A., {S{\'a}ez Casares}, I., {Branchini}, E., {et~al.} 2021,
  \mnras, 507, 1184

\bibitem[{{Vlah} {et~al.}(2016){Vlah}, {Seljak}, {Yat Chu}, \&
  {Feng}}]{VlahEtal2016}
{Vlah}, Z., {Seljak}, U., {Yat Chu}, M., \& {Feng}, Y. 2016, JCAP, 03, 057

\bibitem[{{Wang}(2008)}]{Wang2008}
{Wang}, Y. 2008, \prd, 77, 123525

\bibitem[{{Wang} {et~al.}(2017){Wang}, {Zhao}, {Chuang}, {Ross}, {Percival},
  {Gil-Mar{\'{\i}}n}, {Cuesta}, {Kitaura}, {Rodriguez-Torres}, {Brownstein},
  {Eisenstein}, {Ho}, {Kneib}, {Olmstead}, {Prada}, {Rossi}, {S{\'a}nchez},
  {Salazar-Albornoz}, {Thomas}, {Tinker}, {Tojeiro}, {Vargas-Maga{\~n}a}, \&
  {Zhu}}]{WangEtal2017}
{Wang}, Y., {Zhao}, G.-B., {Chuang}, C.-H., {et~al.} 2017, \mnras, 469, 3762

\bibitem[{{Zhao} {et~al.}(2017){Zhao}, {Wang}, {Saito}, {Wang}, {Ross},
  {Beutler}, {Grieb}, {Chuang}, {Kitaura}, {Rodriguez-Torres}, {Percival},
  {Brownstein}, {Cuesta}, {Eisenstein}, {Gil-Mar{\'{\i}}n}, {Kneib}, {Nichol},
  {Olmstead}, {Prada}, {Rossi}, {Salazar-Albornoz}, {Samushia}, {S{\'a}nchez},
  {Thomas}, {Tinker}, {Tojeiro}, {Weinberg}, \& {Zhu}}]{ZhaoEtal2017}
{Zhao}, G.-B., {Wang}, Y., {Saito}, S., {et~al.} 2017, \mnras, 466, 762

\end{thebibliography}
